\documentclass[longauth]{aa}  

\usepackage{upgreek}
\usepackage[varg]{txfonts}
\usepackage{booktabs} 
\usepackage{orcidlink}
\usepackage{enumitem}
\makeatletter
\let\do@mlinenumbers\relax
\let\linenumbers\relax
\let\endlinenumbers\relax
\let\linenumbersep\relax
\let\thelinenumber\relax
\let\makeLineNumber\relax
\makeatother

\begin{document} 

   \title{The Hot-Neptune Initiative (HONEI)}
   \subtitle{I. Two hot sub-Neptunes on a close-in, eccentric orbit (TOI-5800\,b) and a farther-out, circular orbit (TOI-5817\,b)} 

   \author{L. Naponiello\orcidlink{0000-0001-9390-0988}\inst{1}
    \and
    S. Vissapragada\orcidlink{0000-0003-2527-1475}\inst{2} 
    \and
    A.\,S. Bonomo\inst{1} 
    \and
    M.-L. Steinmeyer\orcidlink{0000-0003-0605-0263}\inst{3} 
    \and
    S. Filomeno\inst{4,5,6} 
    \and
    V. D'Orazi\inst{5,7,8} 
    \and
    C. Dorn\orcidlink{0000-0001-6110-4610}\inst{3} 
    \and
    A. Sozzetti\inst{1} 
    \and
    L. Mancini\inst{4,1,9} 
    \and
    A.\,F. Lanza\orcidlink{0000-0001-5928-7251}\inst{10} 
    \and
    K. Biazzo\inst{5} 
    \and
    C.\,N. Watkins\orcidlink{0000-0001-8621-6731}\inst{11} 
    \and
    G. H\'{e}brard\inst{12,13} 
    \and
    J.\,J. Lissauer\inst{14} 
    \and
    S.~B. Howell\orcidlink{0000-0002-2532-2853}\inst{14} 
    \and
    D.~R. Ciardi\orcidlink{0000-0002-5741-3047}\inst{15} 
    \and
    G. Mantovan\inst{16,7} 
    \and 
    D. Baker\orcidlink{0000-0002-2970-0532}\inst{17} 
    \and
    V. Bourrier\inst{18} 
    \and
    L.~A. Buchhave\orcidlink{0000-0003-1605-5666}\inst{19} 
    \and
    C.~A. Clark\orcidlink{0000-0002-2361-5812}\inst{15} 
    \and
    K.\,A. Collins\orcidlink{0000-0001-6588-9574}\inst{10} 
    \and
    R. Cosentino\inst{20} 
    \and
    M. Damasso\inst{1} 
    \and
    X. Dumusque\inst{18} 
    \and
    A. Fiorenzano\orcidlink{0000-0002-4272-4272}\inst{20} 
    \and
    T. Forveille\inst{21} 
    \and
    N. Heidari\inst{12} 
    \and
    D.~W. Latham\inst{11} 
    \and
    C. Littlefield\orcidlink{0000-0001-7746-5795}\inst{22,14} 
    \and
    M. L\'{o}pez-Morales\inst{23} 
    \and
    M.~B. Lund\orcidlink{0000-0003-2527-1598}\inst{15} 
    \and
    L. Malavolta\inst{24, 7} 
    \and
    F. Manni\inst{4,1} 
    \and
    D. Nardiello\inst{24, 7} 
    \and
    M. Pinamonti\inst{1} 
    \and
    S.~W. Yee\inst{11} 
    \and
    R. Zambelli\inst{25} 
    \and
    C. Ziegler\inst{26} 
    \and
    T. Zingales\inst{24, 7} 
          %
          }

  \institute{INAF -- Osservatorio Astrofisico di Torino,
              Via Osservatorio 20, 10025 Pino Torinese, Italy\\
              \email{luca.naponiello@inaf.it}
            \and 
            Carnegie Science Observatories, 813 Santa Barbara Street, Pasadena, CA 91101, USA
            \and 
            Institute for Particle Physics and Astrophysics, ETH Zürich, Otto-Stern-Weg 5, 8093 Zürich, Switzerland
            \and 
            INAF -- Osservatorio Astronomico di Roma, Monte P. Catone, Italy
            \and 
            Department of Physics, University of Rome ``Tor Vergata'', Via della Ricerca Scientifica 1, 00133 Rome, Italy
            \and 
            La Sapienza University of Rome, Department of Physics, Piazzale Aldo Moro 2, 00185, Rome, Italy
            \and 
            INAF -- Osservatorio Astronomico di Padova, Vicolo dell’Osservatorio 5, I-35122 Padova, Italy
            \and 
            Fulbright Visiting Research Scholar, Dept. of Astronomy, The University of Texas at Austin, Speedway 2515, Austin, Texas, USA
            \and 
            Max-Planck-Institut für Astronomie, Heidelberg, Germany
            \and 
            INAF -- Osservatorio Astrofisico di Catania, Via S. Sofia 78, 95123 Catania, Italy
            \and 
            Center for Astrophysics \textbar \ Harvard \& Smithsonian, 60 Garden Street, Cambridge, MA 02138, USA
            \and 
            Institut d'astrophysique de Paris, UMR7095 CNRS, Universit\'e Pierre \& Marie Curie, 98bis boulevard Arago, 75014 Paris, France 
            \and 
            Observatoire de Haute-Provence, CNRS, Universit\'e d'Aix-Marseille, 04870 Saint-Michel-l'Observatoire, France
            \and 
            NASA Ames Research Center, Moffett Field, CA 94035, USA
            \and 
            NASA Exoplanet Science Institute, IPAC, California Institute of Technology, Pasadena, CA 91125 USA
            \and 
            Centro di Ateneo di Studi e Attivit\`a Spaziali ``G. Colombo'' -- Universit\`a di Padova, Via Venezia 15, IT-35131, Padova, Italy
            \and 
            Physics Department, Austin College, Sherman, TX 75090, USA
            \and 
            Observatoire Astronomique de l’Universit\'e de Gen\'eve, Chemin Pegasi 51, Versoix 1290, Switzerland
            \and 
            DTU Space,  Technical University of Denmark, Elektrovej 328, DK-2800 Kgs. Lyngby, Denmark
            \and 
            Fundaci\'{o}n Galileo Galilei - INAF, Rambla Jos\'{e} Ana Fernandez P\'{e}rez 7, 38712 Bre\~{n}a Baja, TF, Spain
            \and 
            Universit\'e Grenoble Alpes, CNRS, IPAG, 38000 Grenoble, France
            \and 
            Bay Area Environmental Research Institute, Moffett Field, CA 94035, USA
            \and 
            Space Telescope Science Institute, 3700 San Martin Drive, Baltimore, MD 21218, USA
            \and 
            Dipartimento di Fisica e Astronomia ``Galileo Galilei'', Università di Padova, Vicolo dell’Osservatorio 3, I-35122 Padova, Italy
            \and 
            Società Astronomica Lunae, Castelnuovo Magra, Italy
            \and 
            Dept. of Physics, Engineering and Astronomy, Stephen F. Austin State University, 1936 North St, Nacogdoches, TX 75962, USA
            %
             }

   \date{Received 15, May 2025; accepted 28, July 2025}

 
  \abstract
    {Neptune-sized exoplanets are key targets for atmospheric studies, yet their formation and evolution remain poorly understood due to their diverse characteristics and limited sample size. 
    The so-called ``Neptune desert'', a region of parameter space with a dearth of short-period sub- to super-Neptunes, is a critical testbed for theories of atmospheric escape and migration.}
    {The HONEI program aims to confirm and characterize the best Neptune-sized candidates for composition, atmospheric and population studies. By measuring planetary masses with high precision, we want to provide the community with optimal targets whose atmosphere can be effectively explored with the James Webb Space Telescope or by ground-based high-resolution spectroscopy.}
    {For this purpose, we started a radial velocity follow-up campaign, using the twin high-precision spectrographs HARPS and HARPS-N, to measure the masses of TESS Neptune-sized candidates and confirm their planetary nature.}
    {In this first paper of the series, we confirm the planetary nature of two candidates: TOI-5800\,b and TOI-5817\,b. TOI-5800\,b is a hot sub-Neptune ($R_{\rm p}=2.46^{+0.18}_{-0.16}\,R_{\oplus}$, $M_{\rm p}=9.5^{+1.7}_{-1.9}\,M_{\oplus}$, $\rho=3.46^{+1.02}_{-0.90}$\,g\,cm$^{-3}$, $T_{\rm eq}=1108\pm20$ K) located at the lower edges of the Neptune desert ($P=2.628$\,days) and is the most eccentric planet ($e\sim0.3$) ever found within $P<3$\,d. TOI-5800\,b is expected to be still in the tidal migration phase with its parent star, a K3\,V dwarf ($V=9.6$\,mag), although its eccentricity could arise from interactions with another object in the system. Having a high-transmission spectroscopy metric ($\mathrm{TSM}=103^{+35}_{-22}$), it represents a prime target for future atmospheric characterization. TOI-5817\,b is a relatively hot sub-Neptune ($R_{\rm p}=3.08\pm0.14\,R_{\oplus}$, $M_{\rm p}=10.3^{+1.4}_{-1.3}\,M_{\oplus}$, $\rho=1.93^{+0.41}_{-0.34}$\,g\,cm$^{-3}$, $T_{\rm eq}=950^{+21}_{-18}$ K) located in the Neptune savanna ($P=15.610$\,d), on a circular orbit around a bright G2\,IV-V star ($V=8.7$\,mag). Despite a lower $\mathrm{TSM}=56^{+11}_{-9}$, it is a potential target for atmospheric follow-up in the context of sub-Neptunes with $P>15$\,days. Finally, we find that if the difference in the planet densities are mainly due to different gas mass fractions, there would be an order of magnitude difference in the predicted atmospheric carbon-to-oxygen ratios, a prediction that can be tested with atmospheric observations follow-up.}
    {}


   \maketitle
%

\section{Introduction}\label{sec:intro}
Neptune-sized exoplanets ($3\,R_\oplus \lesssim R_{\rm p} \lesssim 7\,R_\oplus$) are unique probes of planetary evolution. Planets in this size range appear to belong to one of three populations, with relatively clear distinctions in system properties, depending on their orbital periods: 
\begin{itemize}[leftmargin=*, label=$\bullet$, itemsep=0pt, parsep=0pt]
    \item[--] Short-period ($P\lesssim3$~d) Neptunes are rare and manifest in a pronounced ``Neptune desert'' in the planetary $R_{\rm p}-P$ and $M_{\rm p}-P$ planes \citep{Latham2011, Szabo2011, Lundkvist2016, Mazeh2016}, which is attributed to both photoevaporative mass loss and dynamical migration \citep{Matsakos2016, Owen2018, Vissapragada2022}. Planets in the Neptune desert tend to orbit metal-rich stars \citep{Vissapragada2025, Doyle2025} and to be exceptionally dense, with extreme examples, such as TOI-849\,b, TOI-332\,b, and TOI-1853\,b, reaching even $\rho\sim10$~g~cm$^{-3}$ \citep{Armstrong2020, Osborn2023, Naponiello2023}. \\ [-10pt]
    \item[--] Between $3~\mathrm{d}\lesssim P\lesssim6~\mathrm{d}$ the occurrence of these planets sharply peaks in a ``Neptune ridge'' \citep{Castro2024}, somewhat analogous to the ``3-day pileup'' in hot Jupiters \citep{Dawson2018}. Planets in the Neptune ridge also tend to orbit metal-rich stars \citep{Dong2018, Petigura2018, Petigura2022, Vissapragada2025}, but with more modest densities $\rho\lesssim2$~g\,cm$^{-3}$ \citep{Castro2024b}, greater eccentricities (i.e. not fully damped by tides; \citealt{Correia2020}), and misaligned orbits with respect to the stellar spin axis \citep{Bourrier2023}. \\ [-10pt]
    \item[--] For $P\gtrsim6$~d the occurrence flattens off into the ``Neptune savanna'' \citep{Bourrier2023, Castro2024}, where planets do \textit{not} exhibit a host star metallicity preference \citep{Dong2018, Petigura2018, Dai2021, Vissapragada2025}, have relatively low eccentricities \citep{Correia2020}, and have even lower densities than the ridge planets, i.e. $\rho\lesssim1$~g\,cm$^{-3}$ \citep{Castro2024b}.
\end{itemize}
These sharp distinctions in occurrence and physical/orbital properties are suggestive of specific evolutionary histories, making Neptunes highly compelling targets for atmospheric characterization. One hypothesis is that the savanna planets may have formed relatively close to their current positions and/or experienced a relatively quiescent disk-driven migration, whereas ridge and desert planets are the product of high-eccentricity migration, with the desert planets suffering catastrophic envelope loss in this process \citep{Bourrier2018,Vissapragada2025}. If this is true, planets in the desert, ridge, and savanna should have distinct atmospheric abundances reflecting their divergent evolutionary trajectories \citep{Oberg2011, Madhusudhan2017, Booth2017, Reggiani2022, Penzlin2024, Kempton2024, Kirk2025}.

To expand the sample of planets with strong predicted atmospheric signals for testing atmospheric difference hypotheses, we recently launched the Hot Neptune Initiative (HONEI), which aims to confirm Neptune-sized candidates suitable for atmospheric observations. The selection of the targets for HONEI was based on the full list of TESS Objects of Interest (TOI) with estimated sizes of $3\,R_\oplus \leq R_{\rm p} \leq 7\,R_\oplus$ within 1$\sigma$ uncertainties on $R_{\rm p}$.
Our size range of interest includes some planets with a radius smaller than that of Neptune, as some of these can have masses greatly exceeding that of Neptune \citep[similar to the dense planets found in the Neptune desert;][]{Armstrong2020, Osborn2023, Naponiello2023}. We restricted our mass measurement follow-up efforts to targets with Transmission Spectroscopy Metric \citep[TSM;][]{Kempton2018} values greater than 70\footnote{As estimated on the \href{https://exofop.ipac.caltech.edu/tess/view_toi.php}{ExoFOP webpage}.} and orbiting bright stars ($J < 11$\,mag), as these are the best targets for follow-up observations with the spectrographs aboard JWST and ground-based high-resolution spectrographs such as CRIRES+ and WINERED \citep{Dorn2023, Otsubo2024}. We vetted all light curves to remove likely false positives and decided to only pursue targets that had been cleared for precision radial velocity (RV) work by small-aperture telescopes. For targets that pass all of these criteria, our aim is to measure the planetary masses with precision better than 20\%, as this is required for precise constraints on atmospheric parameters \citep{Batalha2019} and internal composition models \citep{dorn_generalized_2017}.

In this paper, we report the first results from the HONEI program, confirming the two smallest planets in our target list: a sub-Neptune orbiting TOI-5800 and another sub-Neptune around TOI-5817. Both candidate planets had estimated sizes of $3\,R_\oplus \leq R_{\rm p} \leq 7\,R_\oplus$ (within 1$\sigma$ uncertainties on $R_{\rm p}$) and estimated TSM > 70 at the beginning of the program, although we found that TOI-5800\,b is somewhat smaller ($2.44^{+0.29}_{-0.20}\,R_\oplus$), and TOI-5817\,b has a somewhat lower TSM ($56^{+11}_{-9}$). The structure of the paper is as follows. In Sect.\,\ref{sec:obs}, we present the TESS and all the ground-based observations we have collected. Sect.\,\ref{sec:analysis} details the characterization of the two stars and their planets. In Sect.\,\ref{sec:discussion}, we discuss the properties of the systems and, finally, we summarize our conclusions in Sect.\,\ref{sec:conclusions}.


\section{Observations and data reduction}\label{sec:obs}
\subsection{TESS photometry}\label{sec:tess}

\begin{figure}
\centering
\includegraphics[width=0.262\textwidth]{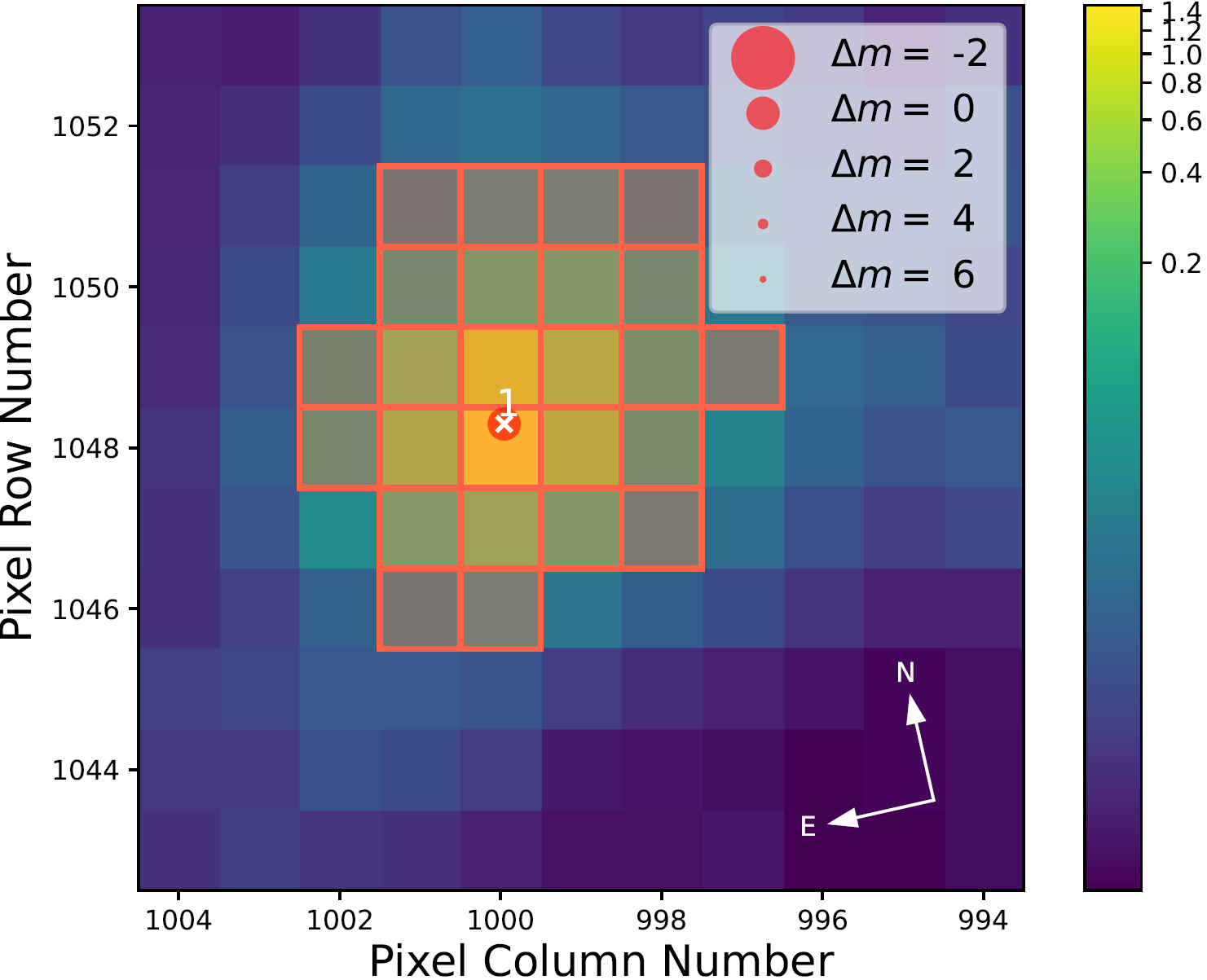}
\includegraphics[width=0.222\textwidth]{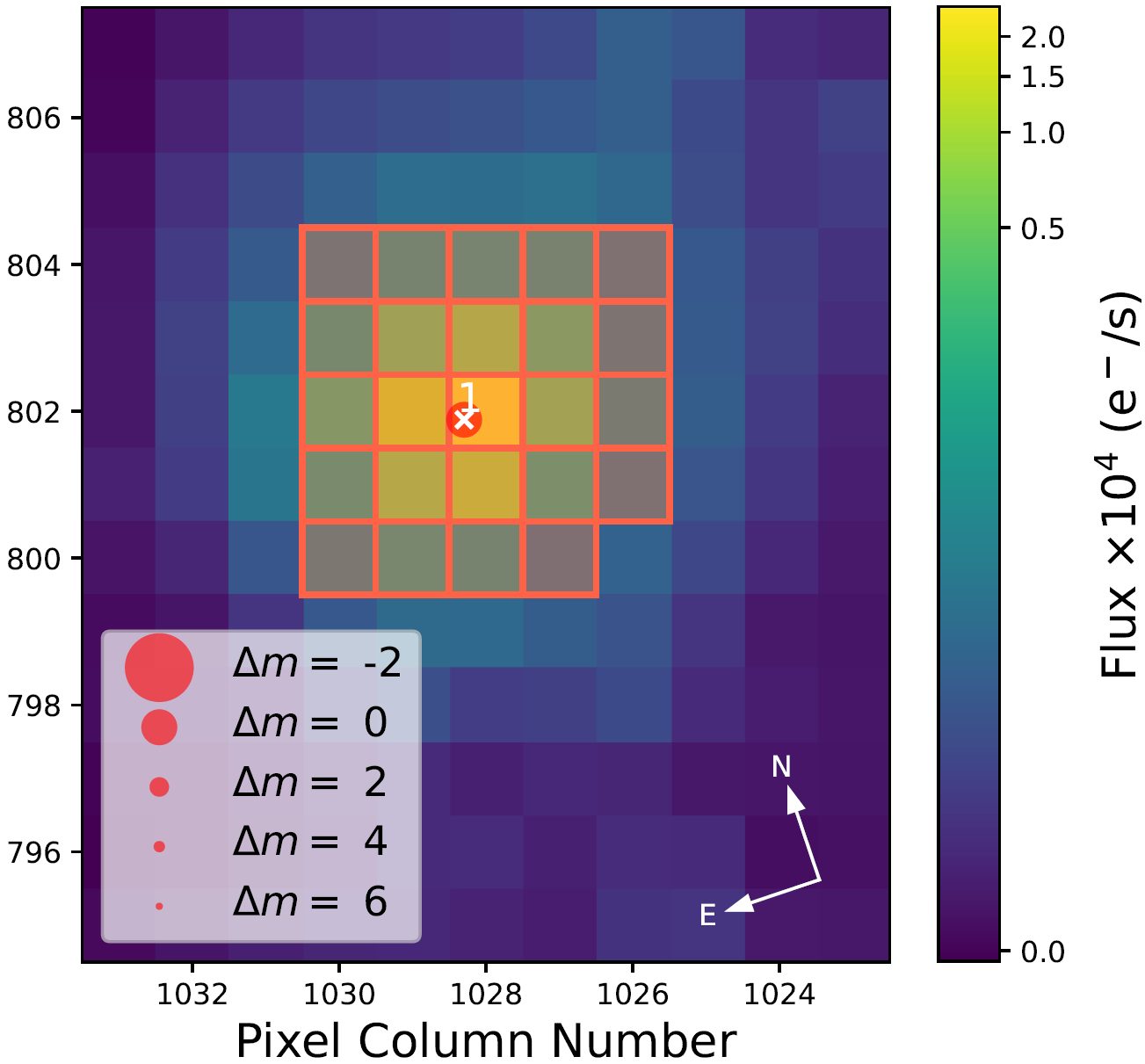}
\caption{Target pixel file from the TESS observation of Sector~54 and Sector~55, centered respectively on TOI-5800 (left panel) and TOI-5817 (right panel). The SPOC pipeline aperture is shown by shaded red squares. Both fields appear to be uncontaminated by stars within 6 {\it Gaia} magnitudes, according to the {\it Gaia} DR3 catalog \citep{Gaia2023}.}\label{fig:tpfs}
\end{figure}

\begin{figure*}
\centering
\includegraphics[width=0.9\textwidth]{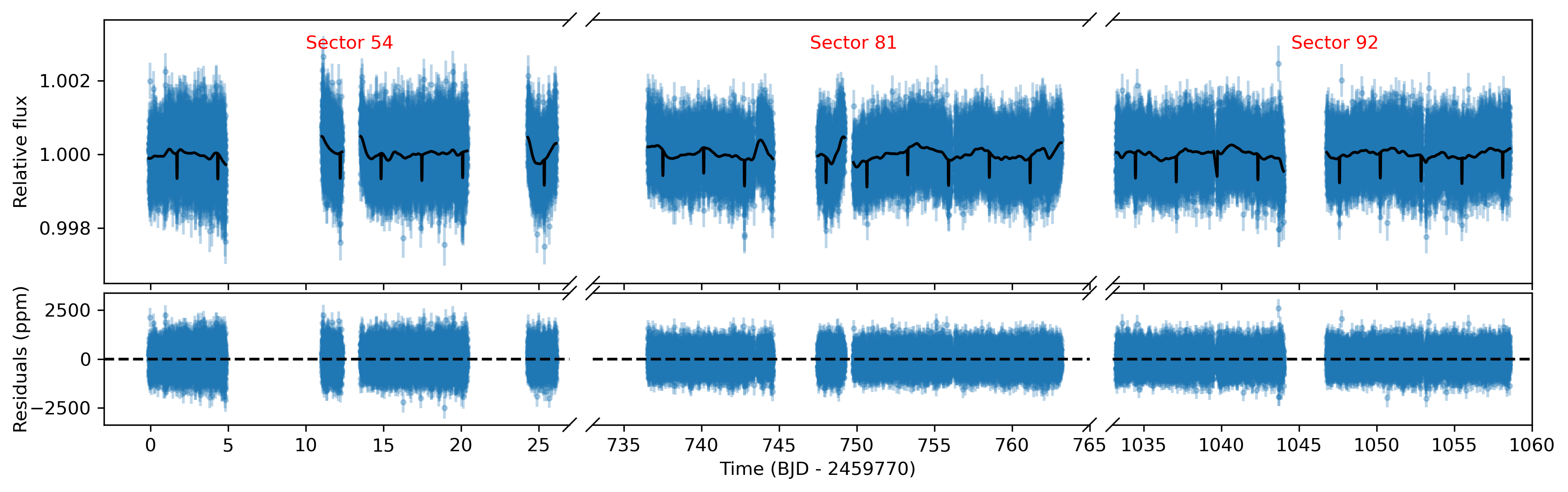}
\includegraphics[width=0.6\textwidth]{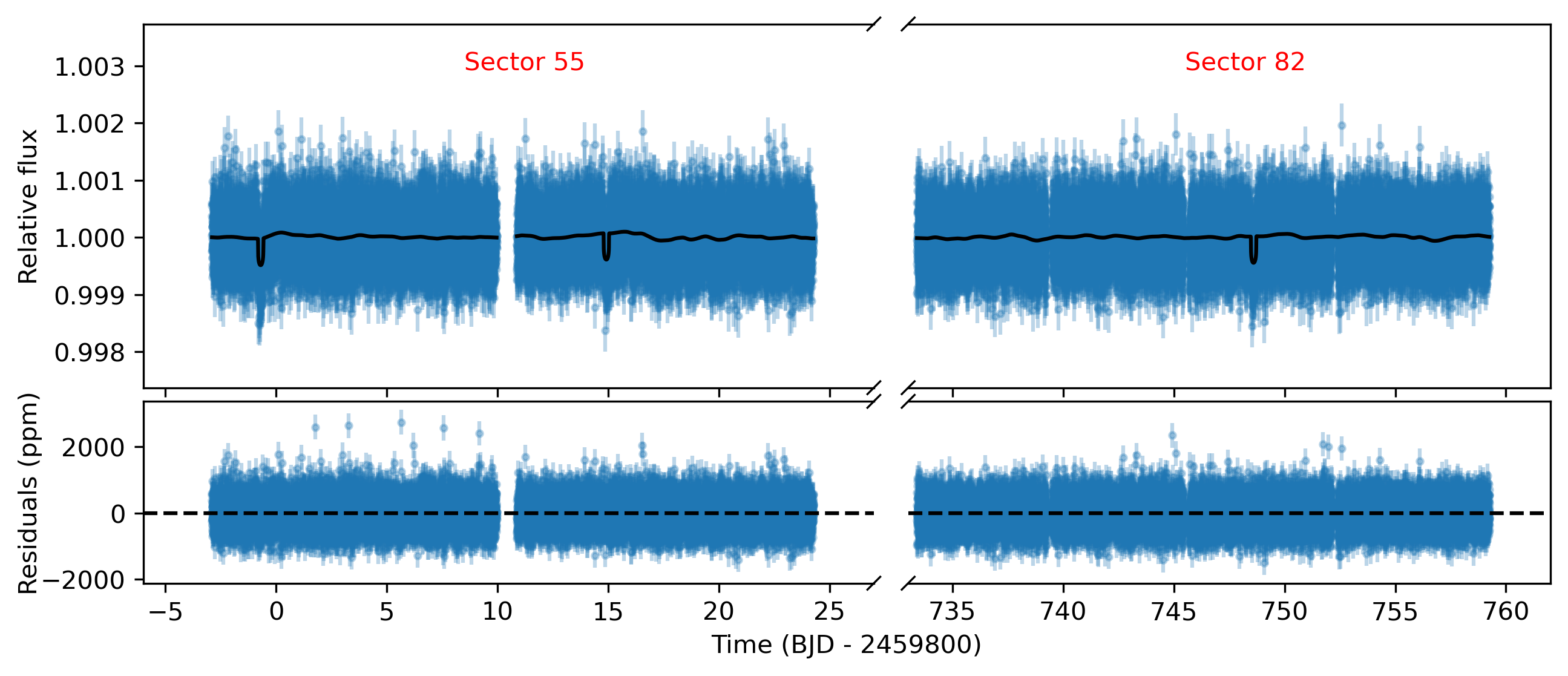}
\caption{Light curves from the PDC-SAP pipeline of TOI-5800 (top panel) and TOI-5817 (bottom panel), as collected by TESS with a 2-minute cadence (or binned to the equivalent of 2-minutes, for Sectors 81 and 92). The black line represents our best-fit transit model and detrending (Sect.\,\ref{sec:joint_analysis}), while the respective bottom panels show the residuals of the best-fit model in parts per million.}\label{fig:lc_clean}
\end{figure*}

TOI-5800 (aka HD\,193396) and TOI-5817 (aka HD\,204660) have been observed by TESS, respectively, in Sectors~54, 81, 92 (in July-August 2022, August 2024 and May 2025) for a total of 23 transits, and in Sectors~55, 82 (in August-September 2022 and 2024) for a total of only 3 transits. For the analysis detailed in Sect.\,\ref{sec:joint_analysis}, we employed the Presearch Data Conditioning Simple Aperture Photometry (PDC-SAP; \citealt{Stumpe2012, Stumpe2014}, \citealt{Smith2012}) 2-minute cadence (Sectors 54, 55, 82) and 20-second cadence (Sectors 81, 92) photometry, which are provided by the TESS Science Processing Operations Center (SPOC) \citep{Jenkins2016} pipeline and retrieved via the Python package \texttt{lightkurve} \citep{lightkurve} from the Mikulski Archive for Space Telescopes\footnote{In order to search for possible activity modulation in the photometry, we have also used the uncorrected SAP light curve.}. As no other star within 6 magnitudes of the target stars is present in both apertures (see Fig.~\ref{fig:tpfs}), the dilution coefficients for these light curves are negligible, though the PDC-SAP has already been corrected for dilution\footnote{We note that, as reported in Sect.\,\ref{sec:multiplicity}, TOI-5817 has a close companion with $\Delta T=6.4$ mag. The flux of the companion has already been taken into account for the PDC-SAP dilution correction, but, in any case, it would impact the radius of the candidate by only $\sim0.1\%$, which is very well within the uncertainty of its measurement.}. Both light curves, along with the transits of the candidates TOI-5800.01 and TOI-5817.01, are plotted in Fig.~\ref{fig:lc_clean}.

\subsection{Ground-based photometry}\label{sec:ground}

We observed a full transit window of TOI-5800.01 continuously for 299 minutes through the Johnson/Cousins $I$ band on August 6, 2024, from the Adams Observatory at Austin College in Sherman, TX. We used a $f/8$ Ritchey-Chr\'{e}tien 0.6\,m telescope, which has a focal length of 4880\,m and is equipped with an FLI ProLine $4096\times 4096$ pixel detector, with a pixel size of $9\,\mu$m. Therefore, the plate scale is $0\farcs38$ pixel$^{-1}$, resulting in a $26\arcmin\times26\arcmin$ field of view. The differential photometric data were extracted using {\tt AstroImageJ} \citep{Collins2017}. The {\tt TESS Transit Finder}, which is a customized version of the {\tt Tapir} software package \citep{Jensen2013}, was used to schedule our transit observations. We observed another full transit window of TOI-5800.01 continuously for 299 minutes in Pan-STARRS $z_{\rm s}$ band on October 23, 2024, from the Las Cumbres Observatory Global Telescope (LCOGT) \citep{Brown2013} 1\,m network node at the Teide Observatory, on the island of Tenerife (TEID). The 1\,m telescope has a focal length of 8100\,mm and is equipped with a $4096\times4096$ SINISTRO camera, with a pixel size of $15\,\mu$m. The plate scale is $0\farcs389$ per pixel, also resulting in a $26\arcmin\times26\arcmin$ field of view. The images were calibrated using the standard LCOGT {\tt BANZAI} pipeline \citep{McCully2018} and differential photometric data were extracted using {\tt AstroImageJ}. We used circular $7\farcs8$ photometric apertures that excluded all of the flux from the nearest known neighbor in the Gaia DR3 catalog (Gaia DR3 4216062669395306240), which is $12\farcs2$ north of TOI-5800.

The TOI-5800.01 SPOC pipeline transit depth of 731\,ppm is generally too shallow to reliably detect with ground-based observations, so we instead checked for possible nearby eclipsing binaries (NEBs) that could be contaminating the TESS photometric aperture and causing the TESS detection. To account for possible contamination from the wings of neighboring star PSFs, we searched for NEBs up to $2\farcm5$ from TOI-5800 in both the Adams Observatory and LCO-TEID observations. If fully blended in the SPOC aperture, a neighboring star that is fainter than the target star by 7.9 magnitudes in the TESS-band could produce the SPOC-reported flux deficit at mid-transit (assuming a 100\% eclipse). To account for possible TESS magnitude uncertainties and possible delta-magnitude differences between the TESS-band and both the Pan-STARRS $z_{\rm s}$ and Johnson/Cousins $I$ bands, we included an extra 0.5 magnitudes fainter (down to \textit{TESS}-band magnitude 16.4). We calculated the RMS of each of the 31 nearby star light curves (binned in 10-minute bins) that meet our search criteria and found that the values are smaller by at least a factor of 5 compared to the required NEB depth in each respective star. We then visually inspected each neighboring star's light curve to ensure no obvious eclipse-like signal. Our analysis ruled out an NEB blend as the cause of the TOI-5800.01 detection of the SPOC pipeline in the \textit{TESS} data. 

Furthermore, we analyzed the frequency content of the light curves ($\textit{g}$-band) taken from the ASAS-SN database\footnote{https://asas-sn.osu.edu/ -- \citealt{Shappee2014, Kochanek2017}} through generalized Lomb-Scargle (GLS; \citealt{Zechmeister2009}) periodograms. 
We found no sign of modulation for TOI-5800, while for TOI-5817 we found a significant peak occurring at $\sim$29.7\, days, and then a second one in the residuals at $\sim$14.8\,d, which do not appear in the time series of other activity diagnostics (see Sects. \ref{sec:hosts} and \ref{sec:joint_analysis}), making it unlikely that the signals are stellar in origin. We believe that, being almost coincident with 1$\times$ and 0.5$\times$ the synodic month, they are due to moonlight contamination.

\subsection{HARPS spectra}\label{sec:harps}

Between 30 June 2024 and 13 August 2024, we collected a total of 26 high-resolution spectra of TOI-5800 (Table\,\ref{tab:5800_RV}) with the High Accuracy Radial velocity Planet Searcher (HARPS; \citealt{Mayor2003}), which is operated at the ESO 3.6m telescope in the La Silla Observatory (Chile), using exposure times of 15\,min. We employed the {\tt serval} (v.2021-03-31; \citealt{Zechmeister2018}) and the {\tt actin} (v.2.0\_beta\_11; \citealt{Gomes2018}) pipelines for the RVs and the stellar activity indices extractions, respectively. In particular, the RV measurements have an average error of $1.46$\,m\,s$^{-1}$, and a median S/N\,$\approx$\,41, measured at a reference wavelength of 5500 {\AA}.

\subsection{HARPS-N spectra}\label{sec:harpsn}

Between 12 May 2023 and 7 December 2024, we collected a total of 54 high-resolution spectra of TOI-5817 (Table\,\ref{tab:5817_RV_HN}) with the High Accuracy Radial velocity Planet Searcher for the Northern hemisphere (HARPS-N; \citealt{Cosentino2012}) using an exposure time of 10\,min. The first 14 spectra were obtained within the Global Architecture of Planetary Systems (GAPS) Neptune project \citep{Naponiello2022,Damasso2023,Naponiello2025}, all the others under the HONEI program. The RVs and activity indices were extracted from HARPS-N spectra, which were reduced using version 3.0.1 of the HARPS-N Data Reduction Software ({\tt DRS}, \citealt{Dumusque2021}), available on the Data Analysis Center for Exoplanets (DACE) web platform. 
Overall, the measurements have an average uncertainty of $1.16$\,m\,s$^{-1}$, and a S/N\,$\approx$\,105, measured at a reference wavelength of 5500 {\AA}.

\subsection{SOPHIE spectra}\label{sec:sophie}
SOPHIE is a stabilized \'{e}chelle spectrograph dedicated to high-precision RV measurements \citep{Perruchot2008,Bouchy2009,Bouchy2013} mounted at the 1.93-m telescope of the Observatoire de Haute-Provence in France. HD\,204660 was part of a volume-limited survey for giant extrasolar planets (see, e.g., \citealt{Hebrard2016}) performed with SOPHIE, before being identified as the TESS transit-host TOI-5817.01. For the observations of TOI-5817, secured in 2010 and 2024, we used its high-resolution mode (resolving power $R=75\,000$) and fast readout~mode. However, a significant SOPHIE upgrade occurred in-between \citep{Bouchy2013}, which introduced a significant improvement in the accuracy of the RV measurements, as well as a possible systematic RV offset whose upper limit is of the order of 15~m/s \citep{Hebrard2016,Demangeon2021}. 

As in other studies, we actually consider the two datasets separately: SOPHIE and SOPHIE+, depending on whether the data were secured before or after the improvement, respectively. Removing a few observations with low precision, for TOI-5817 we have a data set of 11 and 29 measurements with SOPHIE and SOPHIE+, with average uncertainties $\sim5.5$\,m\,s$^{-1}$ ($4.5\times$ that of HARPS-N) and $\sim3$\,m\,s$^{-1}$ ($2.5\times$), respectively. 
The RVs were extracted with the SOPHIE pipeline, as presented by \citet{Bouchy2009} and refined by \citet{Heidari2024,Heidari2025}. 
SOPHIE RV measurements are reported in Table~\ref{tab:5817_RV_SOPHIE}, though they were not included in the final analysis of this work because they do not improve the solution. 

\subsection{High-resolution imaging}
As part of the validation and characterization process for transiting exoplanets, high-resolution imaging is one of the critical assets required. The presence of a close companion star, whether truly bound or line of sight, will provide ``third-light'' contamination of the observed transit, leading to incorrect derived properties for both the exoplanet and the host star \citep{Ciardi2015,Furlan2017,Furlan2020}. In addition, it has been shown that the presence of a close companion dilutes small-planet transits ($R_{\rm p}<1.2\,R_{\oplus}$) to the point of non-detection, thereby possibly yielding false occurrence rate predictions \citep{Lester2021}. Given that nearly half of FGK stars are in binary or multiple star systems \citep{Matson2018}, high-resolution imaging yields vital information toward our understanding of each discovered exoplanet as well as more global information on exoplanetary formation, dynamics and evolution \citep{Howell2021}.

\subsubsection{Gemini}\label{sec:gemini}
TOI-5800 was observed on July 2, 2023, using the Zorro speckle instrument on Gemini South \citep{Scott2021}. Zorro provides simultaneous speckle imaging in two bands (562\,nm and 832\,nm), delivering reconstructed images and high-precision contrast curves that constrain the presence of close companions \citep{Howell2011}. The data were reduced using our standard pipeline \citep{Howell2011}. Fig.\,\ref{fig:gemini} presents the $5\sigma$ contrast curves from the observations, along with the 832\,nm reconstructed speckle image. The contrast limits show that companions brighter than a delta magnitude of 5 at 0.02\,arcsec, and up to a delta magnitude of 9 at 1.2\,arcsec, are excluded. TOI-5800 is thus consistent with being a single star within these detection limits. At the star’s distance of 42.9\,pc, the angular resolution corresponds to projected separations ranging from $\sim$\,0.9 to 51.5\,au.

\subsubsection{Palomar}\label{sec:palomar}
Observations of TOI\,5817 were made on 2023-Jun-30 with the PHARO instrument \citep{hayward2001} on the Palomar Hale (5\,m) behind the P3K natural guide star AO system \citep{dekany2013}. The pixel scale for PHARO is $0.025$\arcsec. The Palomar data were collected in a standard 5-point quincunx dither pattern in the Br-$\gamma$ filter and $H_{cont}$ filter. The reduced science frames were combined into a single mosaic-ed image with final resolutions of $\sim 0.094$\arcsec, and $\sim 0.083$\arcsec, respectively.

The sensitivity of the final combined AO images were determined by injecting simulated sources azimuthally around the primary target every $20^\circ$ at separations of integer multiples of the central source's Full Width at Half Maximum (FWHM) \citep{furlan2017a}. The brightness of each injected source was scaled until standard aperture photometry detected it with $5\sigma$ significance. The final $5\sigma$ limit at each separation was determined from the average of all of the determined limits at that separation and the uncertainty on the limit was set by the rms dispersion of the azimuthal slices at a given radial distance.

The nearby stellar companion (TIC\,2001155314 = Gaia\,DR3\,1790807028048264448) -- located $2.88\pm0.02$\arcsec to the east ($84.0^\circ\pm0.1^\circ$) -- was detected in both near-infrared images, but no additional close-in stars were detected in agreement with the other imaging techniques.  The companion is fainter than the primary target by $\Delta H_{cont} = 4.81\pm0.01$mag and $\Delta Br_{\gamma} = 4.64\pm0.02$mag (See Fig.\,\ref{fig:palomar}).

\subsubsection{SOAR}\label{sec:soar}
We also searched for stellar companions to TOI-5817 with speckle imaging on the 4.1-m Southern Astrophysical Research (SOAR) telescope \citep{Tokovinin2018} on 4 November 2022 UT, observing in Cousins I-band, a similar visible band-pass as TESS. This observation was sensitive with $5\sigma$ detection to a 5.2-magnitude fainter star at an angular distance of 1 arcsec from the target. More details of the observations within the SOAR TESS survey are available in \citealt{Ziegler2020}. The $5\sigma$ detection sensitivity and speckle auto-correlation functions from the observations are shown in Fig.\ref{fig:soar}. 

No nearby stars were firmly detected within 3\arcsec of TOI-5817 in the SOAR observations. In particular, the 2.8\arcsec separated companion resolved by infrared observing Palomar-AO was not detected, despite delta-magnitude sensitivity limits in that separation range of approximately 6.8-magnitudes. This is consistent with a bound late-type companion star, which would be significantly fainter compared to the primary star in the visible pass band of the speckle observation.

\section{Analysis}\label{sec:analysis}
\subsection{Host-stars characterization}\label{sec:hosts}

Stellar atmospheric parameters, including effective temperature ($T_{\text{eff}}$), surface gravity ($\log{g}$), microturbulence velocity ($\xi$), and iron abundance ([Fe/H]), were derived using the Python wrapper {\tt pyMOOGi} \citep{Adamow2017} for the MOOG code \citep{sneden1973}. We utilized MARCS plane-parallel model atmospheres \citep{Gustafsson2008} along with the line lists taken from \cite{DOrazietal2020} and \cite{Biazzoetal2022}. The equivalent widths of the iron lines were measured using ARES \citep{Sousaetal2015} and subsequently inputted into the MOOG driver ({\it abfind}) to obtain abundances through force fitting. The effective temperature was determined by minimizing the slope between the excitation potential of \ion{Fe}{i} lines and the corresponding abundances, while the microturbulence was refined by minimizing the slope between abundances and reduced equivalent widths. Surface gravity was established by enforcing the ionization equilibrium and ensuring that the average abundance derived from \ion{Fe}{i} lines equaled that from \ion{Fe}{ii} lines within a tolerance of 0.05\,dex. To quantify uncertainties, we considered errors associated with the slopes mentioned above, which contributed to uncertainties in $T_{\rm eff}$ and $\xi$. For surface gravity, uncertainties were assessed by adjusting $\log{g}$ until the ionization equilibrium condition was no longer satisfied, resulting in an abundance difference greater than 0.05\,dex. We refer the reader to our previous publications for further details (e.g., \citealt{DOrazietal2020, Biazzoetal2022}). Taking into account the same line lists, codes, and grids of models mentioned above, we also measured the elemental abundances of two refractory elements (i.e. magnesium and silicon). The final values are reported in Table\,\ref{tab:star} and highlight a slight overabundance of the elemental ratios [Mg/Fe] and [Si/Fe].

After deriving the stellar atmospheric parameters, we used the {\tt EXOFASTv2} tool \citep{2017ascl.soft10003E, Eastman2019} to determine the stellar mass, radius, and age in a differential evolution Markov chain Monte Carlo Bayesian framework, by imposing Gaussian priors on $T_{\rm eff}$, [Fe/H], and the Gaia DR3 parallax. Specifically, we simultaneously modeled the spectral energy distribution (SED) of the two host stars and the MESA Isochrones and Stellar Tracks (MIST, \citealt{Paxton2015}) (see \citealt{Eastman2019} and \citealt{Naponiello2025} for more details). To sample the SED, we used the available Tycho-2 $B_{\rm T}$ and $V_{\rm T}$, 2MASS $J$, $H$, and $K_{\rm s}$, and WISE $W1$, $W2$, $W3$, and $W4$ magnitudes for both stars, and the APASS Johnson $B$ magnitude for TOI-5800 only (see Table~\ref{tab:star}). The SEDs of both TOI-5800 and TOI-5817 are shown in Fig.~\ref{fig:stellarSED} together with their best fit. 
We found $M_\star=0.778^{+0.037}_{-0.032}\, M_\odot$, $R_\star=0.773 \pm 0.023\,R_\odot$, and age $t=8.1^{+4.0}_{-4.8}$\,Gyr for TOI-5800, and $M_\star=0.970^{+0.072}_{-0.055}\,M_\odot$, $R_\star=1.427^{+0.043}_{-0.040}\,R_\odot$, and age $t=9.7^{+2.3}_{-2.5}$\,Gyr for TOI-5817 (see Table~\ref{tab:star}).

\subsection{Alternative estimation of TOI-5800 age}\label{sec:age_chromo}

While the advanced age of TOI-5817 is well constrained and consistent with the star being almost on the verge of leaving the main sequence ($\log{g_{\star}}=4.15\pm0.05$ is lower than solar gravity $\log{g_{\sun}}=4.44$, even though $M_\star \sim 1\,M_\sun$), the age of TOI-5800 is practically unconstrained from the MIST tracks, as expected for K dwarfs, owing to their slow evolution while on the main sequence. However, even though we keep the value retrieved in Sect.\,\ref{sec:hosts} as a final estimation, we can also use the chromospheric index $R^{\prime}_{\rm HK}$ to make an alternative estimation of TOI-5800 age. According to Eq.~(3) of \citet{MamajekHillenbrand08}, we obtain $\sim 4.5 \pm 2.0$~Gyr, where the uncertainty is evaluated by considering a semi-amplitude for the variation in the chromospheric index $ \Delta \log{R^{\prime}_{\rm HK}} = \pm \,0.125$ along the activity cycle, a value based on the observed variation along the solar cycle. Moreover, the chromospheric index allows us to estimate the Rossby number $Ro$ of TOI-5800 according to Eq.~(5) of \citet{MamajekHillenbrand08} that gives $Ro \sim 1.9$. Using the convective turnover time given by Eq.~(4) of \citet{Noyesetal84}, we estimate a rotation period $P_{\rm rot} \sim 42 \pm 9$~days for TOI-5800, where the uncertainty comes from the assumed amplitude of the variation in the chromospheric index along an activity cycle. The age estimated from the gyrochronology relationship of \citet{Barnes07} (his Eq. 3) is $4.45 \pm 2.0$~Gyr for the above rotation period and uncertainty.

Such an age estimate is consistent with both the projected rotational velocity, $v\sin{i_{\star}}$ (Table~\ref{tab:star}), and the age derived from the kinematics of the star according to the method proposed by \citet{Almeida2018}. Specifically, it gives an expected age of $3.7 \pm 3.0$~Gyr using the spatial velocity components of TOI-5800 that we found to be $U=-25.0 \pm 0.3$, $V = -24.6 \pm 0.2$, and $W = -0.53 \pm 0.15$~km~s$^{-1}$ computed with the method of \citet{JohnsonSoderblom87} with $U$ positive towards the Galactic centre.

\begin{table}
\centering %
\caption{Stellar parameters of TOI-5800 and TOI-5817.}\label{tab:star} %
\resizebox{1.02\hsize}{!}{
\begin{tabular}{lcccc}
\hline %
\hline  \\[-8pt]
 & Unit & Value & Value & Source \\
\hline  \\[-6pt]
TOI \dotfill & \dotfill & 5800 & 5817 & TOI catalog \\
TIC \dotfill & \dotfill & 151759246 & 418604868 & TIC \\
HD \dotfill & \dotfill & 193396 & 204660 & HD \\
HIP \dotfill & \dotfill& ... & 106097 & HIP \\
Tycho-2 \dotfill & \dotfill&  5174-53-1 & 1676-735-1 & Tycho-2 \\
2MASS \dotfill & \dotfill& {\tiny J20201568-0724422} & {\tiny J21293069+2128296} & 2MASS \\
{\it Gaia}  \dotfill & \dotfill& {\tiny 4216059714460162944} & {\tiny 1790807028049001600} &{\it Gaia}~DR3 \\ [6pt] %
$\alpha$\,(J2016.0) \dotfill & h & \,\,\,20:20:15.73 & \,\,\,21:29:30.66 & {\it Gaia}~DR3 \\
$\delta$\,(J2016.0)  \dotfill & deg & $-$07:24:42.95 & +21:28:28.25 & {\it Gaia}~DR3 \\
$\pi$ \dotfill & mas & $23.394 \pm 0.019$ & $12.447 \pm 0.022$ & {\it Gaia}~DR3 \\
$\mu_\alpha \cos{\delta}$ \dotfill & mas/yr  & $51.531\pm 0.018$ & $-10.391\pm 0.023$\,\,\,\, & {\it Gaia}~DR3 \\
$\mu_\delta$ \dotfill & mas/yr  & $-52.722\pm 0.015$\,\,\,\, & $-89.868\pm 0.013$\,\,\,\, & {\it Gaia}~DR3 \\
$d$ \dotfill & pc  & $42.745\pm 0.033$ & $80.34\pm 0.14$ & This work$^1$ \\ [6pt] %
$B_{\rm T}$ \dotfill & mag & $10.745\pm 0.048$\,\,\, & $9.404\pm 0.018$ & Tycho-2 \\ 
$B$ \dotfill & mag & $10.460\pm0.030$\,\,\, & - & APASS \\
$V_{\rm T}$ \dotfill & mag & $9.658\pm 0.028$ & $8.713\pm 0.013$ & Tycho-2 \\ 
$G$ \dotfill & mag & $9.21824\pm0.00017$ & $8.48343\pm0.00024$ & {\it Gaia}~DR3 \\
$J$ \dotfill & mag & $7.771\pm0.021$ & $7.420\pm0.023$ & 2MASS \\
$H$ \dotfill & mag & $7.284\pm0.026$ & $7.190\pm0.038$ & 2MASS \\
$K_{\rm S}$ \dotfill  & mag & $7.191\pm0.020$ & $7.084\pm0.023$ & 2MASS \\ 
$W1$ \dotfill & mag & $7.111\pm0.044$ & $6.975\pm0.054$ & AllWISE \\
$W2$ \dotfill & mag & $7.188\pm0.021$ & $7.069\pm0.020$ & AllWISE \\     
$W3$ \dotfill & mag & $7.156\pm0.017$ & $7.045\pm0.017$ & AllWISE \\ 
$W4$ \dotfill & mag & $7.095\pm0.115$ & $7.133\pm0.106$ & AllWISE \\ 
$A_V$ \dotfill & mag & $<0.087$ & $<0.099$ & This work$^1$ \\ [6pt] %
Spectral type \dotfill & & K3\,V & G2\,IV-V & This work$^3$ \\
$L_{\star}$ \dotfill & $L_{\sun}$ & $0.309\pm0.015$ & $2.05^{+0.12}_{-0.10}$ & This work$^1$ \\
$M_{\star}$ \dotfill & $M_{\sun}$ & $0.778^{+0.037}_{-0.032}$ & $0.970^{+0.072}_{-0.055}$ & This work$^1$ \\
$R_{\star}$ \dotfill & $R_{\sun}$ & $0.773 \pm 0.023$ & $1.427^{+0.043}_{-0.040}$ & This work$^1$ \\
$T_{\rm eff}$ \dotfill & K & $4850\pm78$ & $5770\pm50$ & This work$^2$\\ 
$v\sin{i_{\star}}$ \dotfill & km\,s$^{-1}$ & $1.3\pm 0.5$ & $1.2\pm 0.7$ & This work$^4$ \\
$\log g_{\star}$ \dotfill & cgs & $4.56\pm0.03$ & $4.12\pm0.04$ & This work$^1$ \\
$\log g_{\star}$ \dotfill & cgs & $4.48\pm0.08$ & $4.15\pm0.05$ & This work$^2$ \\
$\xi$ \dotfill & km\,s$^{-1}$ & $0.65\pm0.10$ & $1.05\pm0.08$ & This work$^2$ \\
${\rm [Fe/H]}$ \dotfill & dex & $-0.04\pm0.13$ & $-0.24\pm0.09$ & This work$^2$ \\
${\rm [Mg/H]}$ \dotfill & dex & $0.09\pm0.08$ & $-0.10\pm0.08$ & This work$^2$ \\
${\rm [Si/H]}$ \dotfill & dex & $0.08\pm0.07$ & $-0.13\pm0.07$ & This work$^2$ \\
$\rho_{\star}$\dotfill & g\,cm$^{-3}$ & $2.41\pm0.21$ & $0.47\pm0.05$ & This work$^1$ \\
$\log R^{\prime}_{\rm HK}$\dotfill & dex & $-4.885\pm0.007$ & $-4.950\pm0.004$ & This work$^2$ \\
Age\dotfill & Gyr & $8.1^{+4.0}_{-4.8}$ & $9.7^{+2.3}_{-2.5}$ & This work$^1$ \\
\hline %
\end{tabular}
}
\tablebib{TESS Primary Mission TOI catalog \citep{Guerrero2021}; TIC \citep{Stassun2018,Stassun2019}; HD \citep{Cannon1924}; HIP \citep{Perryman1997}; {\it Gaia} DR3 \citep{Gaia2023}; Tycho-2 \citep{hog}; APASS \citep{Henden2016}; 2MASS \citep{Cutri2003}; AllWISE \citep{Cutri2013}.}
\begin{flushleft}
\footnotemark[1]{\small From the {\tt EXOFASTv2} modeling (this work).} \\
\footnotemark[2]{\small From the HARPS-N spectral analysis (this work).} \\
\footnotemark[3]{\small According to the \citet{PecautMamajek2013} calibration [v. 2022].} \\
\footnotemark[4]{\small Refer to \citealt{Naponiello2025} for the details on its derivation.
} \\
\end{flushleft}
\end{table}

\subsection{Stellar multiplicity}\label{sec:multiplicity}

While for TOI-5800 we do not detect any stellar companion, TOI-5817 has a stellar companion, as detected in the infrared imaging, which is separated by $2.88\arcsec\pm0.02\arcsec$ at a position angle of $84.0^\circ\pm0.1^\circ$. Because of the brightness of the primary target ($K_{\rm 2MASS}=7.08$ mag), the companion star is undetected in the 2MASS survey but has been resolved by Gaia in both DR2 and DR3, with $\Delta G = 6.8$ mag. Deblending the 2MASS H and K band magnitudes, the companion star has real apparent magnitudes of $H_{\rm 2MASS} = 12.01\pm0.05$ mag and $K_{s,{\rm \,2MASS}} = 11.74\pm0.05$ mag, corresponding to an $H-K_s = 0.27\pm0.07$ mag. The companion star has the same parallax (within 1$\sigma$) of the primary target. Despite the highly significant difference 
in the declination component of proper motion, the standard requirement that the scalar proper motion difference is within 10\% of the total proper motion
($\Delta\mu < 0.1\mu$, see e.g., \citealt{Smart2019}) remains satisfied, clearing any doubts on the boundedness of the system. These objects also appear in the Gaia wide binary catalog by \citet{El-Badry2021}. The infrared magnitudes and colors of the companion star are consistent with those of an M3.5V dwarf \citep{PecautMamajek2013} separated by approximately 230 au (corresponding to an orbital period much longer than our observational baseline, which makes its RV signal hardly detectable).

Close binary system (separation $<1000$\,au) provide key tests for planet formation models due to the influence of the companion star on the protoplanetary disc. The gravitational pull of the secondary star truncates the disc, reducing its mass and lifetime \citep{Artymowicz1994,Zagaria2021}, which limits material available for planet formation. These conditions can significantly alter the formation and evolution of planets compared to single-star systems \citep{Kraus2016,Hirsch2021}, by shortening the timescale for gas accretion, reducing solid mass budgets, and changing the distribution of solids, potentially suppressing the formation of sub-Neptunes \citep{Kraus2012,Moe2021}.

\subsection{Transit and RV combined fit}\label{sec:joint_analysis}
For both datasets, we first computed the GLS periodograms of the RVs, using the Python package \texttt{astropy} v.5.2.2 \citep{Price2018}, and confirmed that the strongest signals of the respective periodograms are associated with the orbital periods of the candidates (refer to Fig.\,\ref{fig:GLS}). Then, we performed a joint transit and RV analysis using the Python wrapper \texttt{juliet}\footnote{\texttt{\url{https://juliet.readthedocs.io/en/latest/}}} \citep{Espinoza2019}, following the same approach of \citet{Naponiello2022}. Basically, for the transit-related priors we used the parameters of the data validation report produced by the TESS SPOC pipeline at the NASA Ames Research Center through the Transiting Planet Search \citep{jenkins2002,jenkins2010,Jenkins2020b} and data validation (\citealt{Twicken2018}, \citealt{Li2019}) modules (Table\,\ref{tab:priors}). Then, we mainly tested two models: one with a fixed eccentricity and another with a uniform distribution of eccentricity values, using the parametrization $S_1=\sqrt{e}\sin{\omega_*}$,\, $S_2=\sqrt{e}\cos{\omega_*}$ (see \citealt{Eastman2012}). We also tested models including additional Keplerians with or without linear and parabolic trends, with no meaningful results.

\begin{figure}
\centering
\includegraphics[width=0.42\textwidth]{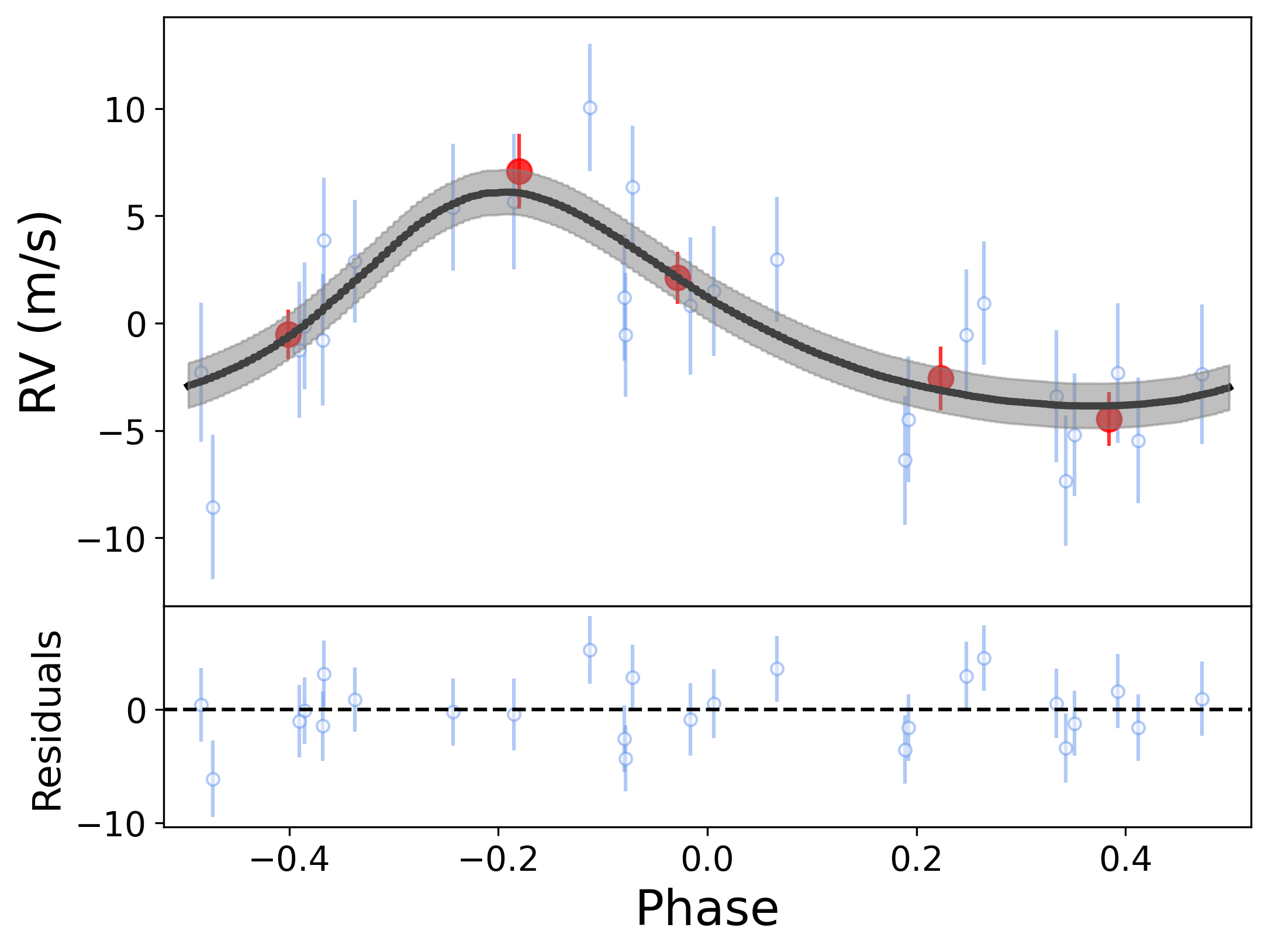}
\includegraphics[width=0.42\textwidth]{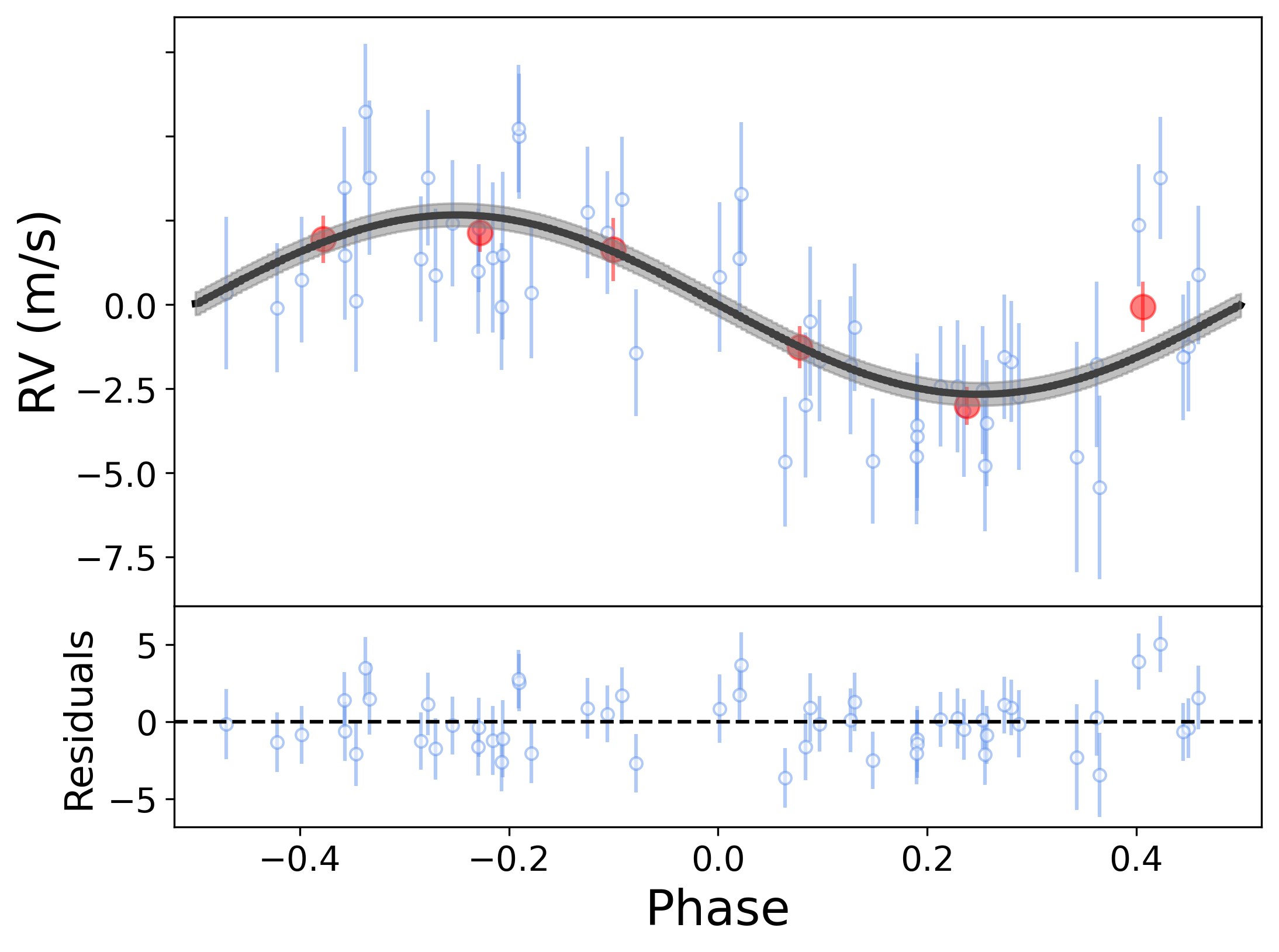}
\caption{Phased HARPS RVs of TOI-5800\,b (top) and phased HARPS-N RVs of TOI-5817\,b (bottom) along with the fitted models, in black, and their residuals below each panel. The red circles represent the average of $\sim5$ and $\sim9$ RVs respectively, while the gray areas represent the $1\sigma$ deviation from the model.}
\label{fig:5800_5817_phaseRV}
\end{figure}

\begin{figure}
\centering
\includegraphics[width=0.42\textwidth]{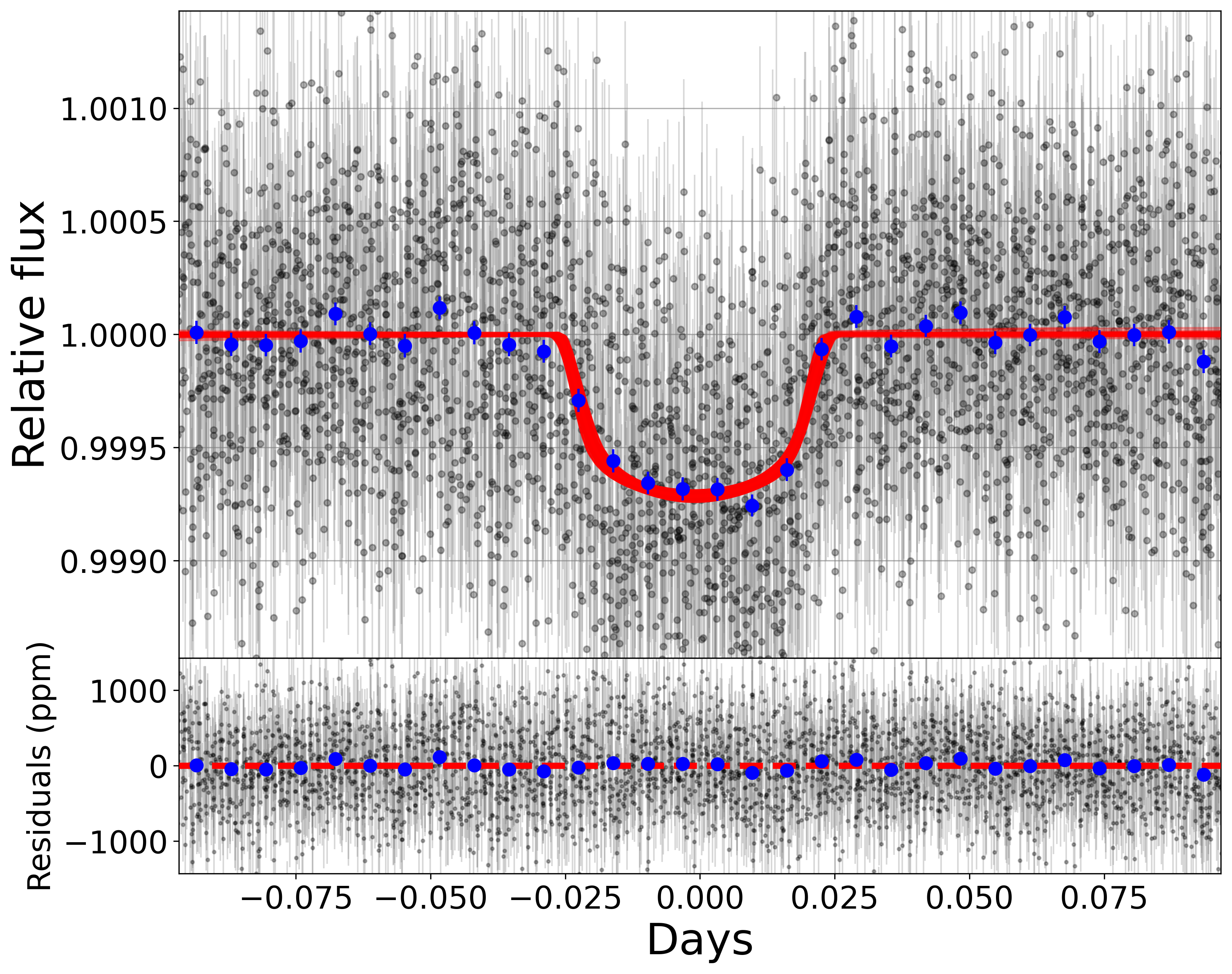}
\includegraphics[width=0.42\textwidth]{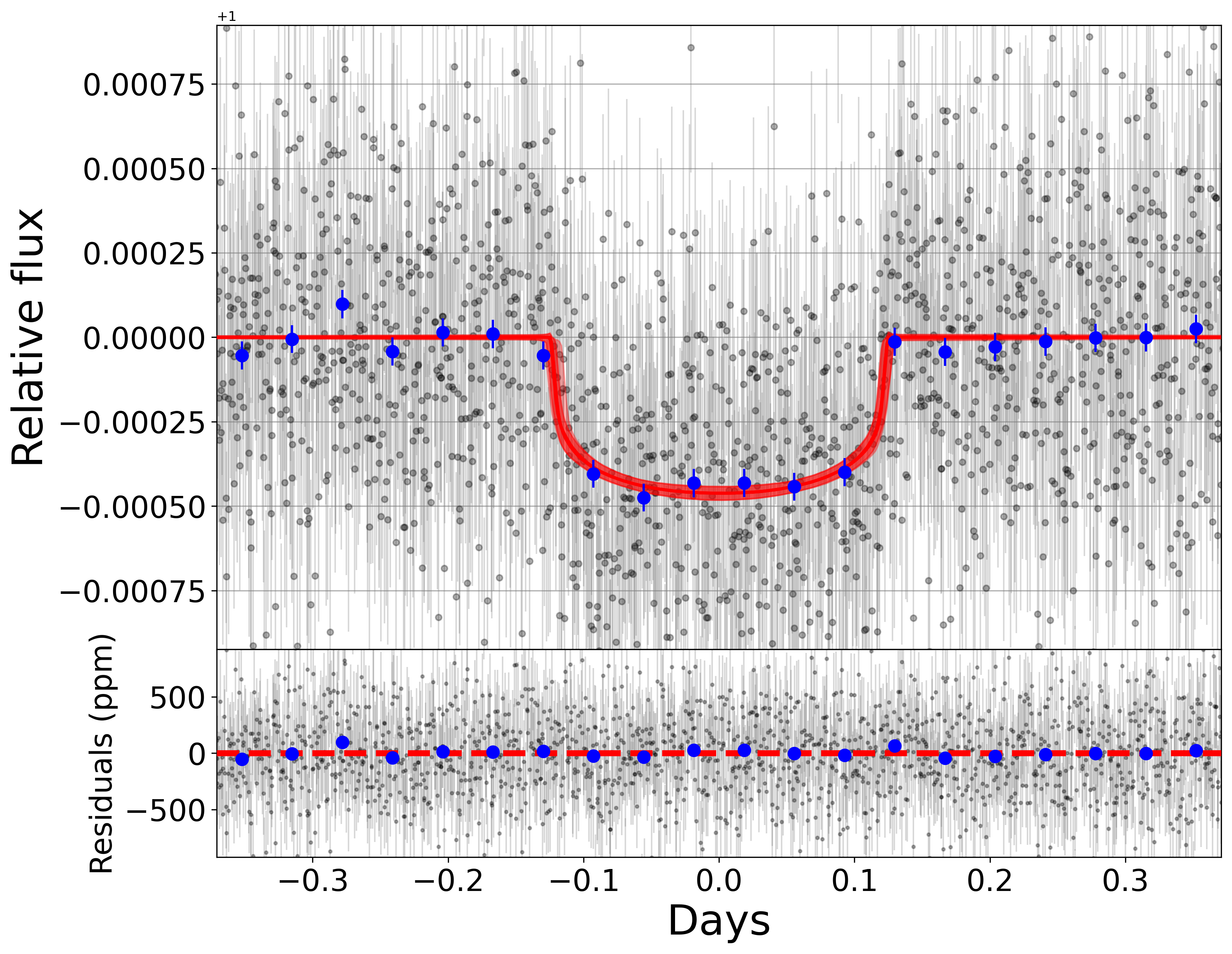}
\caption{The phase-folded TESS transits of TOI-5800\,b (top panel) and TOI-5817\,b (bottom panel), along with the best fit models, in red, and the residuals below each panel. The thickness of the red line represents the $1\sigma$ spread. The blue circles represent the average of $\sim8$ and $\sim45$ minutes respectively.}
\label{fig:5800_5817_phaseLC}
\end{figure}

Our analyses clearly confirm the planetary nature of TOI-5800\,b and TOI-5817\,b. The phased models of both the RVs and the transits for each planet are plotted in Figs.\,\ref{fig:5800_5817_phaseRV} and \ref{fig:5800_5817_phaseLC}, while the RV models of their datasets are displayed in Fig.\,\ref{fig:5800_5817_fullRV}. As final parameters for the planets (Table\,\ref{tab:pparameters}), we decided to adopt the ones from the eccentric fit for TOI-5800\,b and those from the circular fit for TOI-5817\,b, because in the first case $\Delta\ln{\mathcal{Z}^{e\geq0}_{e=0}}=3.5$, while in the second $\Delta\ln{\mathcal{Z}^{e\geq0}_{e=0}}=-1.6$, where $\mathcal{Z}$ is the Bayesian evidence of a model fit (note that the threshold for a \textit{very strong} detection is $\Delta\ln{\mathcal{Z}}\sim5$, according to \citealt{Kass1995}, while $3<\Delta\ln{\mathcal{Z}}<5$ is considered \textit{strong}). Indeed, for TOI-5800\,b the retrieved eccentricity is significant at a $\sim3.5\sigma$ level\footnote{We note that these significance levels assume Gaussian posteriors, while the quoted uncertainties correspond to the 68\% credible interval derived from the posterior distributions. Due to the limited number of samples, we were unable to directly estimate high-significance levels from the tails of the distributions. A more accurate characterization of the tails would be significantly more computationally expensive. We therefore adopt the Gaussian approximation as a practical compromise and acknowledge its limitations.} ($e=0.300^{+0.083}_{-0.082}$), while for TOI-5817\,b the eccentricity is compatible with zero within $\sim2\sigma$ ($e=0.12\pm0.07$)\footnote{As a reference, the uncertainties corresponding to 95.5\% intervals (or $2\sigma$), are respectively $e=0.30^{+0.16}_{-0.18}$ and $e=0.12^{+0.14}_{-0.11}$. We further acknowledge that, for TOI-5800\,b, the uncertainty mainly arises from $\sqrt{e}\sin\omega_*$, since $\sqrt{e}\cos\omega_*$ is actually constrained to $>4\sigma$.}. For TOI-5800\,b we tested alternative eccentricity priors that favor low values while still allowing a broad range. The inferred eccentricity remained consistent across these choices. We also repeated the analysis using Gaussian Processes to model correlated noise, with no significant change in the results.

Furthermore, neither star display any sign of activity, with no peak in the GLS of the activity indices\footnote{In particular, we have explored the FWHM, the Bisector (BIS) and the Contrast of the cross correlation function, along with the $\log R^{\prime}_{\rm HK}$. The values are reported in Tables\,\ref{tab:5800_RV} and \ref{tab:5817_RV_HN}.} below a False Alarm Probability (FAP) of 10\% (evaluated with the bootstrap method), no sign of significant modulation in the SAP light curve and no significant peak in the GLS of the RV residuals after the removal of the planet signals. This is consistent with both the old ages estimated for the stars and their low values of $\log R^{\prime}_{\rm HK}$ (Table\,\ref{tab:star}). We note that there is a small-amplitude, $\sim$5\% FAP signal in the residuals of TOI-5800 at $\sim10$\,days which will eventually require more observations to be confirmed. We also verified that there is no discernible modulation in the transit time variations (TTV) of TOI-5800\,b (see Sect.\ref{sec:TTV}).

\begin{table}
\centering
\caption{Final parameters from the global fits.}\label{tab:pparameters}
\renewcommand{\arraystretch}{1.2}
\resizebox{0.85\hsize}{!}{\begin{tabular}{lcc}
    \hline\hline
     Parameters & TOI-5800\,b & TOI-5817\,b \\
\hline \\[-6pt]%
\multicolumn{1}{l}{\large{Transit and orbital}} \\[2pt]
$K$ (m\,s$^{-1}$)\dotfill & $5.40^{+1.00}_{-1.05}$ & $2.69\pm0.35$ $^{(\dagger)}$ \\
$P_{\rm orb}$ (d)\dotfill & $2.6278838\pm0.0000037$ & $15.61031^{+0.00012}_{-0.00010}$ \\
$T_{\rm 0}$ (BJD - $2\,459\,000$)\dotfill & $771.7145 \pm 0.0013$ & $799.3065 \pm 0.0029$ \\
$T_{\rm 14}$ (h)\dotfill & $1.20\pm0.05$ & $6.00^{+0.17}_{-0.15}$ \\
$R_{\rm p}/R_{\star}$\dotfill & $0.0292^{+0.0020}_{-0.0017}$ & $0.0198 \pm 0.0007$ \\ 
$b$\dotfill & $0.903^{+0.029}_{-0.044}$ & $0.41^{+0.09}_{-0.13}$ \\ 
$i$ (deg)\dotfill & $85.08^{+0.44}_{-0.56}$ & $88.72^{+0.43}_{-0.34}$ \\ 
$a/R_{\star}$\dotfill & $9.58^{+0.27}_{-0.29}$ & $18.44^{+0.69}_{-0.76}$ \\
$q_1$ (TESS)\dotfill & $0.47^{+0.27}_{-0.24}$ & $0.52^{+0.32}_{-0.30}$ \\
$q_2$ (TESS)\dotfill & $0.48^{+0.33}_{-0.32}$ & $0.28^{+0.31}_{-0.20}$ \\
$\sqrt{e}\sin\omega_*$\dotfill & $-0.31^{+0.28}_{-0.18}$ & -- \\
$\sqrt{e}\cos\omega_*$\dotfill & $0.30\pm0.08$ & -- \\[2pt]
\multicolumn{1}{l}{\large{Derived}} \\[2pt]
$M_{\rm p}$ ($M_{\oplus}$)\dotfill & $9.5^{+1.7}_{-1.9}$ & $10.3^{+1.4}_{-1.3}$ \\
$R_{\rm p}$ ($R_{\oplus}$)\dotfill & $2.46^{+0.18}_{-0.16}$ & $3.08\pm0.14$ \\
$\rho_{\rm p}$ (g\,cm$^{-3}$)\dotfill & $3.46^{+1.02}_{-0.90}$ & $1.93^{+0.41}_{-0.34}$ \\ [2pt]
$\log{g_{\rm p}}$ (cgs)\dotfill & $15^{+4}_{-3}$ & $11 \pm 2$ \\
$a$ (au)\dotfill & $0.0344 \pm 0.0014$ & $0.1223 \pm 0.0060$ \\ 
$T_{\rm eq}^{(\ddagger)}$ (K)\dotfill & $1108 \pm 20$ & $950^{+21}_{-18}$ \\
$e$\dotfill & $0.300^{+0.083}_{-0.082}$ & $<0.20$ \\ [2pt]
$\omega_*$ (deg)\dotfill & $-36^{+33}_{-19}$ & unconstrained \\ 
TSM\dotfill & $103^{+35}_{-22}$ & $56^{+11}_{-9}$ \\ [2pt]
$u_1$ (TESS)\dotfill & $0.63^{+0.56}_{-0.44}$ & $0.41^{+0.53}_{-0.30}$ \\
$u_2$ (TESS)\dotfill & $0.04^{+0.46}_{-0.44}$ & $0.30^{+0.37}_{-0.39}$ \\
\multicolumn{1}{l}{\large{Instrumental}} \\[2pt]
$\sigma_{\textsf{TESS}}$ (ppt)\dotfill & $315\pm15$ & $143\pm4$ \\
$\overline{\mu}_{\textsf{HARPS}}$ (m\,s$^{-1}$)\dotfill & $-2.21^{+0.65}_{-0.61}$ & -- \\
$\overline{\mu}_{\textsf{HARPS-N}}$ (m\,s$^{-1}$)\dotfill & -- & $-20801.85\pm0.28$ \\
$\sigma_{\textsf{w,HARPS}}$ (m\,s$^{-1}$)\dotfill & $2.70^{+0.61}_{-0.52}$ & -- \\
$\sigma_{\textsf{w,HARPS-N}}$ (m\,s$^{-1}$)\dotfill & -- & $1.63^{+0.24}_{-0.21}$ \\

    \bottomrule
\end{tabular}
}
\tablefoot{Best-fit median values, with upper and lower 68\% credibility bands as errors, of the fitted and derived parameters for TOI-5800\,b and TOI-5817\,b, as extracted from the posterior distribution of the relative models. $^{(\dagger)}$ This value is comparable to the formal uncertainties of SOPHIE RVs, but it's $\sim3$ times smaller than their actual spread, which justifies their exclusion from the global fit. $^{(\ddagger)}$ This is the equilibrium temperature for a zero Bond albedo and uniform heat redistribution to the night side. The eccentricity upper limit on TOI-5817\,b is constrained at the confidence level of 1$\sigma$.}
\end{table}

\subsection{Detection sensitivities}\label{sec:sensitivity}

We estimated the completeness of both the HARPS (TOI-5800) and HARPS-N (TOI-5817) RV timeseries by performing injection-recovery simulations. We injected synthetic RVs with fake planetary signals at the times of our observations, using the original error bars and the estimated stellar jitter (Table\,\ref{tab:pparameters}). We simulated the signals across a logarithmic $30 \times 30$ grid in planetary mass ($M_{\rm p}$) and semi-major axis ($a$), covering the ranges 0.01--20\,$M_{\rm Jup}$ (or Jupiter masses) and 0.01--10\,au. Similarly to \citet{Bonomo2023}, for each location of the grid we generated 200 synthetic planetary signals, drawing $a$ and $M_{\rm p}$ from a log-uniform distribution inside the cell, $T_0$ from a uniform distribution in $[0,P]$, the orbital inclination $i$ from a uniform $\cos{i}$ distribution in $[0,1]$, $\omega_*$ from a uniform distribution in $[0,2\pi]$, and $e$ from a beta distribution in $[0.0, 0.8]$ \citep{Kipping2013}. We fitted the signals by employing either Keplerian orbits or linear and quadratic terms, in order to take into account long-period signals, which would not be correctly identified as Keplerian due to the short time span of the RV observations (45 and 575\,d respectively). We then adopted the Bayesian information criterion (BIC) to compare the fitted models with a constant model and considered a planet significantly detected only when $\Delta \text{BIC} > 10$ in favor of the planet-induced one. The detection fraction was finally computed as the portion of detected signals for each grid element, as illustrated in Fig.\,\ref{fig:completeness}. In particular, we confirm that any Jupiter-mass planet within 1 au and 5 au would have been found around TOI-5800 and TOI-5817, respectively.

In principle, absolute astrometry can provide additional constraints on the presence of orbiting companions. For both stars, the Gaia Data Release 3 (DR3) archive reports values of the re-normalized unit weight error (RUWE), a diagnostic of the departure from a
good single-star fit to Gaia-only astrometry, of ${\rm RUWE}=1.09$ and ${\rm RUWE}=1.11$, respectively. These numbers are below the threshold value of 1.4 typically adopted to signify that a single-star model fails to satisfactorily describe the data (e.g., \citealt{Lindegren2018,Lindegren2021}). Using a Monte Carlo simulation to assess the possible companion masses and separations that could result in RUWE values in excess of the ones reported in the Gaia DR3 archive (see \citealt{Sozzetti2023} for details) we found that Gaia DR3 astrometry alone can exclude, with $99\%$ confidence, the presence of $\sim4$ and $\sim13$ Jupiter-mass companions within roughly $1-2.5$ au around TOI-5800 and TOI-5817, respectively. In the case of TOI-5800, the sensitivity limits from RVs and Gaia DR3 astrometry are comparable, while for TOI-5817 the RVs can rule out the presence of companions over an order of magnitude smaller than Gaia absolute astrometry. 

Furthermore, the catalogues of Hipparcos-Gaia DR3 astrometric accelerations \citep{Brandt2021,Kervella2022} report no statistically significant values of proper motion anomaly for TOI-5817. In the $3-10$ au range of orbital separations, which corresponds to the sweet spot of sensitivity for the technique, the Hipparcos-Gaia absolute astrometry rules out companions with typical masses above $2.5 \,M_\mathrm{Jup}$ \citep{Kervella2022}. Only at around 10\,au is this limit comparable to that provided by the RV time-series, which instead typically indicates no planetary companions with masses around 1 M$_\mathrm{Jup}$ or lower are present in the system in the same separation range.

\section{Discussion}\label{sec:discussion}
\subsection{Planet interior compositions}\label{sec:composition}
\begin{figure}[t!]
\centering
\vspace{-0.5cm}
\includegraphics[width=7.5cm, angle=90]{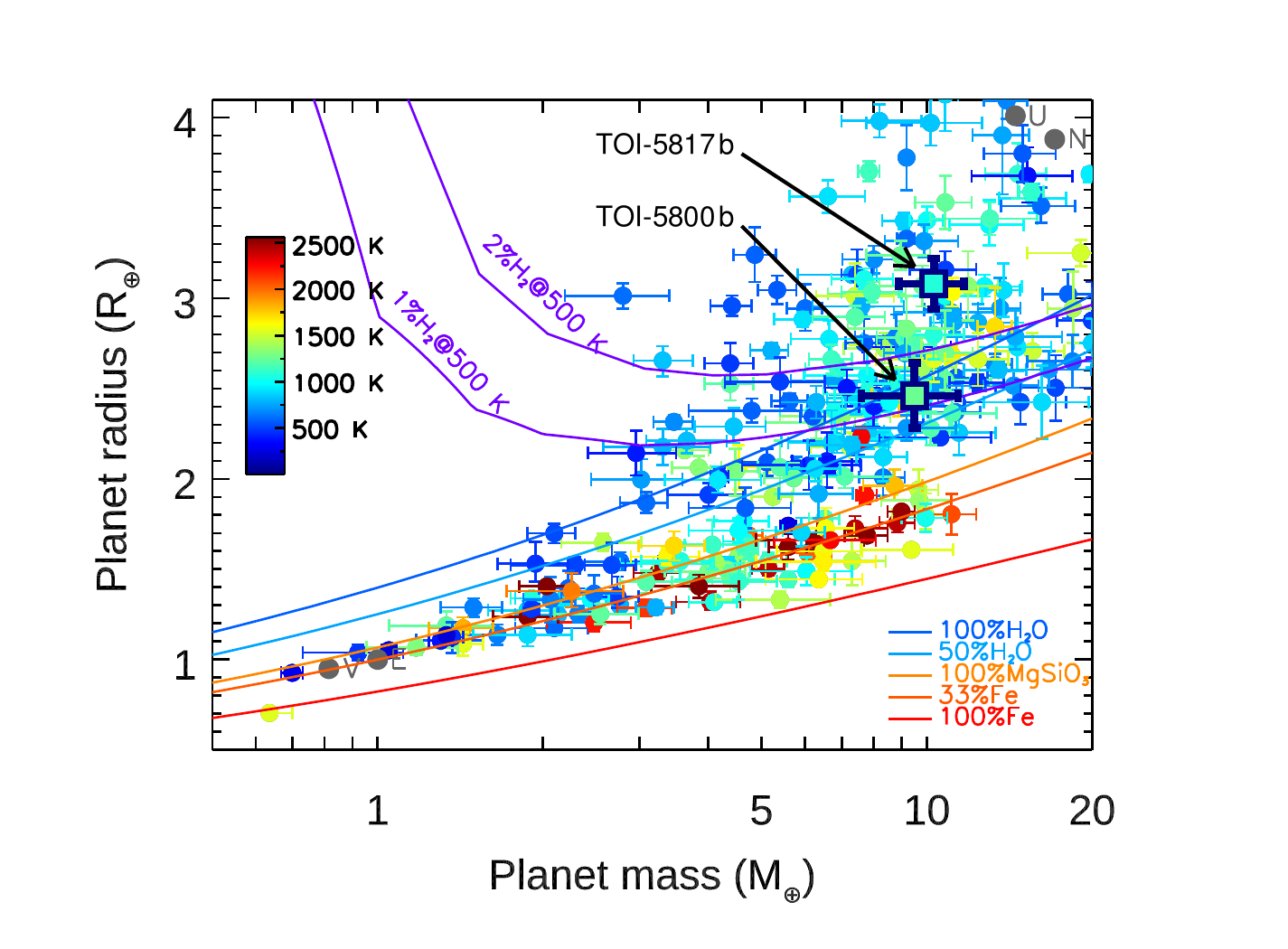}
\vspace{-0.75cm}
\caption{Mass-radius diagram of small ($R_{\rm p} \leq 4~R_\oplus$) planets, color-coded by planet equilibrium temperatures. 
The different solid curves, from bottom to top, correspond to planet compositions of $100\%$ iron, $33\%$ iron core and $67\%$ silicate mantle (Earth-like composition), $100\%$ silicates, $50\%$ rocky interior and $50\%$ water, $100\%$ water, rocky interiors and $1\%$ or $2\%$ hydrogen-dominated atmospheres \citep{Zeng2013}.
The dark gray circles indicate Venus (V), Earth (E), Uranus (U), and Neptune (N). TOI-5800\,b and TOI-5817\,b are indicated with squares.}
\label{fig:massradius}
\end{figure}

\noindent
Fig.~\ref{fig:massradius} shows the mass-radius diagram of small exoplanets with mass and radius determinations better than $4\sigma$ and $10\sigma$, respectively, along with the planet isocomposition curves given in \citet{Zeng2013}. In this diagram TOI-5800\,b is located just below the pure water curve, while TOI-5817\,b lies above it. This implies that the latter has a considerably higher fraction of volatile elements, as expected from its lower bulk density for a comparable planet mass (Table~\ref{tab:pparameters}).

To analyze in more detail their composition, we used an inference model based on \citet{dorn_generalized_2017} with updates in \citet{luo_majority_2024}. The structure model consists of three different layers with an iron core, a silicate mantle, and a volatile layer. The core and mantle are assumed to be adiabatic and can contain both liquid and solid phases. For solid iron we use the equation of state (EOS) for hexagonal close packed iron \citep{hakim_new_2018, miozzi_new_2020} and for liquid iron and iron alloys we use the EOS by \citet{luo_majority_2024}. 
The mantle is composed of three major species: MgO, SiO$_2$, and FeO. For pressures below $\sim 125\,$GPa, the solid mantle mineralogy is modeled using the thermodynamical model \textsc{Perple\_X} \citep{connolly_geodynamic_2009} and the database from \citet{stixrude_thermal_2022}. At higher pressures we define the stable minerals apriori and use their respective EOS from various sources \citep{hemley_constraints_1992,fischer_equation_2011,faik_equation_2018,musella_physical_2019}. As there is no data for the density of liquid MgO in the high-pressure temperature regime available, we model the liquid mantle as an ideal mixture of Mg$_2$SiO$_4$, SiO$_2$, and FeO \citep{melosh_hydrocode_2007,faik_equation_2018,ichikawa_ab_2020,stewart_shock_2020}. The different components are mixed using the additive volume law. 

The volatile layer is modeled as consisting either purely of water or of an uniformly mixed atmosphere of hydrogen, helium, and water. For a pure water layer, the model further takes into account that water can be present in the molten mantle and the iron core \citep{dorn_hidden_2021,luo_majority_2024}. The given water mass fraction in the pure water case is thus the accreted bulk water mass fraction and not the mass of the surface water layer. 
The outer surface water layer until $0.1\,$bar is assumed to be isothermal, while, above $0.1\,$bar, we assume an adiabatic profile. We use the AQUA equation of state for water \citep{haldemann_aqua_2020}, which covers a large P-T space and water phases. Therefore, our model considers the existences of high pressure ice shells in the deep layers of the surface water layer. In the case of an uniformly mixed atmosphere, the atmosphere is split into an irradiated outer atmosphere and an inner atmosphere in radiative-convective equilibrium. The atmosphere is modeled using the analytic description of \citep{guillot_radiative_2010}. The ratio of hydrogen and helium is set to be solar and we use the EOS of \citet{1995_Saumon_EOS} to describe the H$_2$-He mixture. 
The water mass fraction of the atmosphere is given by the metallicity $Z$ and described by the ANEOS EOS \citep{1990_thompson_aneos}. The two components, H$_2$-He and H$_2$O, are again mixed using the additive volume law.  

For the inference, we use a new approach involving surrogate models (paper in prep.), where we use Polynomial Chaos Kriging (PCK) for surrogate modeling \citep{schobi2015polynomial,marelli2014uqlab}, which approximates the global behavior of the full physical forward model and replaces it in an MCMC framework. The advantage is that the computational costs for a full interior inference is only a few minutes. The prior parameter distribution is listed in Table \ref{tab:interiorprior}. We assume that all Si and Mg is in the mantle of the planet. The prior of the Mg/Si mass ratio is thus given by the Mg/Si mass ratio of the host star as measured in this work which corresponds to 0.91 for TOI-5800 and 0.95 for TOI-5817. The upper bound for the Fe/Si mass ratio in the mantle is given the stellar Fe/Si mass ratio, which is 1.28 for TOI-5800 and 1.31 for TOI-5817. In the pure water case, the upper limit of the bulk water content is set to 0.5 as predicted by planet formation models \citep[e.g., ][]{Venturini2020}. For the case of a uniformly mixed atmosphere, the luminosity range is chosen based on the ages of the two systems and the possible atmospheric mass fractions range between 0.1\% and 5\%. The atmospheric mass fraction refers to the total mass of the atmosphere in H$_2$, He, and H$_2$O. The water mass fraction in the atmosphere is given  by the metallicity $Z$.

\begin{table}
\centering
\caption{Inference results for the internal compositions of TOI-5800\,b and TOI-5817\,b. Stated errors represent 84th and 16th-percentiles.}\label{tab:pposterior}
\renewcommand{\arraystretch}{1.2}
\resizebox{0.85\hsize}{!}{\begin{tabular}{lcc}
    \hline\hline
     Parameters & TOI-5800\,b & TOI-5817\,b \\
    \hline \\[-6pt]%
    \multicolumn{1}{l}{\large{H$_2$-He-H$_2$O atmosphere}} \\[2pt]
    $M_\mathrm{atm} \: (M_\oplus)$\dotfill & $0.14^{+0.10}_{-0.11} $ & $0.43^{+0.31}_{-0.34}$ \\
    $M_\mathrm{mantle} \: (M_\oplus)$\dotfill & $5.7 \pm 1.3$ & $6.4\pm1.4$ \\
    $M_\mathrm{core} \: (M_\oplus)$\dotfill & $3.4 \pm 1.2$ & $3.5 \pm 1.2$ \\
    $L \: (\mathrm{erg}\,\mathrm{s}^{-1})$\dotfill & $(2.5\pm0.31)\cdot10^{21}$ & 
    $(2.6 \pm 0.31)\cdot10^{21}$ \\
    $Z$\dotfill & $0.43 \pm 0.24$ & $0.27\pm 0.19$ \\
    \multicolumn{1}{l}{\large{Pure water case}}\\[2pt]
    $M_\mathrm{water}\:(M_\oplus)$\dotfill & $1.54^{+1.00}_{-0.98}$ & $3.29^{+1.97}_{-1.94}$ \\
    $M_\mathrm{mantle} \: (M_\oplus)$\dotfill & $5.6 \pm 1.3$ & $5.5\pm1.8$ \\
    $M_\mathrm{core} \: (M_\oplus)$\dotfill & $2.1 \pm 0.8$ & $1.7 \pm 0.8$ \\
    \bottomrule
\end{tabular}
}
\end{table}
The results of the inference model are summarized in Table\,\ref{tab:pposterior}. 
For both planets a significant amount of volatiles is needed to explain the observed radii given the mass of the two planets. 
In the case of a H$_2$-He-H$_2$O atmosphere, the atmosphere mass fraction of TOI-5800\,b is $1.5\%\pm 1.0\%$, while the atmosphere mass fraction of TOI-5817\,b is $4.3\% \pm 0.3\%$, while both metallicity $Z$ and luminosity $L$ are poorly constrained by the data.
The ratio of the core-to-mantle mass is roughly Earth-like for both planets. This is mainly constrained by the stellar relative element ratios of Fe, Mg, and Si.
Fig.\,\ref{fig:5800_5817_posterior} shows the posterior distribution for the case of a uniformly mixed atmosphere for both planets.  
Nevertheless, we find that both TOI-5800\,b and TOI-5817\,b can possess atmospheres with supersolar metallicities with $Z=0.43\pm0.24$ and $Z=0.27\pm 0.19$ respectively, although metallicity is poorly constrained. This is mainly because we use a uniform prior on $Z$ between 0.02 and 1 and the mass and radius data only carry limited information to put strong constraints on metallicity.

In the hypothetical case of restricting the volatile inventory to water only, the water mass fraction is $16\% \pm 11 \%$ for TOI-5800\,b and $31\% \pm 19\%$ for TOI-5817\,b. This extreme scenario informs us on the maximum amount of water possible in the interiors of both worlds. As we take into account any H and O partitioned in the metal core phase, water dissolution in the silicate mantle and the water in a steam envelope, the inferred amount of water represents the maximum amount of accreted water. The inferred values for TOI-5800\,b of $16\pm11\%$ depart significantly from 50\% mass fraction, which is generally predicted for worlds that mainly formed outside of the water ice-line and migrated inwards during the gas disk lifetime \citep{Venturini2020, burn2024water}. Given the water mass fractions of $31\pm19\%$ for TOI-5817\,b, a formation outside of the water ice-line cannot be excluded.

Inferred volatile mass fractions of TOI-5817\,b are higher than the one for TOI-5800\,b in both scenarios.  Compared to the case with a H$_2$-He-H$_2$O atmosphere, the core mass fraction in the pure water case is lower. However, this mass refers only to the mass of the iron core. A significant fraction of the water will be stored in the core as H and O, raising the total core mass.

\begin{figure}
\centering
\includegraphics[width=1\linewidth]{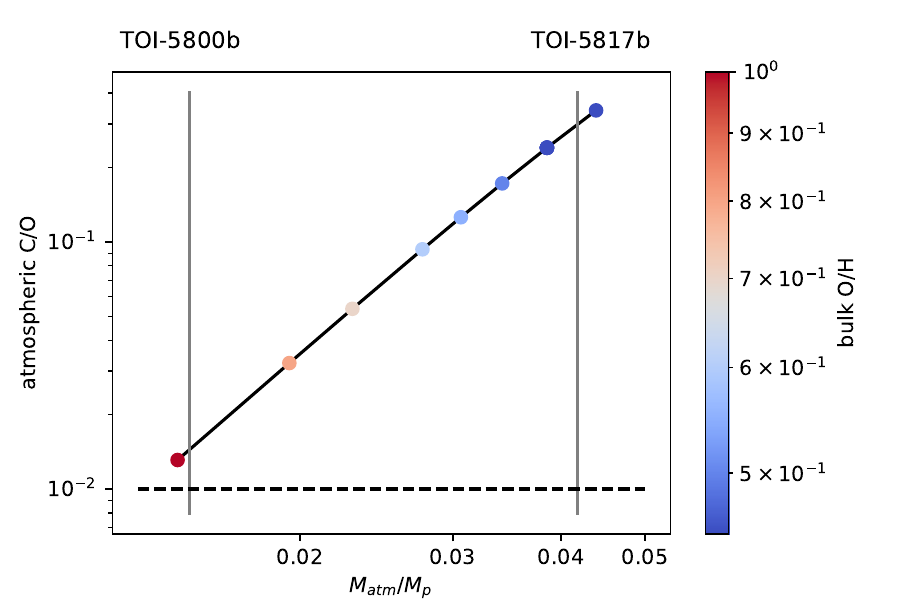}
\caption{Predicted trend how atmospheric C/O values change as a function of atmospheric mass assuming a planet mass of 10 $M_\oplus$, and bulk ratios for C/O = 0.01 (dashed line) and O/Si = 3.3 (Earth-like), and a 3000 K magma ocean surface. Higher atmospheric C/O for TOI-5817\,b is predicted compared to TOI-5800\,b, given their different inferred atmospheric mass fractions (solid gray lines, errors are large and not shown).}
\label{atmCO}
\end{figure}

In order to make a prediction on possible trends of inferred atmospheric C/O ratios from spectroscopy, we employ the model from \citet{schlichting2022chemical} with recent updates \citep{werlen2025atmospheric}. Figure \ref{atmCO} illustrates that an increase in atmospheric mass fraction on top of an Earth-like interior leads to an increase in atmospheric C/O ratios. This can be understood by the fact that the addition of primordial gas leads to a higher production of endogenic water, and water is mainly stored deep in the interior and not in the atmosphere \citep{luo_majority_2024}. As a consequence, a higher atmospheric mass fraction leads to a relative removal of oxygen compared to carbon from the atmosphere. We note that although the trend in atmospheric C/O values holds true, the actual values may vary as we have placed some assumptions that are poorly constrained by mass and radius (e.g., bulk O/Si, bulk C/O, temperature at the magma ocean surface). Overall, if the difference in planet densities between TOI-5800\,b and TOI-5817\,b are mainly due to a difference in gas mass fraction, chemical thermodynamics predict difference in atmospheric C/O ratios on the order of one magnitude, however, we note that the error bars on inferred atmosphere mass fractions are large.

\subsection{Tidal effects in the TOI-5800 system}\label{sec:tidal}
We applied the constant-time lag tidal model by \citet{Leconteetal10} to study the tidal evolution of the system. In the case of the solar-system planets Uranus or Neptune, the tidal dissipation is parametrized by the so-called modified tidal quality factor $Q^{\prime}_{\rm p}$, which is assumed to be of the order of $10^{5}$ \citep[][Sect.~5.4]{Ogilvie14}. Assuming $Q^{\prime}_{\rm p}=10^{5}$ for TOI-5800\,b and expressing the time lag using Eq.~(19) of \citet{Leconteetal10}, we find an $e$-folding timescale for the decay of its eccentricity of $e/|de/dt| \sim 0.6$~Gyr. Therefore, one possibility is that the planet acquired a high-$e$ orbit relatively recently, and we are catching it as it slowly circularizes. The circularization timescale is fairly short compared to the estimated system age ($5-10\%$ given the large uncertainties on system age), so catching a planet in this stage should be somewhat rare -- and indeed, no other sub-Neptunes with such high-$e$, short-period orbits are known. Another possibility is that the orbital eccentricity is excited by another body in the system outside our detection limits (Figure~\ref{fig:completeness}).  Such a third body could be a second planet on an eccentric orbit as in the coplanar two-planet system model by \citet{Mardling07}. For example, assuming an outer planet with the same mass of TOI-5800\,b on a 10-day orbit with an eccentricity of 0.33 (see the end of Sect.\,\ref{sec:analysis}), the $e$-folding decay time of the eccentricity of TOI-5800\,b increases to $\sim 2.9$\,Gyr, which is comparable with the age of the star derived in Section\,\ref{sec:age_chromo}.

We expect significant tidal dissipation in the interior of TOI-5800\,b. In fact, with our assumed $Q^{\prime}_{\rm p}= 10^{5}$, the power dissipated inside TOI-5800\,b planet is $5.3 \times 10^{18}$~W according to Eq.~(13) of \citet{Leconteetal10}. This is already a few percent of the bolometric flux received by the planet ($2.7 \times 10^{20}$~W), and scales with the assumed $Q^{\prime}_{\rm p}$. A smaller $Q^{\prime}_{\rm p}$ -- for example, $Q^{\prime}_{\rm p}\sim10^4$ was recently inferred for WASP-107b \citep{Sing2024, Welbanks2024} -- would result in even greater dissipation. On the other hand, a higher $Q^{\prime}_{\rm p}$ would result in a longer eccentricity damping timescale. In this case, the observed eccentricity could be a remnant of an initially higher value, and the associated tidal heating would be correspondingly lower. Therefore, depending on the tidal quality factor, intense tidal heating could either significantly affect the planet’s thermal balance \citep[and on the atmospheric chemistry;][]{Fortney2020} or have a negligible impact. 


\subsection{Atmospheric characterization prospects}\label{sec:atmo}

With $\mathrm{TSM}=103^{+35}_{-22}$, TOI-5800\,b is an excellent target for atmospheric characterization with \textit{JWST} \citep{Kempton2018}. At $T_\mathrm{eq} = 1108\pm20$~K, TOI-5800\,b is expected to lack CH$_4$ in chemical equilibrium, making it unlikely to host hydrocarbon hazes \citep{Gao2020, Brande2024}. Thus, this warm sub-Neptune could be in an aerosol-free regime that makes it ideal for detecting strong atmospheric features, similarly to TOI-421\,b \citep{Davenport2025}. Furthermore, if the planetary eccentricity is high because it reached its current position recently and is still tidally circularizing, then TOI-5800\,b may not have suffered large atmospheric loss, so its composition could be a relatively pristine record of its formation history. Finally, TOI-5800\,b is one of the top 5 sub-Neptunes for atmospheric spectroscopy orbiting a K dwarf. If strong features are detected in its atmosphere, it could bridge TOI-421 (which orbits a G-type host star and exhibits a low mean-molecular weight atmosphere) and TOI-270\,d and GJ\,9827\,d \citep[which exhibit high mean-molecular weight atmospheres and orbit M dwarfs;][]{Benneke2024, Piaulet-Ghorayeb2024, Davenport2025}. As for TOI-5817\,b, the lower $\mathrm{TSM}=56^{+11}_{-9}$ makes it a less favorable target for \textit{JWST} transmission spectroscopy compared to TOI-5800\,b. However, it may still be a good Tier 1 target for Ariel, which promises to deliver lower-resolution spectroscopy for hundreds of sub-Neptunes \citep{Edwards2022}.

\section{Conclusions}\label{sec:conclusions}

We have confirmed the planetary nature of TOI-5800\,b and TOI-5817\,b, two hot sub-Neptunes ($T_{\rm eq}\sim1110$\,K and $T_{\rm eq}\sim950$\,K), with different orbital periods ($P\sim2.6$\,d and $P\sim15.6$\,d) hosted by low-activity K3\,V and G2\,IV-V stars. In terms of composition, TOI-5800\,b can be explained with an Earth-like interior and either a H$_2$-He-H$_2$O atmosphere of $\sim 1\, \mathrm{wt}\%$ or (as a water world) with a water mass fraction of $\sim 10\%$. Instead, the lower bulk density of TOI-5817\,b requires a higher volatile mass fraction and can be explained with an Earth-like interior and a H$_2$-He-H$_2$O atmosphere of $\sim 4\, \mathrm{wt}\%$ or as a water world with a water mass fraction of $\sim 30\%$. 

In particular, TOI-5800\,b is currently undergoing tidal migration and it stands out as the most eccentric ($e\sim0.3$) planet ever found within $P<3$\,d, with GJ\,436\,b \citep{Lanotte2014} being the distant second ($e\sim0.15$). As it orbits a relatively bright star ($V_{\rm T}=9.7$ mag), it is a prime target for atmospheric follow-up (TSM\,$\gtrsim100$) both with the JWST and from the ground with high-resolution spectroscopy. 
On the other hand, TOI-5817\,b adds to the small but growing sample of well-characterized S-type planets (i.e. it orbits the primary star of a binary system). Studying such planets is essential to test whether the ``radius valley'', the observed gap between rocky super-Earths and gas-rich sub-Neptunes \citep{Fulton2017,Venturini2020}, also applies in binary environments, and how it is shaped by factors like composition, orbit, and stellar properties. Future atmospheric studies of TOI-5817\,b are not out of reach, as it orbits a bright star ($V_{\rm T}=8.7$ mag), and they could help constrain the formation and evolution of planets in these dynamically complex systems.

\begin{acknowledgements}
We thank S. Jenkins and A. Vanderburg for their collegiality.
L.N. acknowledges the financial contribution from the INAF Large Grant 2023 ``EXODEMO''. 
This work has made use of data from the European Space Agency (ESA) mission {\it Gaia} (\url{https://www.cosmos.esa.int/gaia}), processed by the {\it Gaia} Data Processing and Analysis Consortium (DPAC, \url{https://www.cosmos.esa.int/web/gaia/dpac/consortium}). Funding for the DPAC has been provided by national institutions, in particular the institutions participating in the {\it Gaia} Multilateral Agreement. 
This work is based on observations collected at the European Southern Observatory under ESO programme(s) 113.26UJ.001, and on observations made with the Italian Telescopio Nazionale \textit{Galileo} (TNG) operated by the Fundaci\'{o}n Galileo Galilei (FGG) of the Istituto Nazionale di Astrofisica (INAF) at the Observatorio del Roque de los Muchachos (La Palma, Canary Islands, Spain).
We acknowledge the Italian center for Astronomical Archives (IA2, \url{https://www.ia2.inaf.it}), part of the Italian National Institute for Astrophysics (INAF), for providing technical assistance, services and supporting activities of the GAPS collaboration.
This work includes data collected with the TESS mission, obtained from the MAST data archive at the Space Telescope Science Institute (STScI). Funding for the TESS mission is provided by the NASA Explorer Program. STScI is operated by the Association of Universities for Research in Astronomy, Inc., under the NASA contract NAS 5–26555. We acknowledge the use of public TESS data from pipelines at the TESS Science Office and at the TESS Science Processing Operations Center. Resources supporting this work were provided by the NASA High-End Computing (HEC) Program through the NASA Advanced Supercomputing (NAS) Division at Ames Research Center for the production of the SPOC data products. 
This work also includes observations obtained at the Hale Telescope, Palomar Observatory, as part of a collaborative agreement between the Caltech Optical Observatories and the Jet Propulsion Laboratory operated by Caltech for NASA.
Funding for the TESS mission is provided by NASA's Science Mission Directorate. KAC and CNW acknowledge support from the TESS mission via subaward s3449 from MIT.
CD acknowledges support from the Swiss National Science Foundation under grant TMSGI2\_211313. Parts of this work has been carried out within the framework of the NCCR PlanetS supported by the Swiss National Science Foundation under grants 51NF40\_182901 and 51NF40\_205606.
MP acknowledges support from the European Union – NextGenerationEU (PRIN MUR 2022 20229R43BH) and the “Programma di Ricerca Fondamentale INAF 2023”.
TZi acknowledges support from CHEOPS ASI-INAF agreement n. 2019-29-HH.0, NVIDIA Academic Hardware Grant Program for the use of the Titan V GPU card and the Italian MUR Departments of Excellence grant 2023-2027 “Quantum Frontiers”.
JJL received support from NASA's Exoplanet Research Program project 24-XRP24\_2-0020.
This work makes use of observations from the LCOGT network. Part of the LCOGT telescope time was granted by NOIRLab through the Mid-Scale Innovations Program (MSIP). MSIP is funded by NSF.
This research has made use of the Exoplanet Follow-up Observation Program (ExoFOP; DOI: 10.26134/ExoFOP5) website, which is operated by the California Institute of Technology, under contract with the National Aeronautics and Space Administration under the Exoplanet Exploration Program.
This work made use of \texttt{tpfplotter} by J. Lillo-Box (publicly available in www.github.com/jlillo/tpfplotter).
This work uses observations secured with the SOPHIE spectrograph at the 1.93-m telescope of Observatoire Haute-Provence, France, with the support of its staff. This work was  supported by the ``Programme National de Plan\'etologie'' (PNP) of CNRS/INSU, and CNES.
Some of the observations in this paper made use of the High-Resolution Imaging instrument Zorro and were obtained under Gemini LLP Proposal Number: GN/S-2021A-LP-105. Zorro was funded by the NASA Exoplanet Exploration Program and built at the NASA Ames Research Center by Steve B. Howell, Nic Scott, Elliott P. Horch, and Emmett Quigley. Zorro was mounted on the Gemini South telescope of the international Gemini Observatory, a program of NSF’s OIR Lab, which is managed by the Association of Universities for Research in Astronomy (AURA) under a cooperative agreement with the National Science Foundation. on behalf of the Gemini partnership: the National Science Foundation (United States), National Research Council (Canada), Agencia Nacional de Investigación y Desarrollo (Chile), Ministerio de Ciencia, Tecnología e Innovación (Argentina), Ministério da Ciência, Tecnologia, Inovações e Comunicações (Brazil), and Korea Astronomy and Space Science Institute (Republic of Korea).
LM acknowledges financial contribution from PRIN MUR 2022 project 2022J4H55R.
DRC acknowledges partial support from NASA Grant 18-2XRP18\_2-0007. 
\end{acknowledgements}


\bibliographystyle{aa}
\bibliography{58xx}

\begin{thebibliography}{161}
\expandafter\ifx\csname natexlab\endcsname\relax\def\natexlab#1{#1}\fi

\bibitem[{{Adamow}(2017)}]{Adamow2017}
{Adamow}, M.~M. 2017, in American Astronomical Society Meeting Abstracts, Vol. 230, American Astronomical Society Meeting Abstracts \#230, 216.07

\bibitem[{{Almeida-Fernandes} \& {Rocha-Pinto}(2018)}]{Almeida2018}
{Almeida-Fernandes}, F. \& {Rocha-Pinto}, H.~J. 2018, \mnras, 476, 184

\bibitem[{{Armstrong} {et~al.}(2020){Armstrong}, {Lopez}, {Adibekyan}, {Booth}, {Bryant}, {Collins}, {Deleuil}, {Emsenhuber}, {Huang}, {King}, {Lillo-Box}, {Lissauer}, {Matthews}, {Mousis}, {Nielsen}, {Osborn}, {Otegi}, {Santos}, {Sousa}, {Stassun}, {Veras}, {Ziegler}, {Acton}, {Almenara}, {Anderson}, {Barrado}, {Barros}, {Bayliss}, {Belardi}, {Bouchy}, {Brice{\~n}o}, {Brogi}, {Brown}, {Burleigh}, {Casewell}, {Chaushev}, {Ciardi}, {Collins}, {Col{\'o}n}, {Cooke}, {Crossfield}, {D{\'\i}az}, {Delgado Mena}, {Demangeon}, {Dorn}, {Dumusque}, {Eigm{\"u}ller}, {Fausnaugh}, {Figueira}, {Gan}, {Gandhi}, {Gill}, {Gonzales}, {Goad}, {G{\"u}nther}, {Helled}, {Hojjatpanah}, {Howell}, {Jackman}, {Jenkins}, {Jenkins}, {Jensen}, {Kennedy}, {Latham}, {Law}, {Lendl}, {Lozovsky}, {Mann}, {Moyano}, {McCormac}, {Meru}, {Mordasini}, {Osborn}, {Pollacco}, {Queloz}, {Raynard}, {Ricker}, {Rowden}, {Santerne}, {Schlieder}, {Seager}, {Sha}, {Tan}, {Tilbrook}, {Ting}, {Udry}, {Vanderspek}, {Watson}, {West}, {Wilson}, {Winn},
  {Wheatley}, {Villasenor}, {Vines}, \& {Zhan}}]{Armstrong2020}
{Armstrong}, D.~J., {Lopez}, T.~A., {Adibekyan}, V., {et~al.} 2020, \nat, 583, 39

\bibitem[{{Artymowicz} \& {Lubow}(1994)}]{Artymowicz1994}
{Artymowicz}, P. \& {Lubow}, S.~H. 1994, \apj, 421, 651

\bibitem[{{Barnes}(2007)}]{Barnes07}
{Barnes}, S.~A. 2007, \apj, 669, 1167

\bibitem[{{Batalha} {et~al.}(2019){Batalha}, {Lewis}, {Fortney}, {Batalha}, {Kempton}, {Lewis}, \& {Line}}]{Batalha2019}
{Batalha}, N.~E., {Lewis}, T., {Fortney}, J.~J., {et~al.} 2019, \apjl, 885, L25

\bibitem[{{Benneke} {et~al.}(2024){Benneke}, {Roy}, {Coulombe}, {Radica}, {Piaulet}, {Ahrer}, {Pierrehumbert}, {Krissansen-Totton}, {Schlichting}, {Hu}, {Yang}, {Christie}, {Thorngren}, {Young}, {Pelletier}, {Knutson}, {Miguel}, {Evans-Soma}, {Dorn}, {Gagnebin}, {Fortney}, {Komacek}, {MacDonald}, {Raul}, {Cloutier}, {Acuna}, {Lafreni{\`e}re}, {Cadieux}, {Doyon}, {Welbanks}, \& {Allart}}]{Benneke2024}
{Benneke}, B., {Roy}, P.-A., {Coulombe}, L.-P., {et~al.} 2024, submitted to AAS Journals, arXiv:2403.03325

\bibitem[{{Biazzo} {et~al.}(2022){Biazzo}, {D'Orazi}, {Desidera}, {Turrini}, {Benatti}, {Gratton}, {Magrini}, {Sozzetti}, {Baratella}, {Bonomo}, {Borsa}, {Claudi}, {Covino}, {Damasso}, {Di Mauro}, {Lanza}, {Maggio}, {Malavolta}, {Maldonado}, {Marzari}, {Micela}, {Poretti}, {Vitello}, {Affer}, {Bignamini}, {Carleo}, {Cosentino}, {Fiorenzano}, {Giacobbe}, {Harutyunyan}, {Leto}, {Mancini}, {Molinari}, {Molinaro}, {Nardiello}, {Nascimbeni}, {Pagano}, {Pedani}, {Piotto}, {Rainer}, \& {Scandariato}}]{Biazzoetal2022}
{Biazzo}, K., {D'Orazi}, V., {Desidera}, S., {et~al.} 2022, \aap, 664, A161

\bibitem[{{Bonomo} {et~al.}(2023){Bonomo}, {Dumusque}, {Massa}, {Mortier}, {Bongiolatti}, {Malavolta}, {Sozzetti}, {Buchhave}, {Damasso}, {Haywood}, {Morbidelli}, {Latham}, {Molinari}, {Pepe}, {Poretti}, {Udry}, {Affer}, {Boschin}, {Charbonneau}, {Cosentino}, {Cretignier}, {Ghedina}, {Lega}, {L{\'o}pez-Morales}, {Margini}, {Mart{\'\i}nez Fiorenzano}, {Mayor}, {Micela}, {Pedani}, {Pinamonti}, {Rice}, {Sasselov}, {Tronsgaard}, \& {Vanderburg}}]{Bonomo2023}
{Bonomo}, A.~S., {Dumusque}, X., {Massa}, A., {et~al.} 2023, \aap, 677, A33

\bibitem[{{Booth} {et~al.}(2017){Booth}, {Clarke}, {Madhusudhan}, \& {Ilee}}]{Booth2017}
{Booth}, R.~A., {Clarke}, C.~J., {Madhusudhan}, N., \& {Ilee}, J.~D. 2017, \mnras, 469, 3994

\bibitem[{{Bouchy} {et~al.}(2013){Bouchy}, {D{\'\i}az}, {H{\'e}brard}, {Arnold}, {Boisse}, {Delfosse}, {Perruchot}, \& {Santerne}}]{Bouchy2013}
{Bouchy}, F., {D{\'\i}az}, R.~F., {H{\'e}brard}, G., {et~al.} 2013, \aap, 549, A49

\bibitem[{{Bouchy} {et~al.}(2009){Bouchy}, {H{\'e}brard}, {Udry}, {Delfosse}, {Boisse}, {Desort}, {Bonfils}, {Eggenberger}, {Ehrenreich}, {Forveille}, {Lagrange}, {Le Coroller}, {Lovis}, {Moutou}, {Pepe}, {Perrier}, {Pont}, {Queloz}, {Santos}, {S{\'e}gransan}, \& {Vidal-Madjar}}]{Bouchy2009}
{Bouchy}, F., {H{\'e}brard}, G., {Udry}, S., {et~al.} 2009, \aap, 505, 853

\bibitem[{{Bourrier} {et~al.}(2023){Bourrier}, {Attia}, {Mallonn}, {Marret}, {Lendl}, {Konig}, {Krenn}, {Cretignier}, {Allart}, {Henry}, {Bryant}, {Leleu}, {Nielsen}, {Hebrard}, {Hara}, {Ehrenreich}, {Seidel}, {dos Santos}, {Lovis}, {Bayliss}, {Cegla}, {Dumusque}, {Boisse}, {Boucher}, {Bouchy}, {Pepe}, {Lavie}, {Rey Cerda}, {S{\'e}gransan}, {Udry}, \& {Vrignaud}}]{Bourrier2023}
{Bourrier}, V., {Attia}, M., {Mallonn}, M., {et~al.} 2023, \aap, 669, A63

\bibitem[{{Bourrier} {et~al.}(2018){Bourrier}, {Lovis}, {Beust}, {Ehrenreich}, {Henry}, {Astudillo-Defru}, {Allart}, {Bonfils}, {S{\'e}gransan}, {Delfosse}, {Cegla}, {Wyttenbach}, {Heng}, {Lavie}, \& {Pepe}}]{Bourrier2018}
{Bourrier}, V., {Lovis}, C., {Beust}, H., {et~al.} 2018, \nat, 553, 477

\bibitem[{{Brande} {et~al.}(2024){Brande}, {Crossfield}, {Kreidberg}, {Morley}, {Barman}, {Benneke}, {Christiansen}, {Dragomir}, {Fortney}, {Greene}, {Hardegree-Ullman}, {Howard}, {Knutson}, {Lothringer}, \& {Mikal-Evans}}]{Brande2024}
{Brande}, J., {Crossfield}, I. J.~M., {Kreidberg}, L., {et~al.} 2024, \apjl, 961, L23

\bibitem[{{Brandt}(2021)}]{Brandt2021}
{Brandt}, T.~D. 2021, \apjs, 254, 42

\bibitem[{{Brown} {et~al.}(2013){Brown}, {Baliber}, {Bianco}, {Bowman}, {Burleson}, {Conway}, {Crellin}, {Depagne}, {De Vera}, {Dilday}, {Dragomir}, {Dubberley}, {Eastman}, {Elphick}, {Falarski}, {Foale}, {Ford}, {Fulton}, {Garza}, {Gomez}, {Graham}, {Greene}, {Haldeman}, {Hawkins}, {Haworth}, {Haynes}, {Hidas}, {Hjelstrom}, {Howell}, {Hygelund}, {Lister}, {Lobdill}, {Martinez}, {Mullins}, {Norbury}, {Parrent}, {Paulson}, {Petry}, {Pickles}, {Posner}, {Rosing}, {Ross}, {Sand}, {Saunders}, {Shobbrook}, {Shporer}, {Street}, {Thomas}, {Tsapras}, {Tufts}, {Valenti}, {Vander Horst}, {Walker}, {White}, \& {Willis}}]{Brown2013}
{Brown}, T.~M., {Baliber}, N., {Bianco}, F.~B., {et~al.} 2013, \pasp, 125, 1031

\bibitem[{Burn {et~al.}(2024)Burn, Bali, Dorn, Luque, \& Grimm}]{burn2024water}
Burn, R., Bali, K., Dorn, C., Luque, R., \& Grimm, S.~L. 2024, In review at Astronomy \& Astrophysics, arXiv:2411.16879

\bibitem[{Cannon \& Pickering(1918--1924)}]{Cannon1924}
Cannon, A.~J. \& Pickering, E.~C. 1918--1924, Annals of H. College Observatory, Vol. 91--99, The Henry Draper Catalogue (Harvard College Observatory)

\bibitem[{{Castro-Gonz{\'a}lez} {et~al.}(2024{\natexlab{a}}){Castro-Gonz{\'a}lez}, {Bourrier}, {Lillo-Box}, {Delisle}, {Armstrong}, {Barrado}, \& {Correia}}]{Castro2024}
{Castro-Gonz{\'a}lez}, A., {Bourrier}, V., {Lillo-Box}, J., {et~al.} 2024{\natexlab{a}}, \aap, 689, A250

\bibitem[{{Castro-Gonz{\'a}lez} {et~al.}(2024{\natexlab{b}}){Castro-Gonz{\'a}lez}, {Lillo-Box}, {Armstrong}, {Acu{\~n}a}, {Aguichine}, {Bourrier}, {Gandhi}, {Sousa}, {Delgado-Mena}, {Moya}, {Adibekyan}, {Correia}, {Barrado}, {Damasso}, {Winn}, {Santos}, {Barkaoui}, {Barros}, {Benkhaldoun}, {Bouchy}, {Brice{\~n}o}, {Caldwell}, {Collins}, {Essack}, {Ghachoui}, {Gillon}, {Hounsell}, {Jehin}, {Jenkins}, {Keniger}, {Law}, {Mann}, {Nielsen}, {Pozuelos}, {Schanche}, {Seager}, {Tan}, {Timmermans}, {Villase{\~n}or}, {Watkins}, \& {Ziegler}}]{Castro2024b}
{Castro-Gonz{\'a}lez}, A., {Lillo-Box}, J., {Armstrong}, D.~J., {et~al.} 2024{\natexlab{b}}, \aap, 691, A233

\bibitem[{{Ciardi} {et~al.}(2015){Ciardi}, {Beichman}, {Horch}, \& {Howell}}]{Ciardi2015}
{Ciardi}, D.~R., {Beichman}, C.~A., {Horch}, E.~P., \& {Howell}, S.~B. 2015, \apj, 805, 16

\bibitem[{{Collins} {et~al.}(2017){Collins}, {Kielkopf}, {Stassun}, \& {Hessman}}]{Collins2017}
{Collins}, K.~A., {Kielkopf}, J.~F., {Stassun}, K.~G., \& {Hessman}, F.~V. 2017, \aj, 153, 77

\bibitem[{Connolly(2009)}]{connolly_geodynamic_2009}
Connolly, J. A.~D. 2009, The geodynamic equation of state: {What} and how - {Connolly} - 2009 - {Geochemistry}, {Geophysics} - {Wiley} {Online} {Library}

\bibitem[{{Correia} {et~al.}(2020){Correia}, {Bourrier}, \& {Delisle}}]{Correia2020}
{Correia}, A.~C.~M., {Bourrier}, V., \& {Delisle}, J.~B. 2020, \aap, 635, A37

\bibitem[{{Cosentino} {et~al.}(2012){Cosentino}, {Lovis}, {Pepe}, {Collier Cameron}, {Latham}, {Molinari}, {Udry}, {Bezawada}, {Black}, {Born}, {Buchschacher}, {Charbonneau}, {Figueira}, {Fleury}, {Galli}, {Gallie}, {Gao}, {Ghedina}, {Gonzalez}, {Gonzalez}, {Guerra}, {Henry}, {Horne}, {Hughes}, {Kelly}, {Lodi}, {Lunney}, {Maire}, {Mayor}, {Micela}, {Ordway}, {Peacock}, {Phillips}, {Piotto}, {Pollacco}, {Queloz}, {Rice}, {Riverol}, {Riverol}, {San Juan}, {Sasselov}, {Segransan}, {Sozzetti}, {Sosnowska}, {Stobie}, {Szentgyorgyi}, {Vick}, \& {Weber}}]{Cosentino2012}
{Cosentino}, R., {Lovis}, C., {Pepe}, F., {et~al.} 2012, in SPIE Conference Series, Vol. 8446, Ground-based and Airborne Instrumentation for Astronomy IV, ed. I.~S. {McLean}, S.~K. {Ramsay}, \& H.~{Takami}, 84461V

\bibitem[{{Cutri} {et~al.}(2003){Cutri}, {Skrutskie}, {van Dyk}, {Beichman}, {Carpenter}, {Chester}, {Cambresy}, {Evans}, {Fowler}, {Gizis}, {Howard}, {Huchra}, {Jarrett}, {Kopan}, {Kirkpatrick}, {Light}, {Marsh}, {McCallon}, {Schneider}, {Stiening}, {Sykes}, {Weinberg}, {Wheaton}, {Wheelock}, \& {Zacarias}}]{Cutri2003}
{Cutri}, R.~M., {Skrutskie}, M.~F., {van Dyk}, S., {et~al.} 2003, {2MASS All-Sky Catalog of Point Sources}, VizieR On-line Data Catalog: II/246. Published in: University of Massachusetts and Infrared Processing and Analysis Center, (IPAC/California Institute of Technology; 2003)

\bibitem[{{Cutri} {et~al.}(2021){Cutri}, {Wright}, {Conrow}, {Fowler}, {Eisenhardt}, {Grillmair}, {Kirkpatrick}, {Masci}, {McCallon}, {Wheelock}, {Fajardo-Acosta}, {Yan}, {Benford}, {Harbut}, {Jarrett}, {Lake}, {Leisawitz}, {Ressler}, {Stanford}, {Tsai}, {Liu}, {Helou}, {Mainzer}, {Gettngs}, {Gonzalez}, {Hoffman}, {Marsh}, {Padgett}, {Skrutskie}, {Beck}, {Papin}, \& {Wittman}}]{Cutri2013}
{Cutri}, R.~M., {Wright}, E.~L., {Conrow}, T., {et~al.} 2021, {AllWISE Data Release}, VizieR On-line Data Catalog: II/328. Published in: IPAC/Caltech (2013)

\bibitem[{{Dai} {et~al.}(2021){Dai}, {Howard}, {Batalha}, {Beard}, {Behmard}, {Blunt}, {Brinkman}, {Chontos}, {Crossfield}, {Dalba}, {Dressing}, {Fulton}, {Giacalone}, {Hill}, {Huber}, {Isaacson}, {Kane}, {Lubin}, {Mayo}, {Mo{\v{c}}nik}, {Akana Murphy}, {Petigura}, {Rice}, {Robertson}, {Rosenthal}, {Roy}, {Rubenzahl}, {Weiss}, {Zandt}, {Beichman}, {Ciardi}, {Collins}, {Gonzales}, {Howell}, {Matson}, {Matthews}, {Schlieder}, {Schwarz}, {Ricker}, {Vanderspek}, {Latham}, {Seager}, {Winn}, {Jenkins}, {Caldwell}, {Colon}, {Dragomir}, {Lund}, {McLean}, {Rudat}, \& {Shporer}}]{Dai2021}
{Dai}, F., {Howard}, A.~W., {Batalha}, N.~M., {et~al.} 2021, \aj, 162, 62

\bibitem[{{Damasso} {et~al.}(2023){Damasso}, {Locci}, {Benatti}, {Maggio}, {Nardiello}, {Baratella}, {Biazzo}, {Bonomo}, {Desidera}, {D'Orazi}, {Mallonn}, {Lanza}, {Sozzetti}, {Marzari}, {Borsa}, {Maldonado}, {Mancini}, {Poretti}, {Scandariato}, {Bignamini}, {Borsato}, {Capuzzo Dolcetta}, {Cecconi}, {Claudi}, {Cosentino}, {Covino}, {Fiorenzano}, {Harutyunyan}, {Mann}, {Micela}, {Molinari}, {Molinaro}, {Pagano}, {Pedani}, {Pinamonti}, {Piotto}, \& {Stoev}}]{Damasso2023}
{Damasso}, M., {Locci}, D., {Benatti}, S., {et~al.} 2023, \aap, 672, A126

\bibitem[{Davenport {et~al.}(2025)Davenport, Kempton, Nixon, Ih, Deming, Fu, May, Bean, Gao, Rogers, \& Malik}]{Davenport2025}
Davenport, B., Kempton, E. M.-R., Nixon, M.~C., {et~al.} 2025, The Astrophysical Journal Letters, 984, L44

\bibitem[{{Dawson} \& {Johnson}(2018)}]{Dawson2018}
{Dawson}, R.~I. \& {Johnson}, J.~A. 2018, \araa, 56, 175

\bibitem[{{Dekany} {et~al.}(2013){Dekany}, {Roberts}, {Burruss}, {Bouchez}, {Truong}, {Baranec}, {Guiwits}, {Hale}, {Angione}, {Trinh}, {Zolkower}, {Shelton}, {Palmer}, {Henning}, {Croner}, {Troy}, {McKenna}, {Tesch}, {Hildebrandt}, \& {Milburn}}]{dekany2013}
{Dekany}, R., {Roberts}, J., {Burruss}, R., {et~al.} 2013, \apj, 776, 130

\bibitem[{{Demangeon} {et~al.}(2021){Demangeon}, {Dalal}, {H{\'e}brard}, {Nsamba}, {Kiefer}, {Camacho}, {Sahlmann}, {Arnold}, {Astudillo-Defru}, {Bonfils}, {Boisse}, {Bouchy}, {Bourrier}, {Campante}, {Delfosse}, {Deleuil}, {D{\'\i}az}, {Faria}, {Forveille}, {Hara}, {Heidari}, {Hobson}, {Lopez}, {Moutou}, {Rey}, {Santerne}, {Sousa}, {Santos}, {Str{\o}m}, {Tsantaki}, \& {Udry}}]{Demangeon2021}
{Demangeon}, O.~D.~S., {Dalal}, S., {H{\'e}brard}, G., {et~al.} 2021, \aap, 653, A78

\bibitem[{{Dong} {et~al.}(2018){Dong}, {Xie}, {Zhou}, {Zheng}, \& {Luo}}]{Dong2018}
{Dong}, S., {Xie}, J.-W., {Zhou}, J.-L., {Zheng}, Z., \& {Luo}, A. 2018, Proceedings of the National Academy of Science, 115, 266

\bibitem[{{D'Orazi} {et~al.}(2020){D'Orazi}, {Oliva}, {Bragaglia}, {Frasca}, {Sanna}, {Biazzo}, {Casali}, {Desidera}, {Lucatello}, {Magrini}, \& {Origlia}}]{DOrazietal2020}
{D'Orazi}, V., {Oliva}, E., {Bragaglia}, A., {et~al.} 2020, \aap, 633, A38

\bibitem[{Dorn \& Lichtenberg(2021)}]{dorn_hidden_2021}
Dorn, C. \& Lichtenberg, T. 2021, The Astrophysical Journal Letters, 922, L4

\bibitem[{Dorn {et~al.}(2017)Dorn, Venturini, Khan, Heng, Alibert, Helled, Rivoldini, \& Benz}]{dorn_generalized_2017}
Dorn, C., Venturini, J., Khan, A., {et~al.} 2017, Astronomy \& Astrophysics, 597, A37, arXiv:1609.03908 [astro-ph]

\bibitem[{{Dorn} {et~al.}(2023){Dorn}, {Bristow}, {Smoker}, {Rodler}, {Lavail}, {Accardo}, {van den Ancker}, {Baade}, {Baruffolo}, {Courtney-Barrer}, {Blanco}, {Brucalassi}, {Cumani}, {Follert}, {Haimerl}, {Hatzes}, {Haug}, {Heiter}, {Hinterschuster}, {Hubin}, {Ives}, {Jung}, {Jones}, {Kaeufl}, {Kirchbauer}, {Klein}, {Kochukhov}, {Korhonen}, {K{\"o}hler}, {Lizon}, {Moins}, {Molina-Conde}, {Marquart}, {Neeser}, {Oliva}, {Pallanca}, {Pasquini}, {Paufique}, {Piskunov}, {Reiners}, {Schneller}, {Schmutzer}, {Seemann}, {Slumstrup}, {Smette}, {Stegmeier}, {Stempels}, {Tordo}, {Valenti}, {Valenzuela}, {Vernet}, {Vinther}, \& {Wehrhahn}}]{Dorn2023}
{Dorn}, R.~J., {Bristow}, P., {Smoker}, J.~V., {et~al.} 2023, \aap, 671, A24

\bibitem[{{Doyle} {et~al.}(2025){Doyle}, {Armstrong}, {Acu{\~n}a}, {Osborn}, {Sousa}, {Castro-Gonz{\'a}lez}, {Bourrier}, {Alves}, {Barrado}, {Barros}, {Bayliss}, {Cui}, {Demangeon}, {D{\'\i}az}, {Dumusque}, {Eeles-Nolle}, {Gill}, {Hacker}, {Jenkins}, {Keniger}, {Lafarga}, {Lillo-Box}, {Lockley}, {Nielsen}, {Parc}, {Rodrigues}, {Santerne}, {Santos}, \& {Wheatley}}]{Doyle2025}
{Doyle}, L., {Armstrong}, D.~J., {Acu{\~n}a}, L., {et~al.} 2025, \mnras [\eprint[arXiv]{2504.16164}]

\bibitem[{{Dumusque} {et~al.}(2021){Dumusque}, {Cretignier}, {Sosnowska}, {Buchschacher}, {Lovis}, {Phillips}, {Pepe}, {Alesina}, {Buchhave}, {Burnier}, {Cecconi}, {Cegla}, {Cloutier}, {Collier Cameron}, {Cosentino}, {Ghedina}, {Gonz{\'a}lez}, {Haywood}, {Latham}, {Lodi}, {L{\'o}pez-Morales}, {Maldonado}, {Malavolta}, {Micela}, {Molinari}, {Mortier}, {P{\'e}rez Ventura}, {Pinamonti}, {Poretti}, {Rice}, {Riverol}, {Riverol}, {San Juan}, {S{\'e}gransan}, {Sozzetti}, {Thompson}, {Udry}, \& {Wilson}}]{Dumusque2021}
{Dumusque}, X., {Cretignier}, M., {Sosnowska}, D., {et~al.} 2021, \aap, 648, A103

\bibitem[{{Eastman}(2017)}]{2017ascl.soft10003E}
{Eastman}, J. 2017, {EXOFASTv2: Generalized publication-quality exoplanet modeling code}, Astrophysics Source Code Library, record ascl:1710.003

\bibitem[{{Eastman} {et~al.}(2013){Eastman}, {Gaudi}, \& {Agol}}]{Eastman2012}
{Eastman}, J., {Gaudi}, B.~S., \& {Agol}, E. 2013, \pasp, 125, 83

\bibitem[{{Eastman} {et~al.}(2019){Eastman}, {Rodriguez}, {Agol}, {Stassun}, {Beatty}, {Vanderburg}, {Gaudi}, {Collins}, \& {Luger}}]{Eastman2019}
{Eastman}, J.~D., {Rodriguez}, J.~E., {Agol}, E., {et~al.} 2019, \pasp, arXiv:1907.09480

\bibitem[{{Edwards} \& {Tinetti}(2022)}]{Edwards2022}
{Edwards}, B. \& {Tinetti}, G. 2022, \aj, 164, 15

\bibitem[{{El-Badry} {et~al.}(2021){El-Badry}, {Rix}, \& {Heintz}}]{El-Badry2021}
{El-Badry}, K., {Rix}, H.-W., \& {Heintz}, T.~M. 2021, \mnras, 506, 2269

\bibitem[{{Espinoza} {et~al.}(2019){Espinoza}, {Kossakowski}, \& {Brahm}}]{Espinoza2019}
{Espinoza}, N., {Kossakowski}, D., \& {Brahm}, R. 2019, \mnras, 490, 2262

\bibitem[{Faik {et~al.}(2018)Faik, Tauschwitz, \& Iosilevskiy}]{faik_equation_2018}
Faik, S., Tauschwitz, A., \& Iosilevskiy, I. 2018, Computer Physics Communications, 227, 117

\bibitem[{Fischer {et~al.}(2011)Fischer, Campbell, Shofner, Lord, Dera, \& Prakapenka}]{fischer_equation_2011}
Fischer, R.~A., Campbell, A.~J., Shofner, G.~A., {et~al.} 2011, Earth and Planetary Science Letters, 304, 496

\bibitem[{{Fortney} {et~al.}(2020){Fortney}, {Visscher}, {Marley}, {Hood}, {Line}, {Thorngren}, {Freedman}, \& {Lupu}}]{Fortney2020}
{Fortney}, J.~J., {Visscher}, C., {Marley}, M.~S., {et~al.} 2020, \aj, 160, 288

\bibitem[{{Fulton} {et~al.}(2017){Fulton}, {Petigura}, {Howard}, {Isaacson}, {Marcy}, {Cargile}, {Hebb}, {Weiss}, {Johnson}, {Morton}, {Sinukoff}, {Crossfield}, \& {Hirsch}}]{Fulton2017}
{Fulton}, B.~J., {Petigura}, E.~A., {Howard}, A.~W., {et~al.} 2017, \aj, 154, 109

\bibitem[{{Furlan} {et~al.}(2017){Furlan}, {Ciardi}, {Everett}, {Saylors}, {Teske}, {Horch}, {Howell}, {van Belle}, {Hirsch}, {Gautier}, {Adams}, {Barrado}, {Cartier}, {Dressing}, {Dupree}, {Gilliland}, {Lillo-Box}, {Lucas}, \& {Wang}}]{furlan2017a}
{Furlan}, E., {Ciardi}, D.~R., {Everett}, M.~E., {et~al.} 2017, \aj, 153, 71

\bibitem[{{Furlan} \& {Howell}(2017)}]{Furlan2017}
{Furlan}, E. \& {Howell}, S.~B. 2017, \aj, 154, 66

\bibitem[{{Furlan} \& {Howell}(2020)}]{Furlan2020}
{Furlan}, E. \& {Howell}, S.~B. 2020, \apj, 898, 47

\bibitem[{{Gaia Collaboration} {et~al.}(2023){Gaia Collaboration}, {Vallenari}, {Brown}, {Prusti}, {de Bruijne}, {Arenou}, {Babusiaux}, {Biermann}, {Creevey}, {Ducourant}, {Evans}, {Eyer}, {Guerra}, {Hutton}, {Jordi}, {Klioner}, {Lammers}, {Lindegren}, {Luri}, {Mignard}, {Panem}, {Pourbaix}, {Randich}, {Sartoretti}, {Soubiran}, {Tanga}, {Walton}, {Bailer-Jones}, {Bastian}, {Drimmel}, {Jansen}, {Katz}, {Lattanzi}, {van Leeuwen}, {Bakker}, {Cacciari}, {Casta{\~n}eda}, {De Angeli}, {Fabricius}, {Fouesneau}, {Fr{\'e}mat}, {Galluccio}, {Guerrier}, {Heiter}, {Masana}, {Messineo}, {Mowlavi}, {Nicolas}, {Nienartowicz}, {Pailler}, {Panuzzo}, {Riclet}, {Roux}, {Seabroke}, {Sordo}, {Th{\'e}venin}, {Gracia-Abril}, {Portell}, {Teyssier}, {Altmann}, {Andrae}, {Audard}, {Bellas-Velidis}, {Benson}, {Berthier}, {Blomme}, {Burgess}, {Busonero}, {Busso}, {C{\'a}novas}, {Carry}, {Cellino}, {Cheek}, {Clementini}, {Damerdji}, {Davidson}, {de Teodoro}, {Nu{\~n}ez Campos}, {Delchambre}, {Dell'Oro}, {Esquej},
  {Fern{\'a}ndez-Hern{\'a}ndez}, {Fraile}, {Garabato}, {Garc{\'\i}a-Lario}, {Gosset}, {Haigron}, {Halbwachs}, {Hambly}, {Harrison}, {Hern{\'a}ndez}, {Hestroffer}, {Hodgkin}, {Holl}, {Jan{\ss}en}, {Jevardat de Fombelle}, {Jordan}, {Krone-Martins}, {Lanzafame}, {L{\"o}ffler}, {Marchal}, {Marrese}, {Moitinho}, {Muinonen}, {Osborne}, {Pancino}, {Pauwels}, {Recio-Blanco}, {Reyl{\'e}}, {Riello}, {Rimoldini}, {Roegiers}, {Rybizki}, {Sarro}, {Siopis}, {Smith}, {Sozzetti}, {Utrilla}, {van Leeuwen}, {Abbas}, {{\'A}brah{\'a}m}, {Abreu Aramburu}, {Aerts}, {Aguado}, {Ajaj}, {Aldea-Montero}, {Altavilla}, {{\'A}lvarez}, {Alves}, {Anders}, {Anderson}, {Anglada Varela}, {Antoja}, {Baines}, {Baker}, {Balaguer-N{\'u}{\~n}ez}, {Balbinot}, {Balog}, {Barache}, {Barbato}, {Barros}, {Barstow}, {Bartolom{\'e}}, {Bassilana}, {Bauchet}, {Becciani}, {Bellazzini}, {Berihuete}, {Bernet}, {Bertone}, {Bianchi}, {Binnenfeld}, {Blanco-Cuaresma}, {Blazere}, {Boch}, {Bombrun}, {Bossini}, {Bouquillon}, {Bragaglia}, {Bramante}, {Breedt},
  {Bressan}, {Brouillet}, {Brugaletta}, {Bucciarelli}, {Burlacu}, {Butkevich}, {Buzzi}, {Caffau}, {Cancelliere}, {Cantat-Gaudin}, {Carballo}, {Carlucci}, {Carnerero}, {Carrasco}, {Casamiquela}, {Castellani}, {Castro-Ginard}, {Chaoul}, {Charlot}, {Chemin}, {Chiaramida}, {Chiavassa}, {Chornay}, {Comoretto}, {Contursi}, {Cooper}, {Cornez}, {Cowell}, {Crifo}, {Cropper}, {Crosta}, {Crowley}, {Dafonte}, {Dapergolas}, {David}, {David}, {de Laverny}, {De Luise}, {De March}, {De Ridder}, {de Souza}, {de Torres}, {del Peloso}, {del Pozo}, {Delbo}, {Delgado}, {Delisle}, {Demouchy}, {Dharmawardena}, {Di Matteo}, {Diakite}, {Diener}, {Distefano}, {Dolding}, {Edvardsson}, {Enke}, {Fabre}, {Fabrizio}, {Faigler}, {Fedorets}, {Fernique}, {Fienga}, {Figueras}, {Fournier}, {Fouron}, {Fragkoudi}, {Gai}, {Garcia-Gutierrez}, {Garcia-Reinaldos}, {Garc{\'\i}a-Torres}, {Garofalo}, {Gavel}, {Gavras}, {Gerlach}, {Geyer}, {Giacobbe}, {Gilmore}, {Girona}, {Giuffrida}, {Gomel}, {Gomez}, {Gonz{\'a}lez-N{\'u}{\~n}ez},
  {Gonz{\'a}lez-Santamar{\'\i}a}, {Gonz{\'a}lez-Vidal}, {Granvik}, {Guillout}, {Guiraud}, {Guti{\'e}rrez-S{\'a}nchez}, {Guy}, {Hatzidimitriou}, {Hauser}, {Haywood}, {Helmer}, {Helmi}, {Sarmiento}, {Hidalgo}, {Hilger}, {H{\l}adczuk}, {Hobbs}, {Holland}, {Huckle}, {Jardine}, {Jasniewicz}, {Jean-Antoine Piccolo}, {Jim{\'e}nez-Arranz}, {Jorissen}, {Juaristi Campillo}, {Julbe}, {Karbevska}, {Kervella}, {Khanna}, {Kontizas}, {Kordopatis}, {Korn}, {K{\'o}sp{\'a}l}, {Kostrzewa-Rutkowska}, {Kruszy{\'n}ska}, {Kun}, {Laizeau}, {Lambert}, {Lanza}, {Lasne}, {Le Campion}, {Lebreton}, {Lebzelter}, {Leccia}, {Leclerc}, {Lecoeur-Taibi}, {Liao}, {Licata}, {Lindstr{\o}m}, {Lister}, {Livanou}, {Lobel}, {Lorca}, {Loup}, {Madrero Pardo}, {Magdaleno Romeo}, {Managau}, {Mann}, {Manteiga}, {Marchant}, {Marconi}, {Marcos}, {Marcos Santos}, {Mar{\'\i}n Pina}, {Marinoni}, {Marocco}, {Marshall}, {Martin Polo}, {Mart{\'\i}n-Fleitas}, {Marton}, {Mary}, {Masip}, {Massari}, {Mastrobuono-Battisti}, {Mazeh}, {McMillan}, {Messina}, {Michalik},
  {Millar}, {Mints}, {Molina}, {Molinaro}, {Moln{\'a}r}, {Monari}, {Mongui{\'o}}, {Montegriffo}, {Montero}, {Mor}, {Mora}, {Morbidelli}, {Morel}, {Morris}, {Muraveva}, {Murphy}, {Musella}, {Nagy}, {Noval}, {Oca{\~n}a}, {Ogden}, {Ordenovic}, {Osinde}, {Pagani}, {Pagano}, {Palaversa}, {Palicio}, {Pallas-Quintela}, {Panahi}, {Payne-Wardenaar}, {Pe{\~n}alosa Esteller}, {Penttil{\"a}}, {Pichon}, {Piersimoni}, {Pineau}, {Plachy}, {Plum}, {Poggio}, {Pr{\v{s}}a}, {Pulone}, {Racero}, {Ragaini}, {Rainer}, {Raiteri}, {Rambaux}, {Ramos}, {Ramos-Lerate}, {Re Fiorentin}, {Regibo}, {Richards}, {Rios Diaz}, {Ripepi}, {Riva}, {Rix}, {Rixon}, {Robichon}, {Robin}, {Robin}, {Roelens}, {Rogues}, {Rohrbasser}, {Romero-G{\'o}mez}, {Rowell}, {Royer}, {Ruz Mieres}, {Rybicki}, {Sadowski}, {S{\'a}ez N{\'u}{\~n}ez}, {Sagrist{\`a} Sell{\'e}s}, {Sahlmann}, {Salguero}, {Samaras}, {Sanchez Gimenez}, {Sanna}, {Santove{\~n}a}, {Sarasso}, {Schultheis}, {Sciacca}, {Segol}, {Segovia}, {S{\'e}gransan}, {Semeux}, {Shahaf}, {Siddiqui}, {Siebert},
  {Siltala}, {Silvelo}, {Slezak}, {Slezak}, {Smart}, {Snaith}, {Solano}, {Solitro}, {Souami}, {Souchay}, {Spagna}, {Spina}, {Spoto}, {Steele}, {Steidelm{\"u}ller}, {Stephenson}, {S{\"u}veges}, {Surdej}, {Szabados}, {Szegedi-Elek}, {Taris}, {Taylor}, {Teixeira}, {Tolomei}, {Tonello}, {Torra}, {Torra}, {Torralba Elipe}, {Trabucchi}, {Tsounis}, {Turon}, {Ulla}, {Unger}, {Vaillant}, {van Dillen}, {van Reeven}, {Vanel}, {Vecchiato}, {Viala}, {Vicente}, {Voutsinas}, {Weiler}, {Wevers}, {Wyrzykowski}, {Yoldas}, {Yvard}, {Zhao}, {Zorec}, {Zucker}, \& {Zwitter}}]{Gaia2023}
{Gaia Collaboration}, {Vallenari}, A., {Brown}, A.~G.~A., {et~al.} 2023, \aap, 674, A1

\bibitem[{{Gao} {et~al.}(2020){Gao}, {Thorngren}, {Lee}, {Fortney}, {Morley}, {Wakeford}, {Powell}, {Stevenson}, \& {Zhang}}]{Gao2020}
{Gao}, P., {Thorngren}, D.~P., {Lee}, E. K.~H., {et~al.} 2020, Nature Astronomy, 4, 951

\bibitem[{{Gomes da Silva} {et~al.}(2018){Gomes da Silva}, {Figueira}, {Santos}, \& {Faria}}]{Gomes2018}
{Gomes da Silva}, J., {Figueira}, P., {Santos}, N., \& {Faria}, J. 2018, The Journal of Open Source Software, 3, 667

\bibitem[{{Guerrero} {et~al.}(2021){Guerrero}, {Seager}, {Huang}, {Vanderburg}, {Garcia Soto}, {Mireles}, {Hesse}, {Fong}, {Glidden}, {Shporer}, {Latham}, {Collins}, {Quinn}, {Burt}, {Dragomir}, {Crossfield}, {Vanderspek}, {Fausnaugh}, {Burke}, {Ricker}, {Daylan}, {Essack}, {G{\"u}nther}, {Osborn}, {Pepper}, {Rowden}, {Sha}, {Villanueva}, {Yahalomi}, {Yu}, {Ballard}, {Batalha}, {Berardo}, {Chontos}, {Dittmann}, {Esquerdo}, {Mikal-Evans}, {Jayaraman}, {Krishnamurthy}, {Louie}, {Mehrle}, {Niraula}, {Rackham}, {Rodriguez}, {Rowden}, {Sousa-Silva}, {Watanabe}, {Wong}, {Zhan}, {Zivanovic}, {Christiansen}, {Ciardi}, {Swain}, {Lund}, {Mullally}, {Fleming}, {Rodriguez}, {Boyd}, {Quintana}, {Barclay}, {Col{\'o}n}, {Rinehart}, {Schlieder}, {Clampin}, {Jenkins}, {Twicken}, {Caldwell}, {Coughlin}, {Henze}, {Lissauer}, {Morris}, {Rose}, {Smith}, {Tenenbaum}, {Ting}, {Wohler}, {Bakos}, {Bean}, {Berta-Thompson}, {Bieryla}, {Bouma}, {Buchhave}, {Butler}, {Charbonneau}, {Doty}, {Ge}, {Holman}, {Howard}, {Kaltenegger}, {Kane},
  {Kjeldsen}, {Kreidberg}, {Lin}, {Minsky}, {Narita}, {Paegert}, {P{\'a}l}, {Palle}, {Sasselov}, {Spencer}, {Sozzetti}, {Stassun}, {Torres}, {Udry}, \& {Winn}}]{Guerrero2021}
{Guerrero}, N.~M., {Seager}, S., {Huang}, C.~X., {et~al.} 2021, \apjs, 254, 39

\bibitem[{Guillot(2010)}]{guillot_radiative_2010}
Guillot, T. 2010, Astronomy and Astrophysics, 520, A27, arXiv:1006.4702

\bibitem[{{Gustafsson} {et~al.}(2008){Gustafsson}, {Edvardsson}, {Eriksson}, {J{\o}rgensen}, {Nordlund}, \& {Plez}}]{Gustafsson2008}
{Gustafsson}, B., {Edvardsson}, B., {Eriksson}, K., {et~al.} 2008, \aap, 486, 951

\bibitem[{Hakim {et~al.}(2018)Hakim, Rivoldini, Van~Hoolst, Cottenier, Jaeken, Chust, \& Steinle-Neumann}]{hakim_new_2018}
Hakim, K., Rivoldini, A., Van~Hoolst, T., {et~al.} 2018, Icarus, 313, 61

\bibitem[{Haldemann {et~al.}(2020)Haldemann, Alibert, Mordasini, \& Benz}]{haldemann_aqua_2020}
Haldemann, J., Alibert, Y., Mordasini, C., \& Benz, W. 2020, Astronomy \& Astrophysics, 643, A105, arXiv:2009.10098 [astro-ph]

\bibitem[{{Hayward} {et~al.}(2001){Hayward}, {Brandl}, {Pirger}, {Blacken}, {Gull}, {Schoenwald}, \& {Houck}}]{hayward2001}
{Hayward}, T.~L., {Brandl}, B., {Pirger}, B., {et~al.} 2001, \pasp, 113, 105

\bibitem[{{H{\'e}brard} {et~al.}(2016){H{\'e}brard}, {Arnold}, {Forveille}, {Correia}, {Laskar}, {Bonfils}, {Boisse}, {D{\'\i}az}, {Hagelberg}, {Sahlmann}, {Santos}, {Astudillo-Defru}, {Borgniet}, {Bouchy}, {Bourrier}, {Courcol}, {Delfosse}, {Deleuil}, {Demangeon}, {Ehrenreich}, {Gregorio}, {Jovanovic}, {Labrevoir}, {Lagrange}, {Lovis}, {Lozi}, {Moutou}, {Montagnier}, {Pepe}, {Rey}, {Santerne}, {S{\'e}gransan}, {Udry}, {Vanhuysse}, {Vigan}, \& {Wilson}}]{Hebrard2016}
{H{\'e}brard}, G., {Arnold}, L., {Forveille}, T., {et~al.} 2016, \aap, 588, A145

\bibitem[{{Heidari} {et~al.}(2024){Heidari}, {Boisse}, {Hara}, {Wilson}, {Kiefer}, {H{\'e}brard}, {Philipot}, {Hoyer}, {Stassun}, {Henry}, {Santos}, {Acu{\~n}a}, {Almasian}, {Arnold}, {Astudillo-Defru}, {Attia}, {Bonfils}, {Bouchy}, {Bourrier}, {Collet}, {Cort{\'e}s-Zuleta}, {Carmona}, {Delfosse}, {Dalal}, {Deleuil}, {Demangeon}, {D{\'\i}az}, {Dumusque}, {Ehrenreich}, {Forveille}, {Hobson}, {Jenkins}, {Jenkins}, {Lagrange}, {Latham}, {Larue}, {Liu}, {Moutou}, {Mignon}, {Osborn}, {Pepe}, {Rapetti}, {Rodrigues}, {Santerne}, {Segransan}, {Shporer}, {Sulis}, {Torres}, {Udry}, {Vakili}, {Vanderburg}, {Venot}, {Vivien}, \& {Vines}}]{Heidari2024}
{Heidari}, N., {Boisse}, I., {Hara}, N.~C., {et~al.} 2024, \aap, 681, A55

\bibitem[{{Heidari} {et~al.}(2025){Heidari}, {H{\'e}brard}, {Martioli}, {Eastman}, {Jackson}, {Delfosse}, {Jord{\'a}n}, {Correia}, {Sousa}, {Dragomir}, {Forveille}, {Boisse}, {Giacalone}, {D{\'\i}az}, {Brahm}, {Almasian}, {Almenara}, {Bieryla}, {Barkaoui}, {Baker}, {Barros}, {Bonfils}, {Carmona}, {Collins}, {Cort{\'e}s-Zuleta}, {Deleuil}, {Demangeon}, {Edwards}, {Eberhardt}, {Espinoza}, {Eisner}, {Feliz}, {Frommer}, {Fukui}, {Grau}, {Gupta}, {Hara}, {Hobson}, {Henning}, {Howell}, {Jenkins}, {Kiefer}, {LaCourse}, {Laskar}, {Law}, {Mann}, {Murgas}, {Moutou}, {Narita}, {Palle}, {Relles}, {Stassun}, {Serrano Bell}, {Schwarz}, {Srdoc}, {Str{\o}m}, {Safonov}, {Sarkis}, {Schlecker}, {Tala Pinto}, {Pepper}, {Rojas}, {Twicken}, {Trifonov}, {Villanueva}, {Watkins}, {Winn}, \& {Ziegler}}]{Heidari2025}
{Heidari}, N., {H{\'e}brard}, G., {Martioli}, E., {et~al.} 2025, \aap, 694, A36

\bibitem[{Hemley {et~al.}(1992)Hemley, Stixrude, Fei, \& Mao}]{hemley_constraints_1992}
Hemley, R.~J., Stixrude, L., Fei, Y., \& Mao, H.~K. 1992, in High-{Pressure} {Research}: {Application} to {Earth} and {Planetary} {Sciences} (American Geophysical Union; AGU), 183--189

\bibitem[{{Henden} {et~al.}(2016){Henden}, {Templeton}, {Terrell}, {Smith}, {Levine}, \& {Welch}}]{Henden2016}
{Henden}, A.~A., {Templeton}, M., {Terrell}, D., {et~al.} 2016, {AAVSO Photometric All Sky Survey (APASS) DR9}, VizieR On-line Data Catalog: II/336. Published in: 2015AAS...22533616H

\bibitem[{{Hirsch} {et~al.}(2021){Hirsch}, {Rosenthal}, {Fulton}, {Howard}, {Ciardi}, {Marcy}, {Nielsen}, {Petigura}, {de Rosa}, {Isaacson}, {Weiss}, {Sinukoff}, \& {Macintosh}}]{Hirsch2021}
{Hirsch}, L.~A., {Rosenthal}, L., {Fulton}, B.~J., {et~al.} 2021, \aj, 161, 134

\bibitem[{{H{\o}g} {et~al.}(2000){H{\o}g}, {Fabricius}, {Makarov}, {Urban}, {Corbin}, {Wycoff}, {Bastian}, {Schwekendiek}, \& {Wicenec}}]{hog}
{H{\o}g}, E., {Fabricius}, C., {Makarov}, V.~V., {et~al.} 2000, \aap, 355, L27

\bibitem[{{Howell} {et~al.}(2011){Howell}, {Everett}, {Sherry}, {Horch}, \& {Ciardi}}]{Howell2011}
{Howell}, S.~B., {Everett}, M.~E., {Sherry}, W., {Horch}, E., \& {Ciardi}, D.~R. 2011, \aj, 142, 19

\bibitem[{{Howell} {et~al.}(2021){Howell}, {Matson}, {Ciardi}, {Everett}, {Livingston}, {Scott}, {Horch}, \& {Winn}}]{Howell2021}
{Howell}, S.~B., {Matson}, R.~A., {Ciardi}, D.~R., {et~al.} 2021, \aj, 161, 164

\bibitem[{{Ichikawa} \& {Tsuchiya}(2020)}]{ichikawa_ab_2020}
{Ichikawa}, H. \& {Tsuchiya}, T. 2020, Minerals, 10, 59

\bibitem[{{Jenkins}(2002)}]{jenkins2002}
{Jenkins}, J.~M. 2002, \apj, 575, 493

\bibitem[{{Jenkins} {et~al.}(2010){Jenkins}, {Chandrasekaran}, {McCauliff}, {Caldwell}, {Tenenbaum}, {Li}, {Klaus}, {Cote}, \& {Middour}}]{jenkins2010}
{Jenkins}, J.~M., {Chandrasekaran}, H., {McCauliff}, S.~D., {et~al.} 2010, in SPIE Conference Series, Vol. 7740, Software and Cyberinfrastructure for Astronomy, ed. N.~M. {Radziwill} \& A.~{Bridger}, 77400D

\bibitem[{{Jenkins} {et~al.}(2020){Jenkins}, {Tenenbaum}, {Seader}, {Burke}, {McCauliff}, {Smith}, {Twicken}, \& {Chandrasekaran}}]{Jenkins2020b}
{Jenkins}, J.~M., {Tenenbaum}, P., {Seader}, S., {et~al.} 2020, {Kepler Data Processing Handbook: Transiting Planet Search}, Kepler Science Document KSCI-19081-003, id. 9. Edited by Jon M. Jenkins.

\bibitem[{{Jenkins} {et~al.}(2016){Jenkins}, {Twicken}, {McCauliff}, {Campbell}, {Sanderfer}, {Lung}, {Mansouri-Samani}, {Girouard}, {Tenenbaum}, {Klaus}, {Smith}, {Caldwell}, {Chacon}, {Henze}, {Heiges}, {Latham}, {Morgan}, {Swade}, {Rinehart}, \& {Vanderspek}}]{Jenkins2016}
{Jenkins}, J.~M., {Twicken}, J.~D., {McCauliff}, S., {et~al.} 2016, in Society of Photo-Optical Instrumentation Engineers (SPIE) Conference Series, Vol. 9913, Software and Cyberinfrastructure for Astronomy IV, ed. G.~{Chiozzi} \& J.~C. {Guzman}, 99133E

\bibitem[{{Jensen}(2013)}]{Jensen2013}
{Jensen}, E. 2013, {Tapir: A web interface for transit/eclipse observability}, Astrophysics Source Code Library

\bibitem[{{Johnson} \& {Soderblom}(1987)}]{JohnsonSoderblom87}
{Johnson}, D. R.~H. \& {Soderblom}, D.~R. 1987, \aj, 93, 864

\bibitem[{Kass \& Raftery(1995)}]{Kass1995}
Kass, R.~E. \& Raftery, A.~E. 1995, Journal of the American Statistical Association, 90, 773

\bibitem[{{Kempton} {et~al.}(2018){Kempton}, {Bean}, {Louie}, {Deming}, {Koll}, {Mansfield}, {Christiansen}, {L{\'o}pez-Morales}, {Swain}, {Zellem}, {Ballard}, {Barclay}, {Barstow}, {Batalha}, {Beatty}, {Berta-Thompson}, {Birkby}, {Buchhave}, {Charbonneau}, {Cowan}, {Crossfield}, {de Val-Borro}, {Doyon}, {Dragomir}, {Gaidos}, {Heng}, {Hu}, {Kane}, {Kreidberg}, {Mallonn}, {Morley}, {Narita}, {Nascimbeni}, {Pall{\'e}}, {Quintana}, {Rauscher}, {Seager}, {Shkolnik}, {Sing}, {Sozzetti}, {Stassun}, {Valenti}, \& {von Essen}}]{Kempton2018}
{Kempton}, E. M.~R., {Bean}, J.~L., {Louie}, D.~R., {et~al.} 2018, \pasp, 130, 114401

\bibitem[{{Kempton} \& {Knutson}(2024)}]{Kempton2024}
{Kempton}, E. M.~R. \& {Knutson}, H.~A. 2024, Reviews in Mineralogy and Geochemistry, 90, 411

\bibitem[{{Kervella} {et~al.}(2022){Kervella}, {Arenou}, \& {Th{\'e}venin}}]{Kervella2022}
{Kervella}, P., {Arenou}, F., \& {Th{\'e}venin}, F. 2022, \aap, 657, A7

\bibitem[{{Kipping}(2013)}]{Kipping2013}
{Kipping}, D.~M. 2013, \mnras, 434, L51

\bibitem[{{Kirk} {et~al.}(2025){Kirk}, {Ahrer}, {Claringbold}, {Zamyatina}, {Fisher}, {McCormack}, {Panwar}, {Powell}, {Taylor}, {Thorngren}, {Christie}, {Esparza-Borges}, {Tsai}, {Alderson}, {Booth}, {Fairman}, {L{\'o}pez-Morales}, {Mayne}, {Meech}, {Molli{\`e}re}, {Owen}, {Penzlin}, {Sergeev}, {Valentine}, {Wakeford}, \& {Wheatley}}]{Kirk2025}
{Kirk}, J., {Ahrer}, E.-M., {Claringbold}, A.~B., {et~al.} 2025, \mnras [\eprint[arXiv]{2410.08116}]

\bibitem[{{Kochanek} {et~al.}(2017){Kochanek}, {Shappee}, {Stanek}, {Holoien}, {Thompson}, {Prieto}, {Dong}, {Shields}, {Will}, {Britt}, {Perzanowski}, \& {Pojma{\'n}ski}}]{Kochanek2017}
{Kochanek}, C.~S., {Shappee}, B.~J., {Stanek}, K.~Z., {et~al.} 2017, \pasp, 129, 104502

\bibitem[{{Kraus} {et~al.}(2012){Kraus}, {Ireland}, {Hillenbrand}, \& {Martinache}}]{Kraus2012}
{Kraus}, A.~L., {Ireland}, M.~J., {Hillenbrand}, L.~A., \& {Martinache}, F. 2012, \apj, 745, 19

\bibitem[{{Kraus} {et~al.}(2016){Kraus}, {Ireland}, {Huber}, {Mann}, \& {Dupuy}}]{Kraus2016}
{Kraus}, A.~L., {Ireland}, M.~J., {Huber}, D., {Mann}, A.~W., \& {Dupuy}, T.~J. 2016, \aj, 152, 8

\bibitem[{{Lanotte} {et~al.}(2014){Lanotte}, {Gillon}, {Demory}, {Fortney}, {Astudillo}, {Bonfils}, {Magain}, {Delfosse}, {Forveille}, {Lovis}, {Mayor}, {Neves}, {Pepe}, {Queloz}, {Santos}, \& {Udry}}]{Lanotte2014}
{Lanotte}, A.~A., {Gillon}, M., {Demory}, B.~O., {et~al.} 2014, \aap, 572, A73

\bibitem[{{Latham} {et~al.}(2011){Latham}, {Rowe}, {Quinn}, {Batalha}, {Borucki}, {Brown}, {Bryson}, {Buchhave}, {Caldwell}, {Carter}, {Christiansen}, {Ciardi}, {Cochran}, {Dunham}, {Fabrycky}, {Ford}, {Gautier}, {Gilliland}, {Holman}, {Howell}, {Ibrahim}, {Isaacson}, {Jenkins}, {Koch}, {Lissauer}, {Marcy}, {Quintana}, {Ragozzine}, {Sasselov}, {Shporer}, {Steffen}, {Welsh}, \& {Wohler}}]{Latham2011}
{Latham}, D.~W., {Rowe}, J.~F., {Quinn}, S.~N., {et~al.} 2011, \apjl, 732, L24

\bibitem[{{Leconte} {et~al.}(2010){Leconte}, {Chabrier}, {Baraffe}, \& {Levrard}}]{Leconteetal10}
{Leconte}, J., {Chabrier}, G., {Baraffe}, I., \& {Levrard}, B. 2010, \aap, 516, A64

\bibitem[{{Lester} {et~al.}(2021){Lester}, {Matson}, {Howell}, {Furlan}, {Gnilka}, {Scott}, {Ciardi}, {Everett}, {Hartman}, \& {Hirsch}}]{Lester2021}
{Lester}, K.~V., {Matson}, R.~A., {Howell}, S.~B., {et~al.} 2021, \aj, 162, 75

\bibitem[{Li {et~al.}(2019)Li, Tenenbaum, Twicken, Burke, Jenkins, Quintana, Rowe, \& Seader}]{Li2019}
Li, J., Tenenbaum, P., Twicken, J.~D., {et~al.} 2019, \pasp, 131, 1

\bibitem[{{Lightkurve Collaboration} {et~al.}(2018){Lightkurve Collaboration}, {Cardoso}, {Hedges}, {Gully-Santiago}, {Saunders}, {Cody}, {Barclay}, {Hall}, {Sagear}, {Turtelboom}, {Zhang}, {Tzanidakis}, {Mighell}, {Coughlin}, {Bell}, {Berta-Thompson}, {Williams}, {Dotson}, \& {Barentsen}}]{lightkurve}
{Lightkurve Collaboration}, {Cardoso}, J. V. d.~M., {Hedges}, C., {et~al.} 2018, {Lightkurve: Kepler and TESS time series analysis in Python}, Astrophysics Source Code Library, record ascl:1812.013

\bibitem[{{Lindegren} {et~al.}(2018){Lindegren}, {Hern{\'a}ndez}, {Bombrun}, {Klioner}, {Bastian}, {Ramos-Lerate}, {de Torres}, {Steidelm{\"u}ller}, {Stephenson}, {Hobbs}, {Lammers}, {Biermann}, {Geyer}, {Hilger}, {Michalik}, {Stampa}, {McMillan}, {Casta{\~n}eda}, {Clotet}, {Comoretto}, {Davidson}, {Fabricius}, {Gracia}, {Hambly}, {Hutton}, {Mora}, {Portell}, {van Leeuwen}, {Abbas}, {Abreu}, {Altmann}, {Andrei}, {Anglada}, {Balaguer-N{\'u}{\~n}ez}, {Barache}, {Becciani}, {Bertone}, {Bianchi}, {Bouquillon}, {Bourda}, {Br{\"u}semeister}, {Bucciarelli}, {Busonero}, {Buzzi}, {Cancelliere}, {Carlucci}, {Charlot}, {Cheek}, {Crosta}, {Crowley}, {de Bruijne}, {de Felice}, {Drimmel}, {Esquej}, {Fienga}, {Fraile}, {Gai}, {Garralda}, {Gonz{\'a}lez-Vidal}, {Guerra}, {Hauser}, {Hofmann}, {Holl}, {Jordan}, {Lattanzi}, {Lenhardt}, {Liao}, {Licata}, {Lister}, {L{\"o}ffler}, {Marchant}, {Martin-Fleitas}, {Messineo}, {Mignard}, {Morbidelli}, {Poggio}, {Riva}, {Rowell}, {Salguero}, {Sarasso}, {Sciacca}, {Siddiqui}, {Smart},
  {Spagna}, {Steele}, {Taris}, {Torra}, {van Elteren}, {van Reeven}, \& {Vecchiato}}]{Lindegren2018}
{Lindegren}, L., {Hern{\'a}ndez}, J., {Bombrun}, A., {et~al.} 2018, \aap, 616, A2

\bibitem[{{Lindegren} {et~al.}(2021){Lindegren}, {Klioner}, {Hern{\'a}ndez}, {Bombrun}, {Ramos-Lerate}, {Steidelm{\"u}ller}, {Bastian}, {Biermann}, {de Torres}, {Gerlach}, {Geyer}, {Hilger}, {Hobbs}, {Lammers}, {McMillan}, {Stephenson}, {Casta{\~n}eda}, {Davidson}, {Fabricius}, {Gracia-Abril}, {Portell}, {Rowell}, {Teyssier}, {Torra}, {Bartolom{\'e}}, {Clotet}, {Garralda}, {Gonz{\'a}lez-Vidal}, {Torra}, {Abbas}, {Altmann}, {Anglada Varela}, {Balaguer-N{\'u}{\~n}ez}, {Balog}, {Barache}, {Becciani}, {Bernet}, {Bertone}, {Bianchi}, {Bouquillon}, {Brown}, {Bucciarelli}, {Busonero}, {Butkevich}, {Buzzi}, {Cancelliere}, {Carlucci}, {Charlot}, {Cioni}, {Crosta}, {Crowley}, {del Peloso}, {del Pozo}, {Drimmel}, {Esquej}, {Fienga}, {Fraile}, {Gai}, {Garcia-Reinaldos}, {Guerra}, {Hambly}, {Hauser}, {Jan{\ss}en}, {Jordan}, {Kostrzewa-Rutkowska}, {Lattanzi}, {Liao}, {Licata}, {Lister}, {L{\"o}ffler}, {Marchant}, {Masip}, {Mignard}, {Mints}, {Molina}, {Mora}, {Morbidelli}, {Murphy}, {Pagani}, {Panuzzo}, {Pe{\~n}alosa
  Esteller}, {Poggio}, {Re Fiorentin}, {Riva}, {Sagrist{\`a} Sell{\'e}s}, {Sanchez Gimenez}, {Sarasso}, {Sciacca}, {Siddiqui}, {Smart}, {Souami}, {Spagna}, {Steele}, {Taris}, {Utrilla}, {van Reeven}, \& {Vecchiato}}]{Lindegren2021}
{Lindegren}, L., {Klioner}, S.~A., {Hern{\'a}ndez}, J., {et~al.} 2021, \aap, 649, A2

\bibitem[{{Lundkvist} {et~al.}(2016){Lundkvist}, {Kjeldsen}, {Albrecht}, {Davies}, {Basu}, {Huber}, {Justesen}, {Karoff}, {Silva Aguirre}, {van Eylen}, {Vang}, {Arentoft}, {Barclay}, {Bedding}, {Campante}, {Chaplin}, {Christensen-Dalsgaard}, {Elsworth}, {Gilliland}, {Handberg}, {Hekker}, {Kawaler}, {Lund}, {Metcalfe}, {Miglio}, {Rowe}, {Stello}, {Tingley}, \& {White}}]{Lundkvist2016}
{Lundkvist}, M.~S., {Kjeldsen}, H., {Albrecht}, S., {et~al.} 2016, Nature Communications, 7, 11201

\bibitem[{{Luo} {et~al.}(2024){Luo}, {Dorn}, \& {Deng}}]{luo_majority_2024}
{Luo}, H., {Dorn}, C., \& {Deng}, J. 2024, Nature Astronomy, 8, 1399

\bibitem[{{Madhusudhan} {et~al.}(2017){Madhusudhan}, {Bitsch}, {Johansen}, \& {Eriksson}}]{Madhusudhan2017}
{Madhusudhan}, N., {Bitsch}, B., {Johansen}, A., \& {Eriksson}, L. 2017, \mnras, 469, 4102

\bibitem[{{Mamajek} \& {Hillenbrand}(2008)}]{MamajekHillenbrand08}
{Mamajek}, E.~E. \& {Hillenbrand}, L.~A. 2008, \apj, 687, 1264

\bibitem[{{Mardling}(2007)}]{Mardling07}
{Mardling}, R.~A. 2007, \mnras, 382, 1768

\bibitem[{Marelli \& Sudret(2014)}]{marelli2014uqlab}
Marelli, S. \& Sudret, B. 2014, in Vulnerability, uncertainty, and risk (2nd Int. Conf. on Vulnerability, Risk Analysis and Management), 2554--2563

\bibitem[{{Matsakos} \& {K{\"o}nigl}(2016)}]{Matsakos2016}
{Matsakos}, T. \& {K{\"o}nigl}, A. 2016, \apjl, 820, L8

\bibitem[{{Matson} {et~al.}(2018){Matson}, {Howell}, {Horch}, \& {Everett}}]{Matson2018}
{Matson}, R.~A., {Howell}, S.~B., {Horch}, E.~P., \& {Everett}, M.~E. 2018, \aj, 156, 31

\bibitem[{{Mayor} {et~al.}(2003){Mayor}, {Pepe}, {Queloz}, {Bouchy}, {Rupprecht}, {Lo Curto}, {Avila}, {Benz}, {Bertaux}, {Bonfils}, {Dall}, {Dekker}, {Delabre}, {Eckert}, {Fleury}, {Gilliotte}, {Gojak}, {Guzman}, {Kohler}, {Lizon}, {Longinotti}, {Lovis}, {Megevand}, {Pasquini}, {Reyes}, {Sivan}, {Sosnowska}, {Soto}, {Udry}, {van Kesteren}, {Weber}, \& {Weilenmann}}]{Mayor2003}
{Mayor}, M., {Pepe}, F., {Queloz}, D., {et~al.} 2003, The Messenger, 114, 20

\bibitem[{{Mazeh} {et~al.}(2016){Mazeh}, {Holczer}, \& {Faigler}}]{Mazeh2016}
{Mazeh}, T., {Holczer}, T., \& {Faigler}, S. 2016, \aap, 589, A75

\bibitem[{{McCully} {et~al.}(2018){McCully}, {Volgenau}, {Harbeck}, {Lister}, {Saunders}, {Turner}, {Siiverd}, \& {Bowman}}]{McCully2018}
{McCully}, C., {Volgenau}, N.~H., {Harbeck}, D.-R., {et~al.} 2018, in SPIE Conference Series, Vol. 10707, \procspie, 107070K

\bibitem[{{Melosh}(2007)}]{melosh_hydrocode_2007}
{Melosh}, H.~J. 2007, Meteoritics \& Planetary Science, 42, 2079

\bibitem[{Miozzi {et~al.}(2020)Miozzi, Matas, Guignot, Badro, Siebert, \& Fiquet}]{miozzi_new_2020}
Miozzi, F., Matas, J., Guignot, N., {et~al.} 2020, Minerals, 10, 100, number: 2 Publisher: Multidisciplinary Digital Publishing Institute

\bibitem[{{Moe} \& {Kratter}(2021)}]{Moe2021}
{Moe}, M. \& {Kratter}, K.~M. 2021, \mnras, 507, 3593

\bibitem[{Musella {et~al.}(2019)Musella, Mazevet, \& Guyot}]{musella_physical_2019}
Musella, R., Mazevet, S., \& Guyot, F. 2019, Physical Review B, 99, 064110, publisher: American Physical Society

\bibitem[{{Naponiello} {et~al.}(2025){Naponiello}, {Bonomo}, {Mancini}, {Steinmeyer}, {Biazzo}, {Polychroni}, {Dorn}, {Turrini}, {Lanza}, {Sozzetti}, {Desidera}, {Damasso}, {Collins}, {Carleo}, {Collins}, {Colombo}, {D'Arpa}, {Dumusque}, {Gonz{\'a}lez}, {Guilluy}, {Lorenzi}, {Mantovan}, {Nardiello}, {Pinamonti}, {Schwarz}, {Singh}, {Watkins}, \& {Zingales}}]{Naponiello2025}
{Naponiello}, L., {Bonomo}, A.~S., {Mancini}, L., {et~al.} 2025, \aap, 693, A7

\bibitem[{{Naponiello} {et~al.}(2022){Naponiello}, {Mancini}, {Damasso}, {Bonomo}, {Sozzetti}, {Nardiello}, {Biazzo}, {Stognone}, {Lillo-Box}, {Lanza}, {Poretti}, {Lissauer}, {Zeng}, {Bieryla}, {H{\'e}brard}, {Basilicata}, {Benatti}, {Bignamini}, {Borsa}, {Claudi}, {Cosentino}, {Covino}, {de Gurtubai}, {Delfosse}, {Desidera}, {Dragomir}, {Eastman}, {Essack}, {Fiorenzano}, {Giacobbe}, {Harutyunyan}, {Heidari}, {Hellier}, {Jenkins}, {Knapic}, {K{\"o}nig}, {Latham}, {Magazz{\`u}}, {Maggio}, {Maldonado}, {Micela}, {Molinari}, {Molinaro}, {Morgan}, {Moutou}, {Nascimbeni}, {Pace}, {Pagano}, {Pedani}, {Piotto}, {Pinamonti}, {Quintana}, {Rainer}, {Ricker}, {Seager}, {Twicken}, {Vanderspek}, \& {Winn}}]{Naponiello2022}
{Naponiello}, L., {Mancini}, L., {Damasso}, M., {et~al.} 2022, \aap, 667, A8

\bibitem[{{Naponiello} {et~al.}(2023){Naponiello}, {Mancini}, {Sozzetti}, {Bonomo}, {Morbidelli}, {Dou}, {Zeng}, {Leinhardt}, {Biazzo}, {Cubillos}, {Pinamonti}, {Locci}, {Maggio}, {Damasso}, {Lanza}, {Lissauer}, {Collins}, {Carter}, {Jensen}, {Bignamini}, {Boschin}, {Bouma}, {Ciardi}, {Cosentino}, {Crossfield}, {Desidera}, {Dumusque}, {Fiorenzano}, {Fukui}, {Giacobbe}, {Gnilka}, {Ghedina}, {Guilluy}, {Harutyunyan}, {Howell}, {Jenkins}, {Lund}, {Kielkopf}, {Lester}, {Malavolta}, {Mann}, {Matson}, {Matthews}, {Nardiello}, {Narita}, {Pace}, {Pagano}, {Palle}, {Pedani}, {Seager}, {Schlieder}, {Schwarz}, {Shporer}, {Twicken}, {Winn}, {Ziegler}, \& {Zingales}}]{Naponiello2023}
{Naponiello}, L., {Mancini}, L., {Sozzetti}, A., {et~al.} 2023, \nat, 622, 255

\bibitem[{{Noyes} {et~al.}(1984){Noyes}, {Hartmann}, {Baliunas}, {Duncan}, \& {Vaughan}}]{Noyesetal84}
{Noyes}, R.~W., {Hartmann}, L.~W., {Baliunas}, S.~L., {Duncan}, D.~K., \& {Vaughan}, A.~H. 1984, \apj, 279, 763

\bibitem[{{{\"O}berg} {et~al.}(2011){{\"O}berg}, {Murray-Clay}, \& {Bergin}}]{Oberg2011}
{{\"O}berg}, K.~I., {Murray-Clay}, R., \& {Bergin}, E.~A. 2011, \apjl, 743, L16

\bibitem[{{Ogilvie}(2014)}]{Ogilvie14}
{Ogilvie}, G.~I. 2014, \araa, 52, 171

\bibitem[{{Osborn} {et~al.}(2023){Osborn}, {Armstrong}, {Fern{\'a}ndez Fern{\'a}ndez}, {Knierim}, {Adibekyan}, {Collins}, {Delgado-Mena}, {Fridlund}, {Gomes da Silva}, {Hellier}, {Jackson}, {King}, {Lillo-Box}, {Matson}, {Matthews}, {Santos}, {Sousa}, {Stassun}, {Tan}, {Ricker}, {Vanderspek}, {Latham}, {Seager}, {Winn}, {Jenkins}, {Bayliss}, {Bouma}, {Ciardi}, {Collins}, {Col{\'o}n}, {Crossfield}, {Demangeon}, {D{\'\i}az}, {Dorn}, {Dumusque}, {Keniger}, {Figueira}, {Gan}, {Goeke}, {Hadjigeorghiou}, {Hawthorn}, {Helled}, {Howell}, {Nielsen}, {Osborn}, {Quinn}, {Sefako}, {Shporer}, {Str{\o}m}, {Twicken}, {Vanderburg}, \& {Wheatley}}]{Osborn2023}
{Osborn}, A., {Armstrong}, D.~J., {Fern{\'a}ndez Fern{\'a}ndez}, J., {et~al.} 2023, \mnras, 526, 548

\bibitem[{{Otsubo} {et~al.}(2024){Otsubo}, {Sarugaku}, {Takeuchi}, {Ikeda}, {Matsunaga}, {McWilliam}, {Hull}, {Yoshikawa}, {Katoh}, {Kondo}, {Hamano}, {Taniguchi}, \& {Kawakita}}]{Otsubo2024}
{Otsubo}, S., {Sarugaku}, Y., {Takeuchi}, T., {et~al.} 2024, in SPIE Conference Series, Vol. 13096, Ground-based and Airborne Instrumentation for Astronomy X, ed. J.~J. {Bryant}, K.~{Motohara}, \& J.~R.~D. {Vernet}, 1309631

\bibitem[{{Owen} \& {Lai}(2018)}]{Owen2018}
{Owen}, J.~E. \& {Lai}, D. 2018, \mnras, 479, 5012

\bibitem[{{Paxton} {et~al.}(2015){Paxton}, {Marchant}, {Schwab}, {Bauer}, {Bildsten}, {Cantiello}, {Dessart}, {Farmer}, {Hu}, {Langer}, {Townsend}, {Townsley}, \& {Timmes}}]{Paxton2015}
{Paxton}, B., {Marchant}, P., {Schwab}, J., {et~al.} 2015, \apjs, 220, 15

\bibitem[{{Pecaut} \& {Mamajek}(2013)}]{PecautMamajek2013}
{Pecaut}, M.~J. \& {Mamajek}, E.~E. 2013, \apjs, 208, 9

\bibitem[{{Penzlin} {et~al.}(2024){Penzlin}, {Booth}, {Kirk}, {Owen}, {Ahrer}, {Christie}, {Claringbold}, {Esparza-Borges}, {L{\'o}pez-Morales}, {Mayne}, {McCormack}, {Meech}, {Panwar}, {Powell}, {Sergeev}, {Taylor}, {Wheatley}, \& {Zamyatina}}]{Penzlin2024}
{Penzlin}, A. B.~T., {Booth}, R.~A., {Kirk}, J., {et~al.} 2024, \mnras, 535, 171

\bibitem[{{Perruchot} {et~al.}(2008){Perruchot}, {Kohler}, {Bouchy}, {Richaud}, {Richaud}, {Moreaux}, {Merzougui}, {Sottile}, {Hill}, {Knispel}, {Regal}, {Meunier}, {Ilovaisky}, {Le Coroller}, {Gillet}, {Schmitt}, {Pepe}, {Fleury}, {Sosnowska}, {Vors}, {M{\'e}gevand}, {Blanc}, {Carol}, {Point}, {Laloge}, \& {Brunel}}]{Perruchot2008}
{Perruchot}, S., {Kohler}, D., {Bouchy}, F., {et~al.} 2008, in SPIE Conference Series, Vol. 7014, Ground-based and Airborne Instrumentation for Astronomy II, ed. I.~S. {McLean} \& M.~M. {Casali}, 70140J

\bibitem[{{Perryman} {et~al.}(1997){Perryman}, {Lindegren}, {Kovalevsky}, {Hoeg}, {Bastian}, {Bernacca}, {Cr{\'e}z{\'e}}, {Donati}, {Grenon}, {Grewing}, {van Leeuwen}, {van der Marel}, {Mignard}, {Murray}, {Le Poole}, {Schrijver}, {Turon}, {Arenou}, {Froeschl{\'e}}, \& {Petersen}}]{Perryman1997}
{Perryman}, M.~A.~C., {Lindegren}, L., {Kovalevsky}, J., {et~al.} 1997, \aap, 323, L49

\bibitem[{{Petigura} {et~al.}(2018){Petigura}, {Marcy}, {Winn}, {Weiss}, {Fulton}, {Howard}, {Sinukoff}, {Isaacson}, {Morton}, \& {Johnson}}]{Petigura2018}
{Petigura}, E.~A., {Marcy}, G.~W., {Winn}, J.~N., {et~al.} 2018, \aj, 155, 89

\bibitem[{{Petigura} {et~al.}(2022){Petigura}, {Rogers}, {Isaacson}, {Owen}, {Kraus}, {Winn}, {MacDougall}, {Howard}, {Fulton}, {Kosiarek}, {Weiss}, {Behmard}, \& {Blunt}}]{Petigura2022}
{Petigura}, E.~A., {Rogers}, J.~G., {Isaacson}, H., {et~al.} 2022, \aj, 163, 179

\bibitem[{{Piaulet-Ghorayeb} {et~al.}(2024){Piaulet-Ghorayeb}, {Benneke}, {Radica}, {Raul}, {Coulombe}, {Ahrer}, {Kubyshkina}, {Howard}, {Krissansen-Totton}, {MacDonald}, {Roy}, {Louca}, {Christie}, {Fournier-Tondreau}, {Allart}, {Miguel}, {Schlichting}, {Welbanks}, {Cadieux}, {Dorn}, {Evans-Soma}, {Fortney}, {Pierrehumbert}, {Lafreni{\`e}re}, {Acu{\~n}a}, {Komacek}, {Innes}, {Beatty}, {Cloutier}, {Doyon}, {Gagnebin}, {Gapp}, \& {Knutson}}]{Piaulet-Ghorayeb2024}
{Piaulet-Ghorayeb}, C., {Benneke}, B., {Radica}, M., {et~al.} 2024, \apjl, 974, L10

\bibitem[{Price-Whelan {et~al.}(2018)Price-Whelan, Sipőcz, Günther, Lim, Crawford, Conseil, Shupe, Craig, Dencheva, Ginsburg, VanderPlas, Bradley, Pérez-Suárez, de~Val-Borro, Aldcroft, Cruz, Robitaille, Tollerud, Ardelean, Babej, Bach, Bachetti, Bakanov, Bamford, Barentsen, Barmby, Baumbach, Berry, Biscani, Boquien, Bostroem, Bouma, Brammer, Bray, Breytenbach, Buddelmeijer, Burke, Calderone, Rodríguez, Cara, Cardoso, Cheedella, Copin, Corrales, Crichton, D’Avella, Deil, Depagne, Dietrich, Donath, Droettboom, Earl, Erben, Fabbro, Ferreira, Finethy, Fox, Garrison, Gibbons, Goldstein, Gommers, Greco, Greenfield, Groener, Grollier, Hagen, Hirst, Homeier, Horton, Hosseinzadeh, Hu, Hunkeler, Ivezić, Jain, Jenness, Kanarek, Kendrew, Kern, Kerzendorf, Khvalko, King, Kirkby, Kulkarni, Kumar, Lee, Lenz, Littlefair, Ma, Macleod, Mastropietro, McCully, Montagnac, Morris, Mueller, Mumford, Muna, Murphy, Nelson, Nguyen, Ninan, Nöthe, Ogaz, Oh, Parejko, Parley, Pascual, Patil, Patil, Plunkett, Prochaska, Rastogi,
  Janga, Sabater, Sakurikar, Seifert, Sherbert, Sherwood-Taylor, Shih, Sick, Silbiger, Singanamalla, Singer, Sladen, Sooley, Sornarajah, Streicher, Teuben, Thomas, Tremblay, Turner, Terrón, van Kerkwijk, de~la Vega, Watkins, Weaver, Whitmore, Woillez, \& Zabalza}]{Price2018}
Price-Whelan, A.~M., Sipőcz, B.~M., Günther, H.~M., {et~al.} 2018, \aj, 156, 123

\bibitem[{{Reggiani} {et~al.}(2022){Reggiani}, {Schlaufman}, {Healy}, {Lothringer}, \& {Sing}}]{Reggiani2022}
{Reggiani}, H., {Schlaufman}, K.~C., {Healy}, B.~F., {Lothringer}, J.~D., \& {Sing}, D.~K. 2022, in the 21st Cambridge Workshop on cool stars, stellar systems, and the Sun, 64

\bibitem[{{Saumon} {et~al.}(1995){Saumon}, {Chabrier}, \& {van Horn}}]{1995_Saumon_EOS}
{Saumon}, D., {Chabrier}, G., \& {van Horn}, H.~M. 1995, \apjs, 99, 713

\bibitem[{Schlichting \& Young(2022)}]{schlichting2022chemical}
Schlichting, H.~E. \& Young, E.~D. 2022, The Planetary Science Journal, 3, 127

\bibitem[{Schobi {et~al.}(2015)Schobi, Sudret, \& Wiart}]{schobi2015polynomial}
Schobi, R., Sudret, B., \& Wiart, J. 2015, International Journal for Uncertainty Quantification, 5

\bibitem[{{Scott} {et~al.}(2021){Scott}, {Howell}, {Gnilka}, {Stephens}, {Salinas}, {Matson}, {Furlan}, {Horch}, {Everett}, {Ciardi}, {Mills}, \& {Quigley}}]{Scott2021}
{Scott}, N.~J., {Howell}, S.~B., {Gnilka}, C.~L., {et~al.} 2021, Frontiers in Astronomy and Space Sciences, 8, 138

\bibitem[{{Shappee} {et~al.}(2014){Shappee}, {Prieto}, {Grupe}, {Kochanek}, {Stanek}, {De Rosa}, {Mathur}, {Zu}, {Peterson}, {Pogge}, {Komossa}, {Im}, {Jencson}, {Holoien}, {Basu}, {Beacom}, {Szczygie{\l}}, {Brimacombe}, {Adams}, {Campillay}, {Choi}, {Contreras}, {Dietrich}, {Dubberley}, {Elphick}, {Foale}, {Giustini}, {Gonzalez}, {Hawkins}, {Howell}, {Hsiao}, {Koss}, {Leighly}, {Morrell}, {Mudd}, {Mullins}, {Nugent}, {Parrent}, {Phillips}, {Pojmanski}, {Rosing}, {Ross}, {Sand}, {Terndrup}, {Valenti}, {Walker}, \& {Yoon}}]{Shappee2014}
{Shappee}, B.~J., {Prieto}, J.~L., {Grupe}, D., {et~al.} 2014, \apj, 788, 48

\bibitem[{{Sing} {et~al.}(2024){Sing}, {Rustamkulov}, {Thorngren}, {Barstow}, {Tremblin}, {Alves de Oliveira}, {Beck}, {Birkmann}, {Challener}, {Crouzet}, {Espinoza}, {Ferruit}, {Giardino}, {Gressier}, {Lee}, {Lewis}, {Maiolino}, {Manjavacas}, {Rauscher}, {Sirianni}, \& {Valenti}}]{Sing2024}
{Sing}, D.~K., {Rustamkulov}, Z., {Thorngren}, D.~P., {et~al.} 2024, \nat, 630, 831

\bibitem[{{Smart} {et~al.}(2019){Smart}, {Marocco}, {Sarro}, {Barrado}, {Beam{\'\i}n}, {Caballero}, \& {Jones}}]{Smart2019}
{Smart}, R.~L., {Marocco}, F., {Sarro}, L.~M., {et~al.} 2019, \mnras, 485, 4423

\bibitem[{{Smith} {et~al.}(2012){Smith}, {Stumpe}, {Van Cleve}, {Jenkins}, {Barclay}, {Fanelli}, {Girouard}, {Kolodziejczak}, {McCauliff}, {Morris}, \& {Twicken}}]{Smith2012}
{Smith}, J.~C., {Stumpe}, M.~C., {Van Cleve}, J.~E., {et~al.} 2012, \pasp, 124, 1000

\bibitem[{{Sneden}(1973)}]{sneden1973}
{Sneden}, C. 1973, \apj, 184, 839

\bibitem[{{Sousa} {et~al.}(2015){Sousa}, {Santos}, {Adibekyan}, {Delgado-Mena}, \& {Israelian}}]{Sousaetal2015}
{Sousa}, S.~G., {Santos}, N.~C., {Adibekyan}, V., {Delgado-Mena}, E., \& {Israelian}, G. 2015, \aap, 577, A67

\bibitem[{{Sozzetti}(2023)}]{Sozzetti2023}
{Sozzetti}, A. 2023, \aap, 670, L17

\bibitem[{Stassun {et~al.}(2019)Stassun, Oelkers, Paegert, Torres, Pepper, De~Lee, Collins, Latham, Muirhead, Chittidi, Rojas-Ayala, Fleming, Rose, Tenenbaum, Ting, Kane, Barclay, Bean, Brassuer, Charbonneau, Ge, Lissauer, Mann, McLean, Mullally, Narita, Plavchan, Ricker, Sasselov, Seager, Sharma, Shiao, Sozzetti, Stello, Vanderspek, Wallace, \& Winn}]{Stassun2019}
Stassun, K.~G., Oelkers, R.~J., Paegert, M., {et~al.} 2019, \aj, 158, 138

\bibitem[{{Stassun} {et~al.}(2018){Stassun}, {Oelkers}, {Pepper}, {Paegert}, {De Lee}, {Torres}, {Latham}, {Charpinet}, {Dressing}, {Huber}, {Kane}, {L{\'e}pine}, {Mann}, {Muirhead}, {Rojas-Ayala}, {Silvotti}, {Fleming}, {Levine}, \& {Plavchan}}]{Stassun2018}
{Stassun}, K.~G., {Oelkers}, R.~J., {Pepper}, J., {et~al.} 2018, \aj, 156, 102

\bibitem[{Stewart {et~al.}(2020)Stewart, Davies, Duncan, Lock, Root, Townsend, Kraus, Caracas, \& Jacobsen}]{stewart_shock_2020}
Stewart, S.~T., Davies, E.~J., Duncan, M.~S., {et~al.} 2020, American Institute of Physics Conference Series, 080003, arXiv:1910.04687

\bibitem[{Stixrude \& Lithgow-Bertelloni(2022)}]{stixrude_thermal_2022}
Stixrude, L. \& Lithgow-Bertelloni, C. 2022, Geophysical Journal International, 228, 1119

\bibitem[{{Stumpe} {et~al.}(2014){Stumpe}, {Smith}, {Catanzarite}, {Van Cleve}, {Jenkins}, {Twicken}, \& {Girouard}}]{Stumpe2014}
{Stumpe}, M.~C., {Smith}, J.~C., {Catanzarite}, J.~H., {et~al.} 2014, \pasp, 126, 100

\bibitem[{{Stumpe} {et~al.}(2012){Stumpe}, {Smith}, {Van Cleve}, {Twicken}, {Barclay}, {Fanelli}, {Girouard}, {Jenkins}, {Kolodziejczak}, {McCauliff}, \& {Morris}}]{Stumpe2012}
{Stumpe}, M.~C., {Smith}, J.~C., {Van Cleve}, J.~E., {et~al.} 2012, \pasp, 124, 985

\bibitem[{{Szab{\'o}} \& {Kiss}(2011)}]{Szabo2011}
{Szab{\'o}}, G.~M. \& {Kiss}, L.~L. 2011, \apjl, 727, L44

\bibitem[{Thompson(1990)}]{1990_thompson_aneos}
Thompson, S. 1990, Lab. Doc. SAND89-2951

\bibitem[{{Tokovinin}(2018)}]{Tokovinin2018}
{Tokovinin}, A. 2018, \pasp, 130, 035002

\bibitem[{{Twicken} {et~al.}(2018){Twicken}, {Catanzarite}, {Clarke}, {Girouard}, {Jenkins}, {Klaus}, {Li}, {McCauliff}, {Seader}, {Tenenbaum}, {Wohler}, {Bryson}, {Burke}, {Caldwell}, {Haas}, {Henze}, \& {Sanderfer}}]{Twicken2018}
{Twicken}, J.~D., {Catanzarite}, J.~H., {Clarke}, B.~D., {et~al.} 2018, \pasp, 130, 064502

\bibitem[{{Venturini} {et~al.}(2020){Venturini}, {Guilera}, {Haldemann}, {Ronco}, \& {Mordasini}}]{Venturini2020}
{Venturini}, J., {Guilera}, O.~M., {Haldemann}, J., {Ronco}, M.~P., \& {Mordasini}, C. 2020, \aap, 643, L1

\bibitem[{{Vissapragada} \& {Behmard}(2025)}]{Vissapragada2025}
{Vissapragada}, S. \& {Behmard}, A. 2025, \aj, 169, 117

\bibitem[{{Vissapragada} {et~al.}(2022){Vissapragada}, {Knutson}, {Greklek-McKeon}, {Oklop{\v{c}}i{\'c}}, {Dai}, {dos Santos}, {Jovanovic}, {Mawet}, {Millar-Blanchaer}, {Paragas}, {Spake}, {Tinyanont}, \& {Vasisht}}]{Vissapragada2022}
{Vissapragada}, S., {Knutson}, H.~A., {Greklek-McKeon}, M., {et~al.} 2022, \aj, 164, 234

\bibitem[{{Welbanks} {et~al.}(2024){Welbanks}, {Bell}, {Beatty}, {Line}, {Ohno}, {Fortney}, {Schlawin}, {Greene}, {Rauscher}, {McGill}, {Murphy}, {Parmentier}, {Tang}, {Edelman}, {Mukherjee}, {Wiser}, {Lagage}, {Dyrek}, \& {Arnold}}]{Welbanks2024}
{Welbanks}, L., {Bell}, T.~J., {Beatty}, T.~G., {et~al.} 2024, \nat, 630, 836

\bibitem[{Werlen {et~al.}(2025)Werlen, Dorn, Schlichting, Grimm, \& Young}]{werlen2025atmospheric}
Werlen, A., Dorn, C., Schlichting, H.~E., Grimm, S.~L., \& Young, E.~D. 2025, The Astrophysical Journal Letters, 988, L55

\bibitem[{{Zagaria} {et~al.}(2021){Zagaria}, {Rosotti}, \& {Lodato}}]{Zagaria2021}
{Zagaria}, F., {Rosotti}, G.~P., \& {Lodato}, G. 2021, \mnras, 504, 2235

\bibitem[{{Zechmeister} \& {K{\"u}rster}(2009)}]{Zechmeister2009}
{Zechmeister}, M. \& {K{\"u}rster}, M. 2009, \aap, 496, 577

\bibitem[{{Zechmeister} {et~al.}(2018){Zechmeister}, {Reiners}, {Amado}, {Azzaro}, {Bauer}, {B{\'e}jar}, {Caballero}, {Guenther}, {Hagen}, {Jeffers}, {Kaminski}, {K{\"u}rster}, {Launhardt}, {Montes}, {Morales}, {Quirrenbach}, {Reffert}, {Ribas}, {Seifert}, {Tal-Or}, \& {Wolthoff}}]{Zechmeister2018}
{Zechmeister}, M., {Reiners}, A., {Amado}, P.~J., {et~al.} 2018, \aap, 609, A12

\bibitem[{{Zeng} \& {Sasselov}(2013)}]{Zeng2013}
{Zeng}, L. \& {Sasselov}, D. 2013, \pasp, 125, 227

\bibitem[{{Ziegler} {et~al.}(2020){Ziegler}, {Tokovinin}, {Brice{\~n}o}, {Mang}, {Law}, \& {Mann}}]{Ziegler2020}
{Ziegler}, C., {Tokovinin}, A., {Brice{\~n}o}, C., {et~al.} 2020, \aj, 159, 19

\end{thebibliography}

\appendix


\section{TOI-5800\,b TTV analysis}\label{sec:TTV}

\begin{figure}[!h]
\centering
\includegraphics[width=0.5\textwidth]{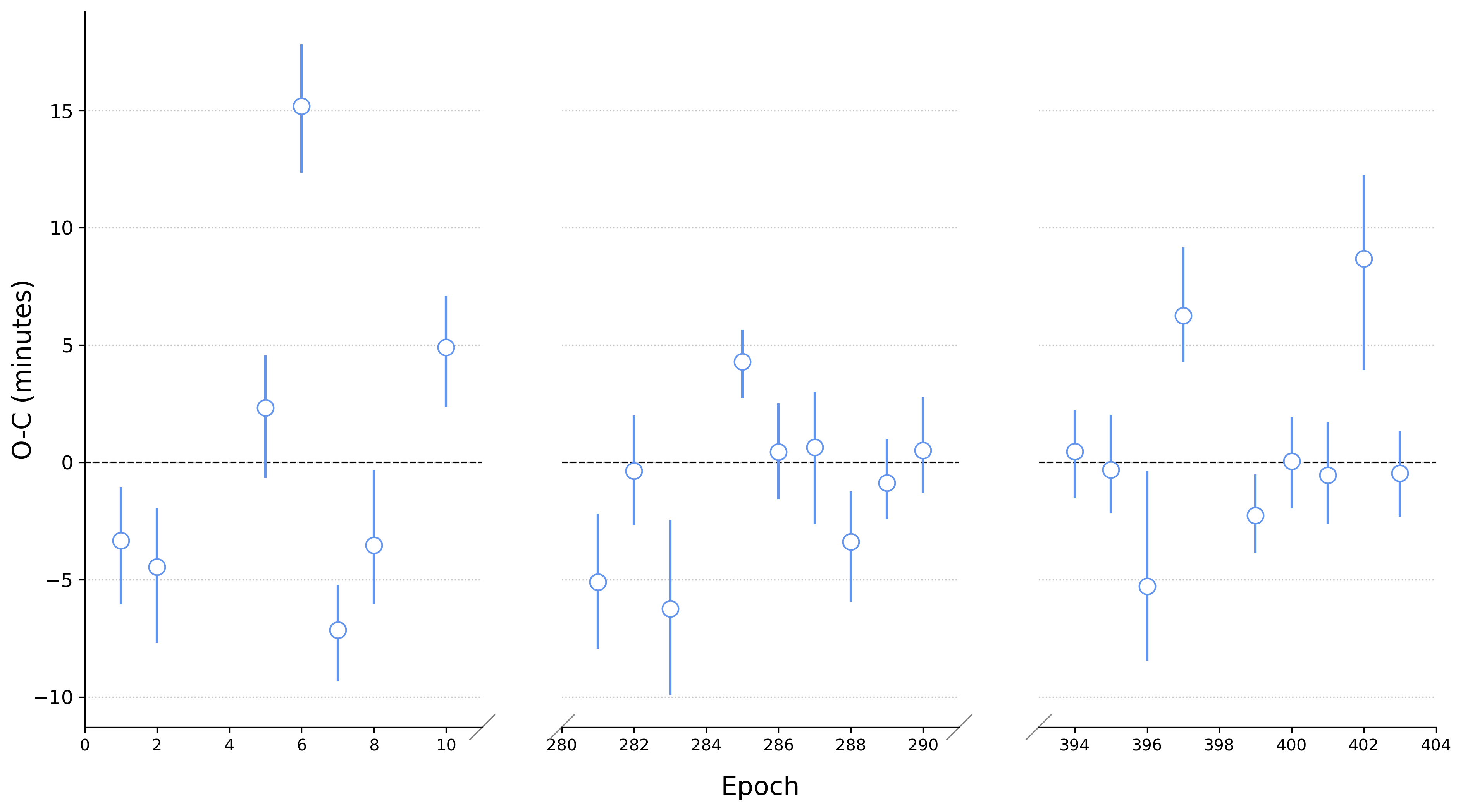}
\caption{Observed minus calculated center times for the transits of TOI-5800\,b (Sectors 54, 81 and 92).}
\label{fig:TTV}
\end{figure}

To investigate the presence of TTVs for TOI-5800\,b, we performed a dedicated analysis of the TESS photometric time series using \texttt{juliet}, fitting each transit independently. The resulting mid-transit times and their associated uncertainties were compiled into a dataset of transit epochs and corresponding O–C (Observed minus Calculated) residuals relative to a linear ephemeris. The analysis revealed no statistically significant TTV signal in the observed transits (Fig.\,\ref{fig:TTV}, Table\,\ref{tab:TTVs}). The posterior distributions of the mid-transit times are consistent with white noise dominated deviations, and no apparent periodic structure is evident in the O–C diagram. We conclude that, within the current timing precision and observational baseline, there is no evidence for significant dynamical perturbations in the orbit of TOI-5800\,b detectable via TTVs.

\begin{table}[h!]
    \centering
    \caption{TTV values and uncertainties for TOI-5800\,b.}
    \label{tab:TTVs}
    {\tiny\renewcommand{\arraystretch}{.8}
    \resizebox{!}{.16\paperheight}{%
    \begin{tabular}{cccc}
        \hline\hline \\[-6pt]
        BJD-2457000 & OC (minutes) & $+1\sigma_{\mathrm{OC}}$ & $-1\sigma_{\mathrm{OC}}$ \\
        \hline \\[-6pt]
        2771.7150833 & -3.33 & 2.28 & 2.73 \\
        2774.3429625 & -4.46 & 2.50 & 3.23 \\
        2782.2266001 & 2.33 & 2.23 & 2.99 \\
        2784.8544793 & 15.19 & 2.64 & 2.84 \\
        2787.4823585 & -7.15 & 1.94 & 2.18 \\
        2790.1102377 & -3.54 & 3.21 & 2.50 \\
        2795.3659961 & 4.90 & 2.19 & 2.55 \\
        3507.5212590 & -5.11 & 2.91 & 2.83 \\
        3510.1491382 & -0.37 & 2.37 & 2.30 \\
        3512.7770174 & -6.24 & 3.79 & 3.66 \\
        3518.0327758 & 4.29 & 1.37 & 1.55 \\
        3520.6606550 & 0.45 & 2.06 & 2.02 \\
        3523.2885342 & 0.64 & 2.37 & 3.29 \\
        3525.9164134 & -3.39 & 2.16 & 2.56 \\
        3528.5442926 & -0.88 & 1.87 & 1.54 \\
        3531.1721718 & 0.52 & 2.27 & 1.83 \\
        3804.4716085 & 0.46 & 1.76 & 2.00 \\
        3807.0994877 & -0.31 & 2.34 & 1.85 \\
        3809.7273669 & -5.29 & 4.93 & 3.16 \\
        3812.3552461 & 6.26 & 2.91 & 2.00 \\
        3817.6110045 & -2.27 & 1.76 & 1.59 \\
        3820.2388837 & 0.05 & 1.87 & 2.02 \\
        3822.8667629 & -0.54 & 2.26 & 2.07 \\
        3825.4946421 & 8.68 & 3.57 & 4.74 \\
        3828.1225213 & -0.47 & 1.82 & 1.85 \\
        \hline
    \end{tabular}
    }}
\end{table}

\section{Additional figures and tables}\label{sec:appendix}

\begin{figure}[!h]
\centering
\includegraphics[width=0.4\textwidth]{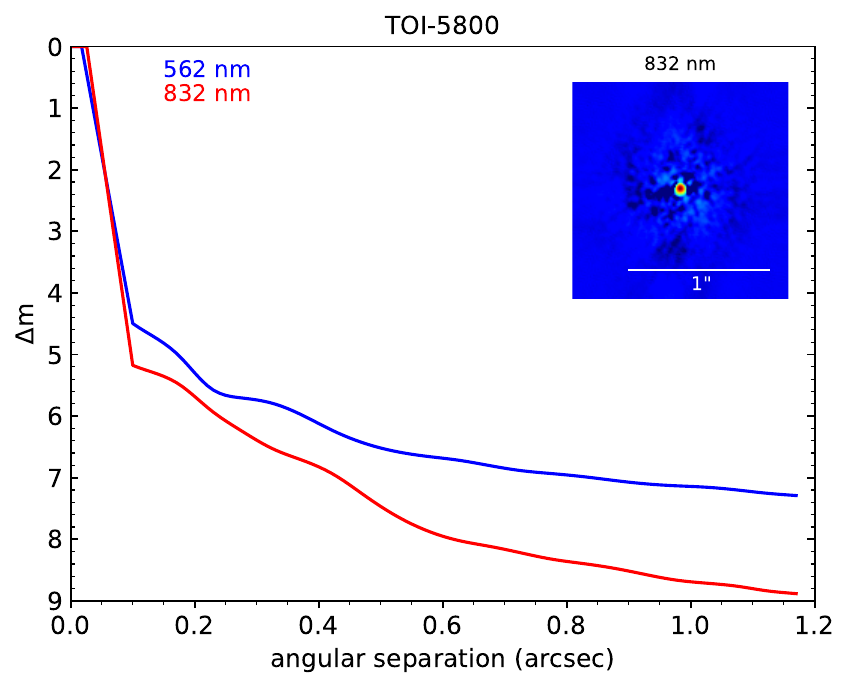}
\caption{The figure shows $5\sigma$ magnitude contrast curves in both filters as a function of the angular separation out to 1.2 arcsec, as taken by Gemini. The inset shows the reconstructed 832 nm image of TOI-5800 with a 1 arcsec scale bar. TOI-5800 was found to have no close companions from the diffraction limit (0.02”) out to 1.2 arcsec to within the contrast levels achieved.}
\label{fig:gemini}
\end{figure}

\begin{figure}[!h]
\centering
\includegraphics[width=0.4\textwidth]{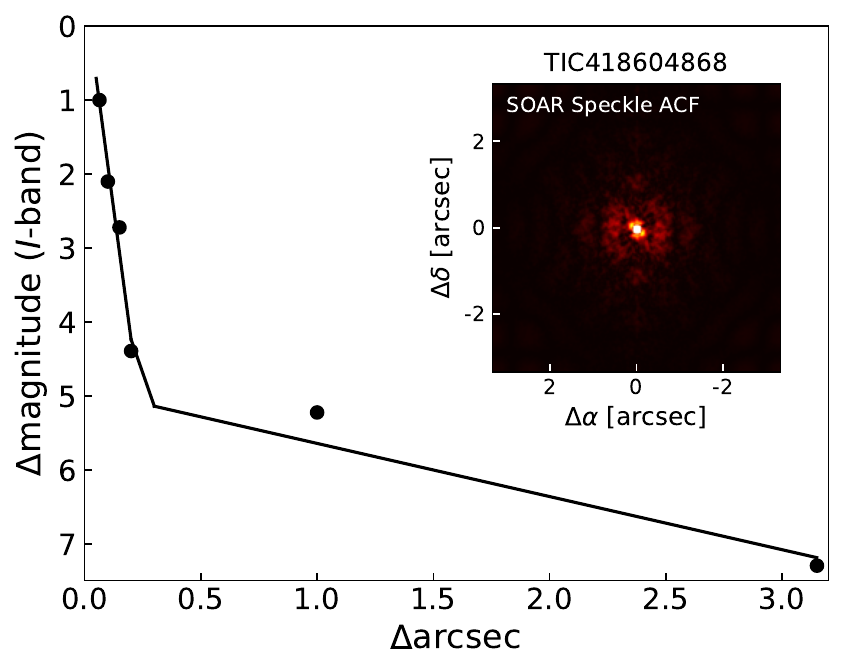}
\caption{The figure shows the $5\sigma$ magnitude contrast curve in Cousins I-band as a function of the angular separation out to 3 arcsec, as taken by SOAR. The inset shows the reconstructed images of TOI-5817 in the same band. No stellar companion to TOI-5817 is seen.}
\label{fig:soar}
\end{figure}

\begin{figure*}[!h]
\centering
\includegraphics[width=0.8\textwidth]{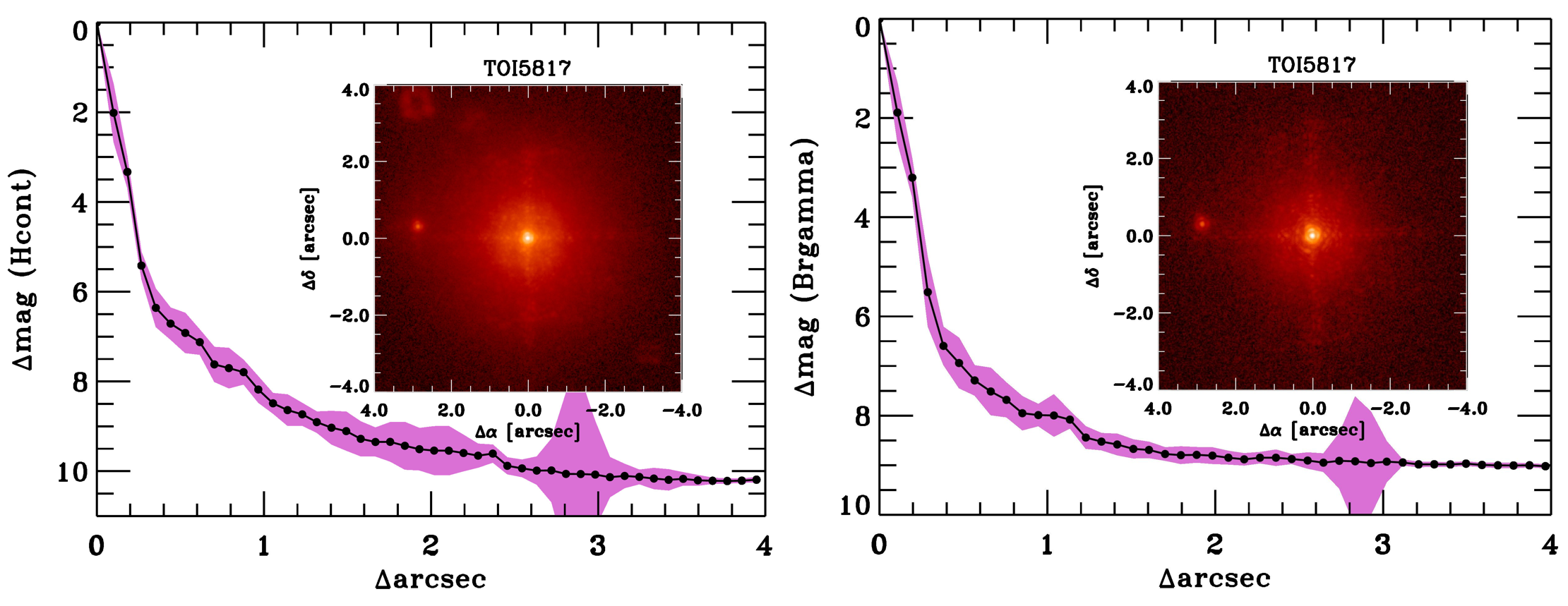}
\caption{The figure shows $5\sigma$ magnitude contrast curves in $H_{cont}$ filter (left) and Br-$\gamma$ filter (right) as a function of the angular separation out to 4 arcsec, as taken by Palomar-AO. The inset shows the reconstructed images of TOI-5817 in the respective filters. A stellar companion of TOI-5817 is clearly spotted, as detailed in Sect.\,\ref{sec:multiplicity}, though no close-in companions were detected in agreement with the other speckle imaging performed for TOI-5817 (Fig.\,\ref{fig:soar}).}
\label{fig:palomar}
\end{figure*}

\begin{figure*}
\centering
\includegraphics[width=0.49\linewidth]{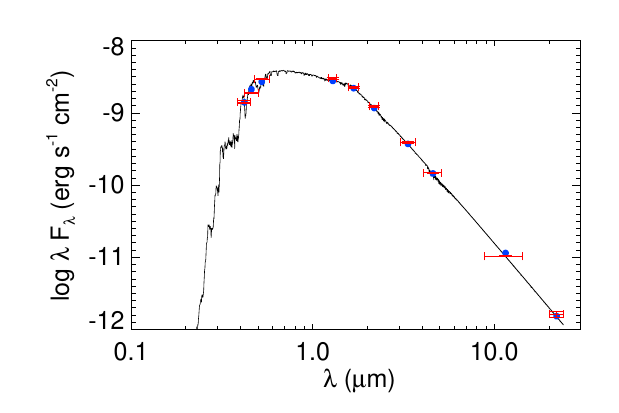}
\includegraphics[width=0.49\linewidth]{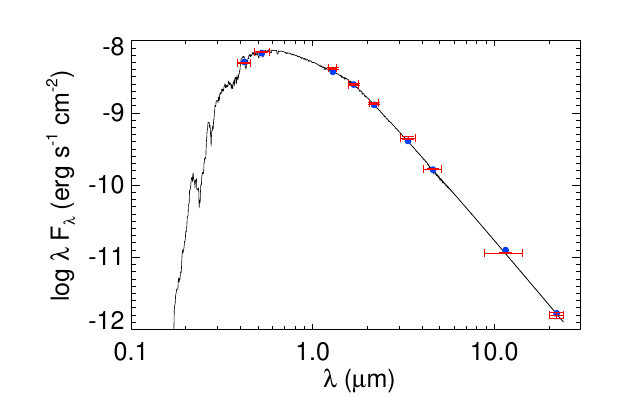}
\caption{Spectral energy distributions of the host stars TOI-5800 (left) and TOI-5817 (right). The broad band measurements from the Tycho, APASS Johnson, 2MASS and WISE magnitudes are shown in red, and the corresponding theoretical values with blue circles. The unaveraged best-fit model is displayed with a black solid line.} 
\label{fig:stellarSED}
\end{figure*}

\begin{figure*}
\centering
\includegraphics[width=0.49\textwidth]{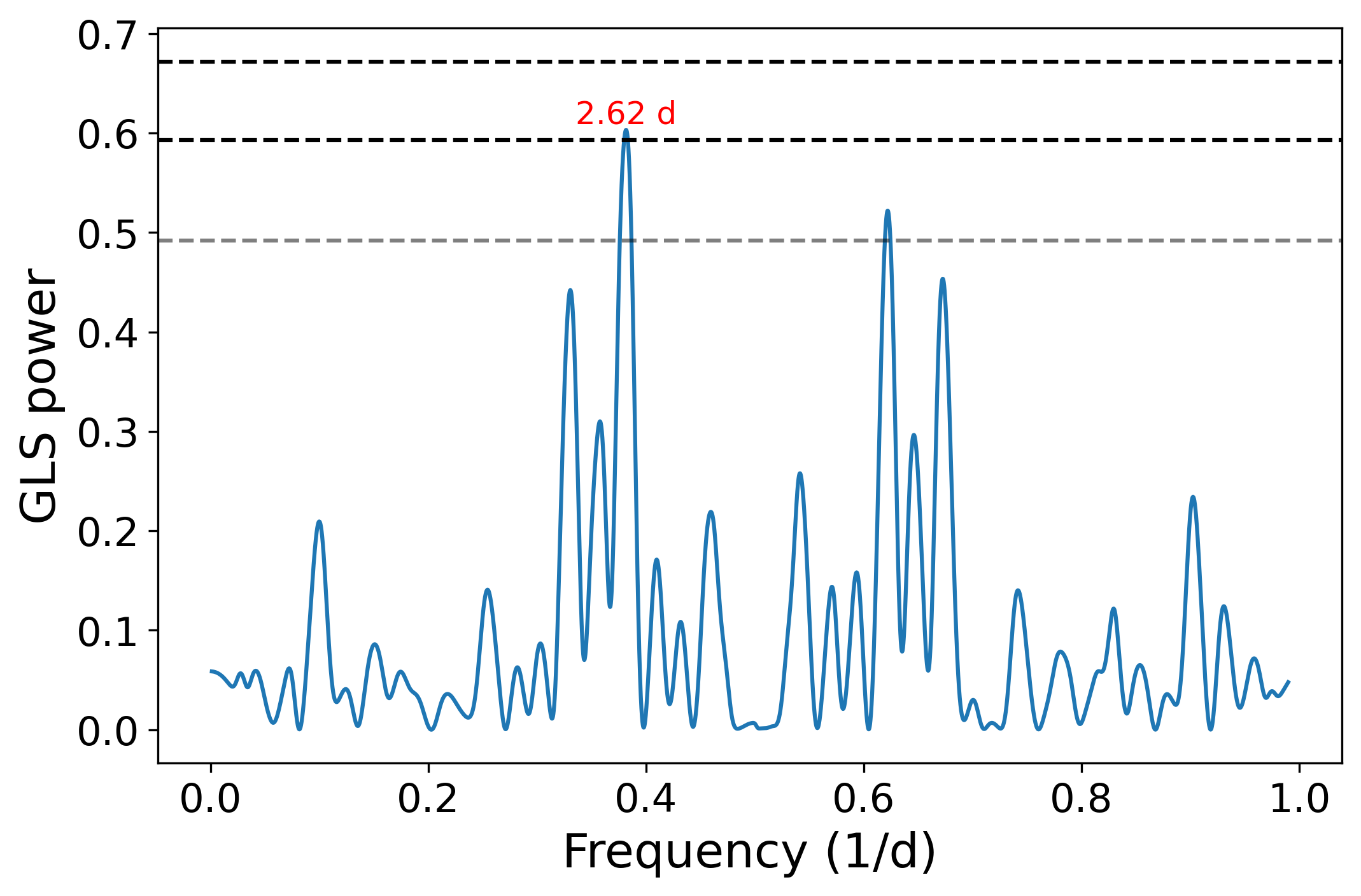}
\includegraphics[width=0.49\textwidth]{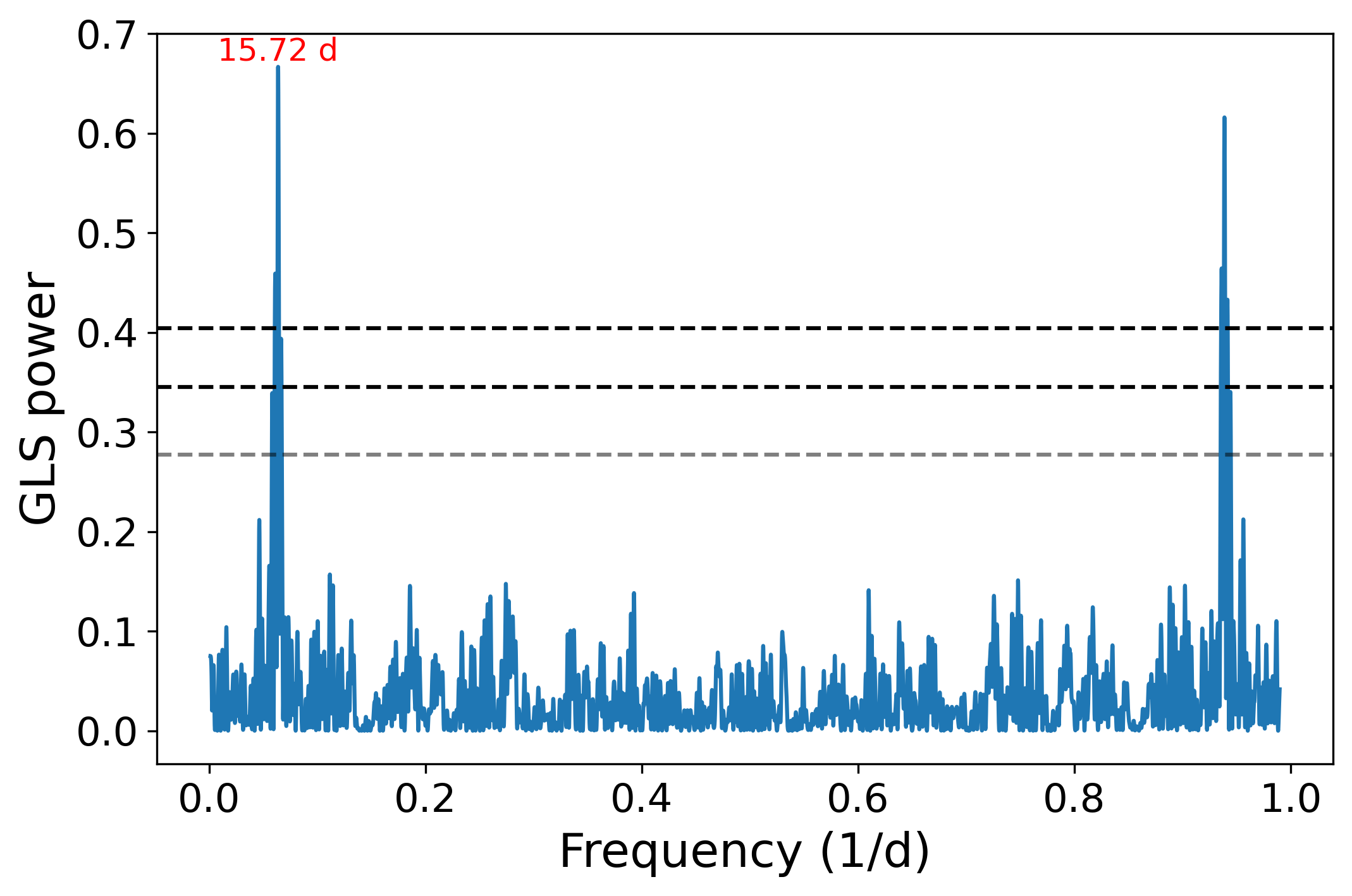}
\caption{GLS periodograms of TOI-5800 (left) and TOI-5817 (right) RVs. The horizontal dotted lines mark, respectively, the $0.1\%$, $1\%$, and $10\%$ false alarm probability levels from top to bottom. In both cases, the second peak falls precisely at the 1\,d alias of the transiting candidate periods (highlighted in red).}
\label{fig:GLS}
\end{figure*}

\begin{figure*}
\centering
\includegraphics[width=0.85\textwidth]{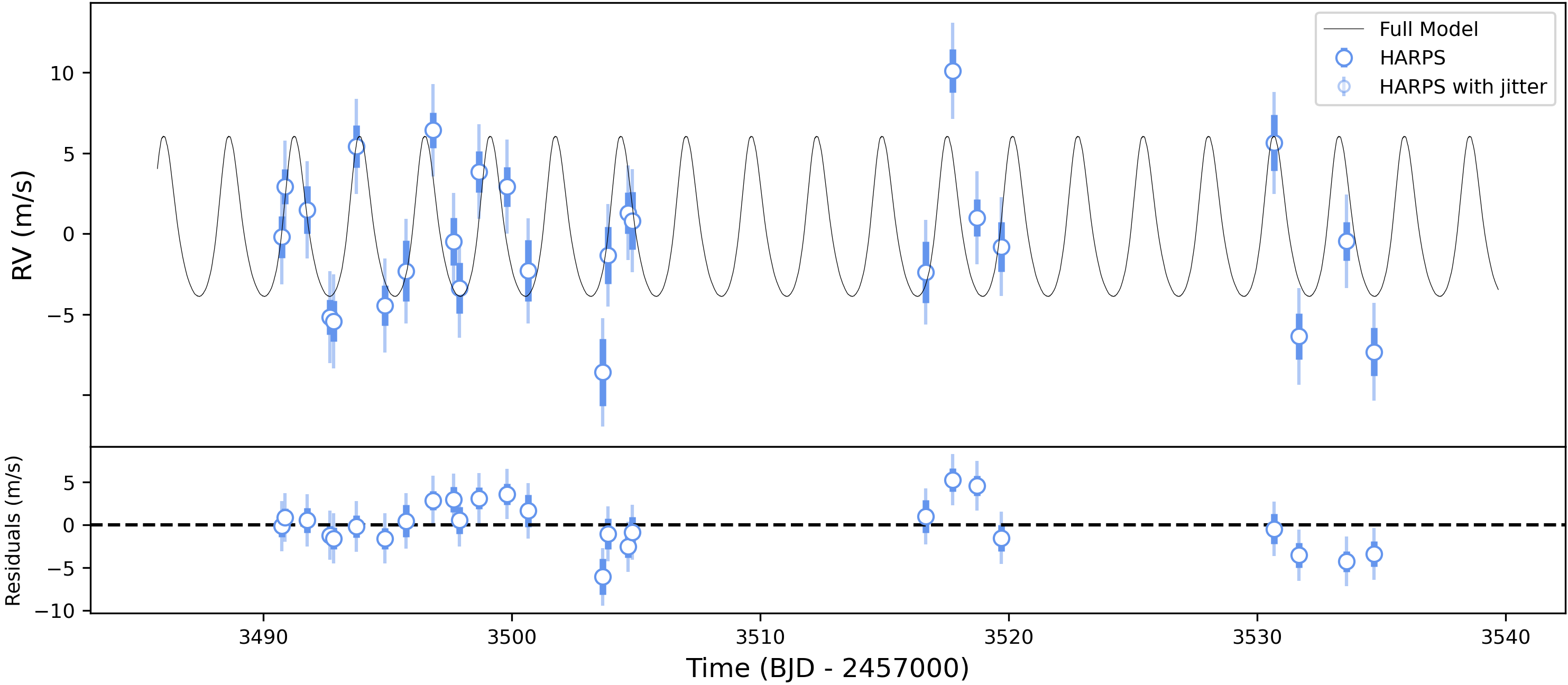}
\includegraphics[width=0.85\textwidth]{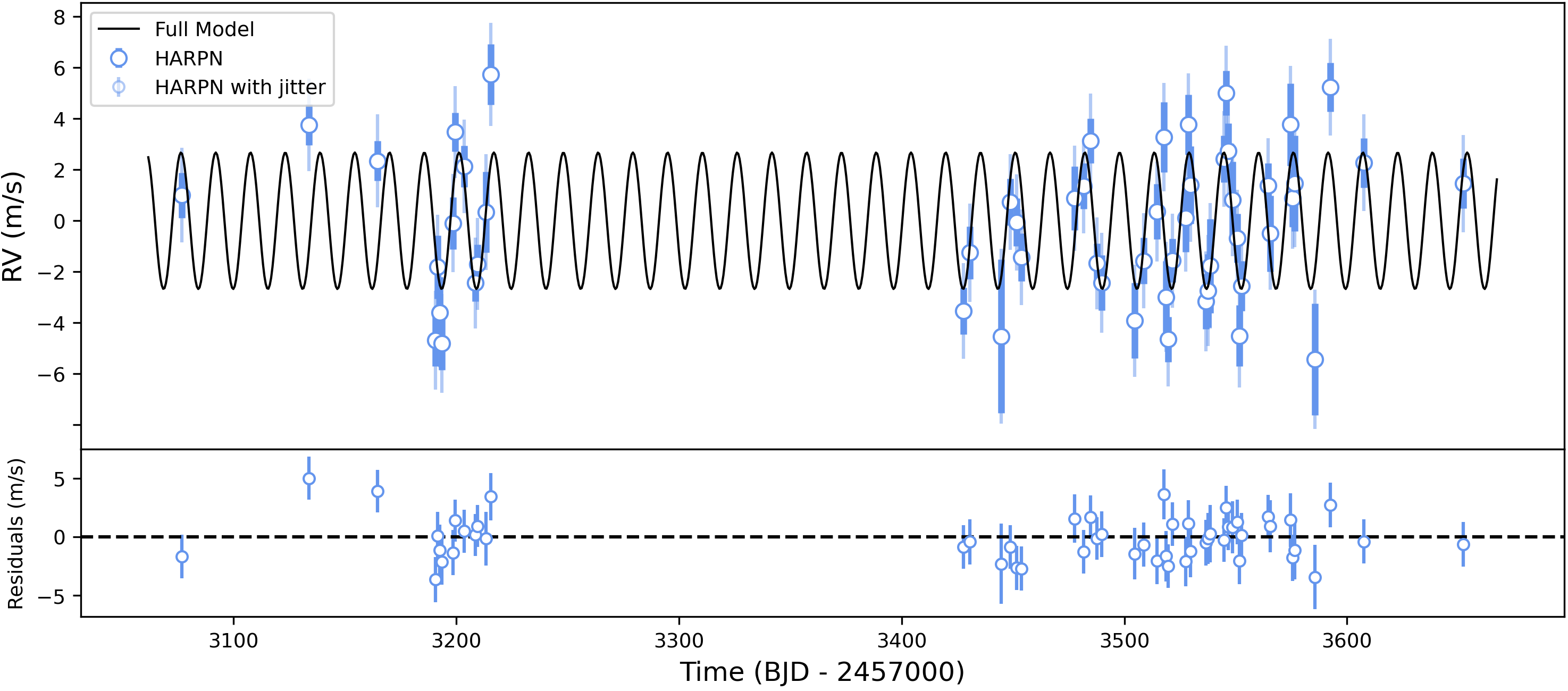}
\caption{HARPS RVs of TOI-5800 (top panel) and HARPS-N RVs of TOI-5817 (bottom panel) along with the best fit models, in black, and their residuals below each panel.}
\label{fig:5800_5817_fullRV}
\end{figure*}

\begin{figure*}
\centering
\includegraphics[width=0.45\textwidth]{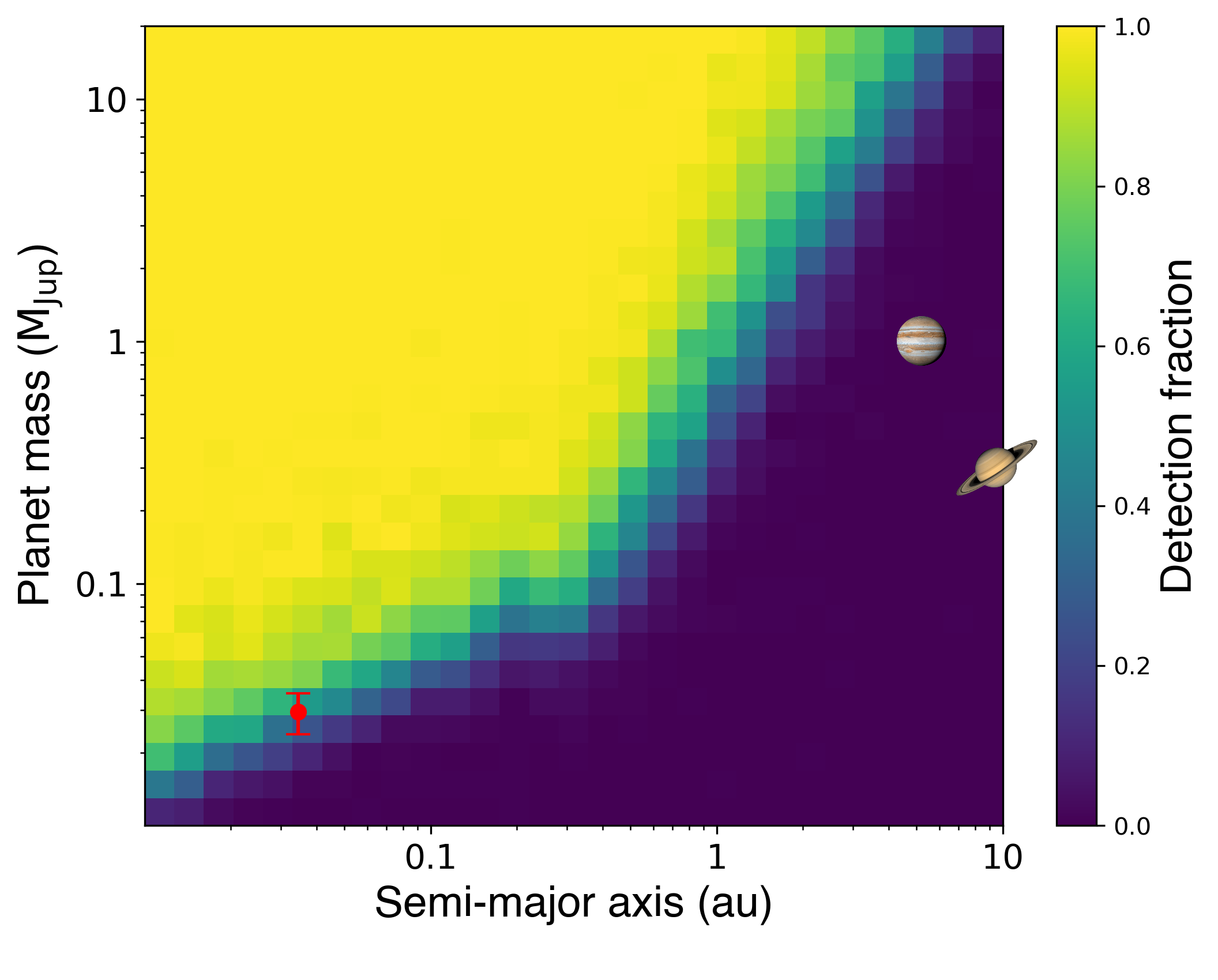}
\includegraphics[width=0.45\textwidth]{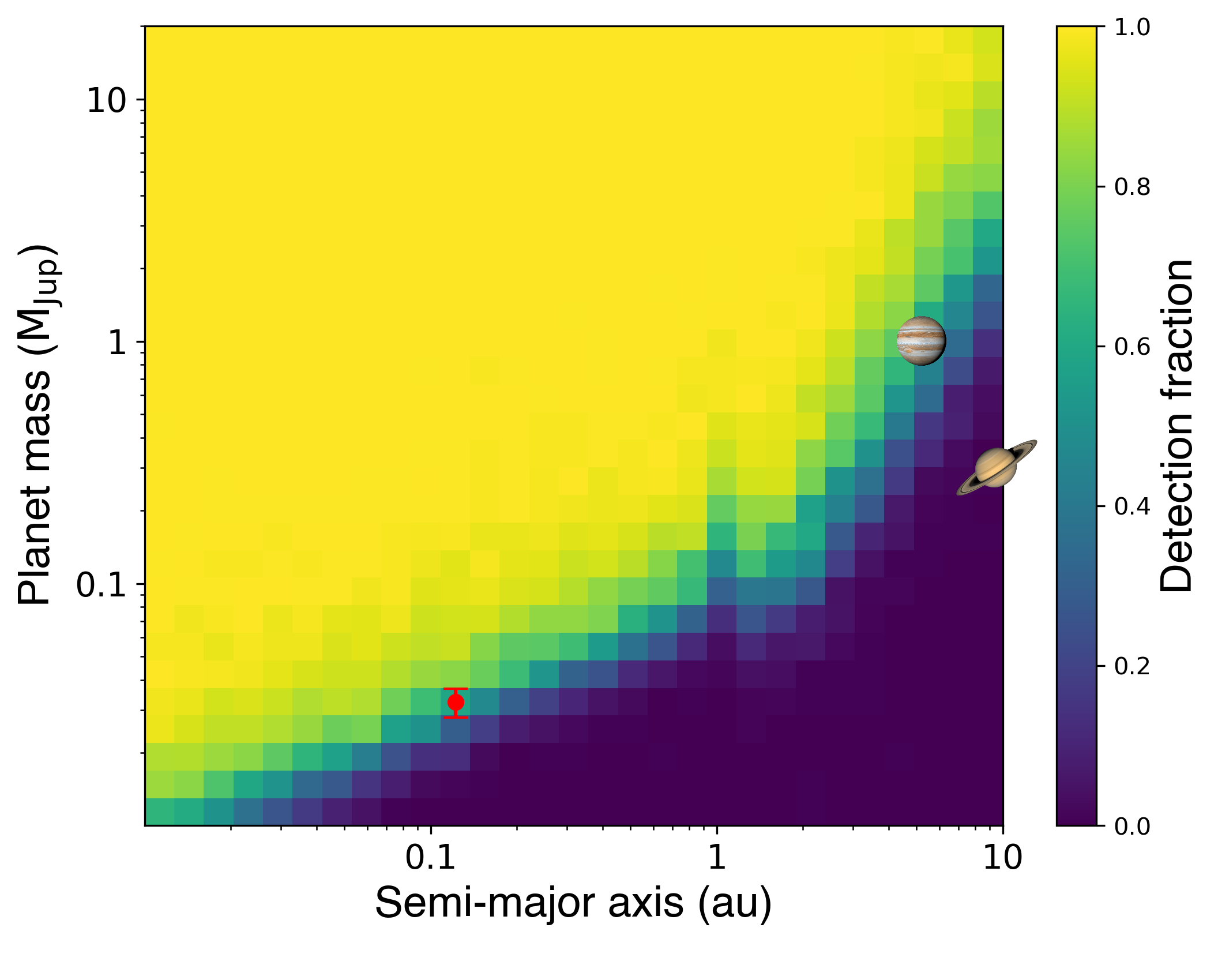}
\caption{HARPS and HARPS-N RV detection maps for TOI-5800 (left) and TOI-5817 (right). The color scale expresses the detection fraction (i.e., the detection probability), while the red circles mark the position of TOI-5800\,b (left) and TOI-5817\,b (right). Jupiter and Saturn are shown for comparison.}
\label{fig:completeness}
\end{figure*}

\begin{figure*}
\centering
\includegraphics[width=0.80\textwidth]{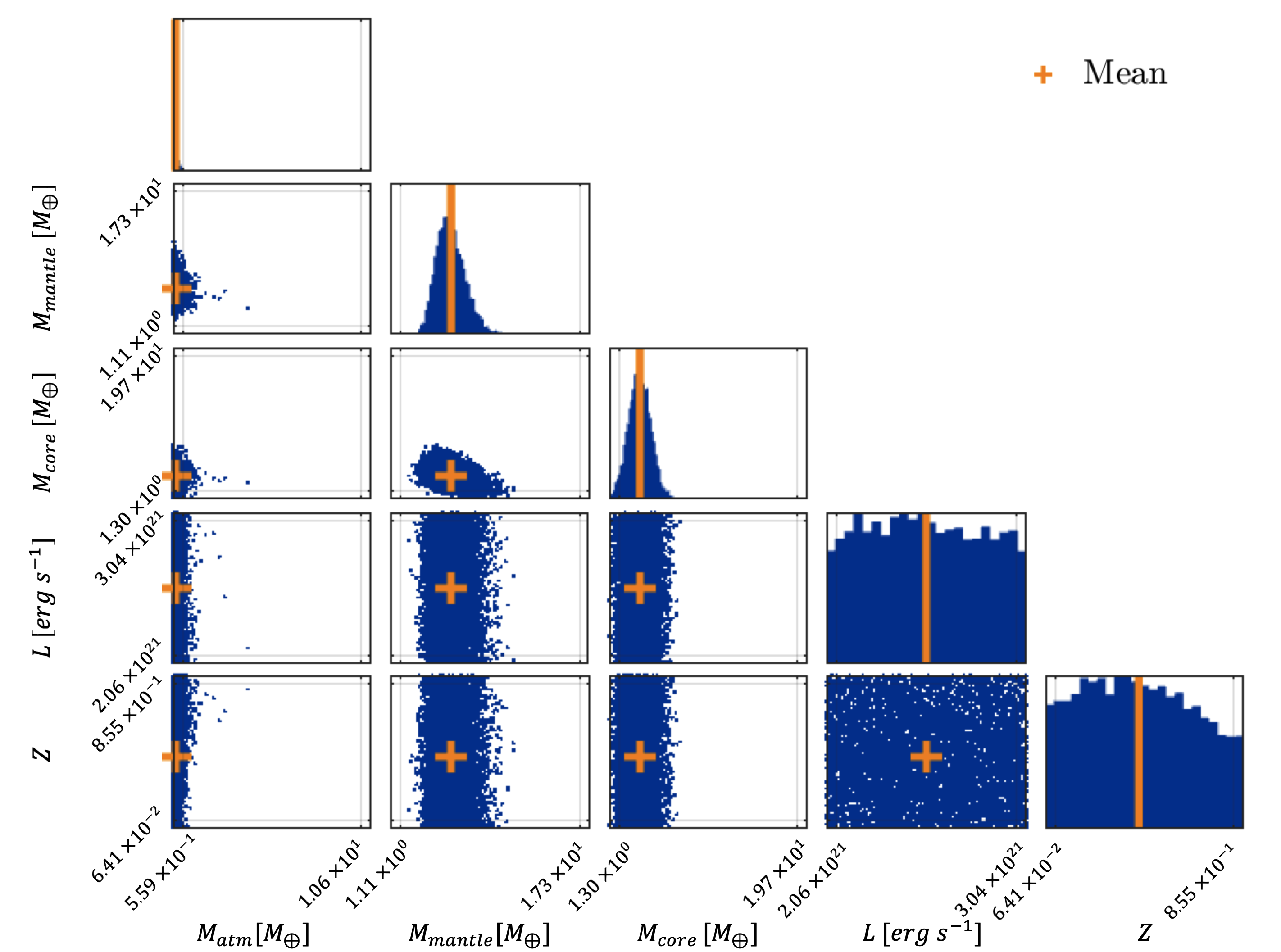}
\includegraphics[width=0.80\textwidth]{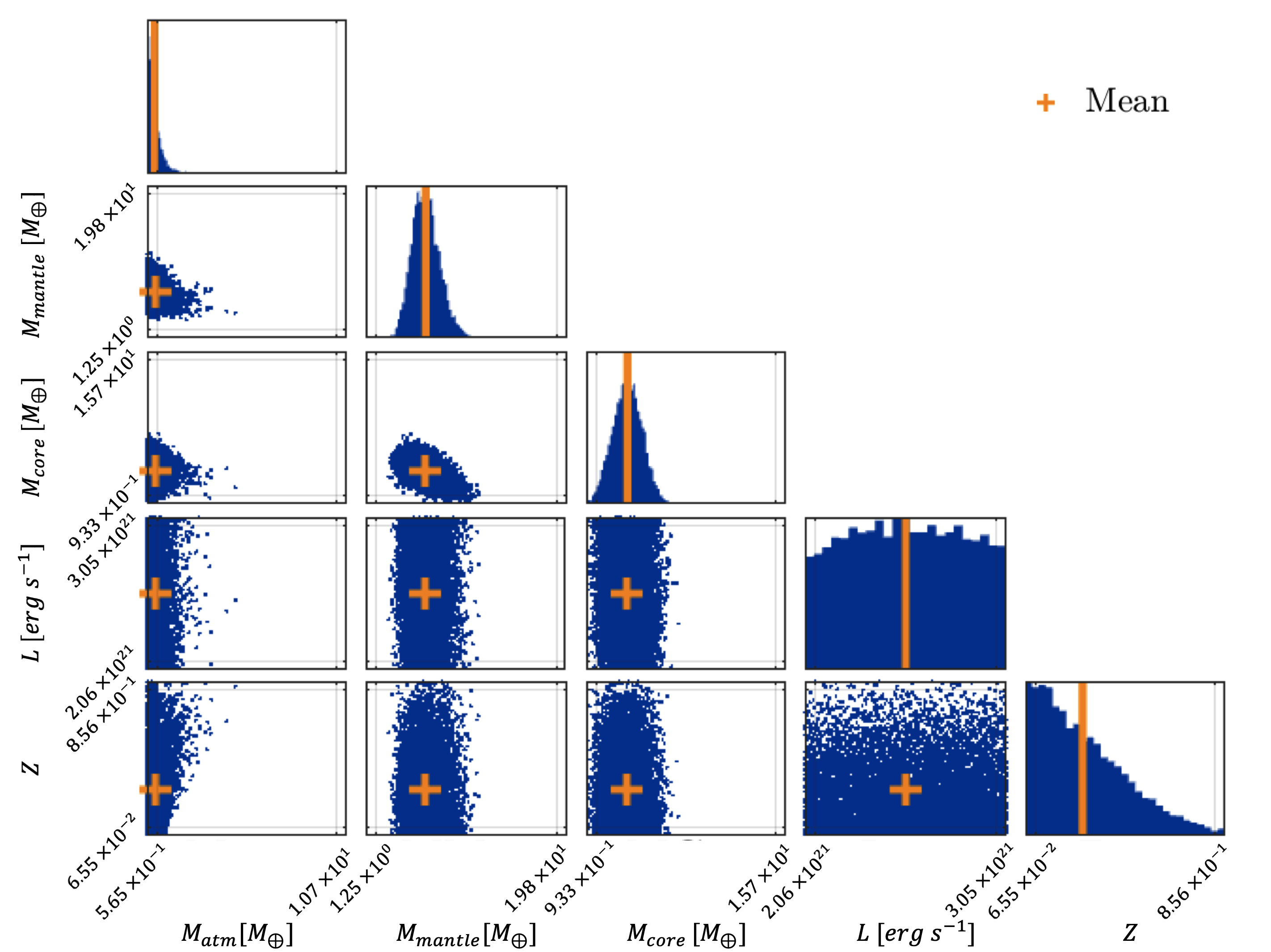}
\caption{Posterior distribution for the internal composition analyses of TOI-5800\,b (top panel) and TOI-5817\,b (bottom panel).}
\label{fig:5800_5817_posterior}
\end{figure*}

\begin{table*}
\centering
\caption[]{TOI-5800 HARPS RV data points and activity indices.}\label{tab:5800_RV}
{\tiny\renewcommand{\arraystretch}{.8}
\resizebox{!}{.12\paperheight}{%
\begin{tabular}{llllllllllll}
    \hline\hline \\[-6pt]
    $\mathrm{BJD_{\textsf{TDB}}}$ & RV & $\pm1\upsigma_{\textsf{RV}}$ & FWHM & $\upsigma_{\textsf{FWHM}}$ & BIS & $\upsigma_{\textsf{BIS}}$ & Contrast & $\upsigma_{\textsf{Cont}}$ & $\log R^{\prime}_{\rm HK}$ & $\upsigma_{\log R^{\prime}_{\rm HK}}$ \\ 
    $-2457000\,[d]$ & $\mathrm{~~[m\,s^{-1}]}$ & $\mathrm{~~[m\,s^{-1}]}$ & & & & & & & & & \rule[-0.8ex]{0pt}{0pt} \\ 
    \hline \\
    3490.7417933 & -2.42  & 1.3  & 6968.2 & 1.21 & -21.0 & 1.21 & 56.49 & 0.1 & -4.890 & 0.005 \\
    3490.8676529 & 0.71   & 1.07 & 6971.6 & 1.59 & -27.0 & 1.59 & 56.56 & 0.1 & -4.865 & 0.005 \\
    3491.7699307 & -0.73  & 1.47 & 6958.6 & 2.02 & -18.4 & 2.02 & 56.59 & 0.1 & -4.866 & 0.006 \\
    3492.6770654 & -7.39  & 1.09 & 6984.5 & 1.67 & -23.2 & 1.67 & 56.13 & 0.1 & -4.858 & 0.005 \\
    3492.8367319 & -7.65  & 1.26 & 6950.2 & 1.79 & -13.4 & 1.79 & 56.64 & 0.1 & -4.831 & 0.005 \\
    3493.7421664 & 3.2    & 1.32 & 6975.0 & 1.25 & -26.8 & 1.25 & 56.37 & 0.1 & -4.910 & 0.006 \\
    3494.8863740 & -6.67  & 1.24 & 6965.9 & 2.20 & -23.9 & 2.20 & 56.42 & 0.1 & -4.903 & 0.007 \\
    3495.7370919 & -4.53  & 1.89 & 6987.8 & 1.13 & -32.5 & 1.13 & 56.24 & 0.1 & -4.814 & 0.009 \\
    3496.8219282 & 4.2    & 1.11 & 6940.2 & 1.47 & -22.5 & 1.47 & 56.74 & 0.1 & -4.847 & 0.005 \\
    3497.6601831 & -2.71  & 1.48 & 6968.4 & 1.55 & -27.1 & 1.55 & 56.61 & 0.1 & -4.914 & 0.006 \\
    3497.8861105 & -5.6   & 1.57 & 6953.0 & 2.81 & -21.0 & 2.81 & 56.53 & 0.1 & -4.900 & 0.007 \\
    3498.6727189 & 1.63   & 1.27 & 6961.3 & 1.39 & -16.1 & 1.39 & 56.41 & 0.1 & -4.888 & 0.005 \\
    3499.8127087 & 0.7    & 1.23 & 6964.5 & 1.30 & -18.8 & 1.30 & 56.40 & 0.1 & -4.911 & 0.006 \\
    3500.6690869 & -4.52  & 1.9  & 6982.2 & 1.34 & -33.7 & 1.34 & 56.28 & 0.1 & -4.907 & 0.009 \\
    3503.6506869 & -10.82 & 2.08 & 6960.6 & 2.33 & -26.3 & 2.33 & 56.39 & 0.1 & -4.873 & 0.008 \\
    3503.8668915 & -3.55  & 1.77 & 6975.0 & 1.84 & -22.2 & 1.84 & 56.48 & 0.1 & -4.876 & 0.008 \\
    3504.6840737 & -0.93  & 1.28 & 6957.5 & 2.39 & -24.3 & 2.39 & 56.49 & 0.1 & -4.864 & 0.006 \\
    3504.8510218 & -1.41  & 1.8  & 6951.6 & 2.17 & -10.7 & 2.17 & 56.59 & 0.1 & -4.996 & 0.011 \\
    3516.6483423 & -4.6   & 1.9  & 6971.7 & 1.15 & -24.9 & 1.15 & 56.44 & 0.1 & -4.895 & 0.008 \\
    3517.7373326 & 7.89   & 1.36 & 6953.3 & 2.44 & -16.0 & 2.44 & 56.64 & 0.1 & -4.855 & 0.005 \\
    3518.7279464 & -1.23  & 1.14 & 6959.5 & 1.45 & -17.6 & 1.45 & 56.50 & 0.1 & -4.856 & 0.005 \\
    3519.6936925 & -3.02  & 1.54 & 6966.2 & 1.20 & -21.9 & 1.20 & 56.60 & 0.1 & -4.882 & 0.007 \\
    3530.6859318 & 3.42   & 1.74 & 6950.9 & 2.01 & -27.0 & 2.01 & 56.65 & 0.1 & -4.869 & 0.009 \\
    3531.6686592 & -8.58  & 1.43 & 6963.4 & 1.12 & -16.2 & 1.12 & 56.54 & 0.1 & -4.895 & 0.008 \\
    3533.5936581 & -2.68  & 1.19 & 6974.6 & 1.53 & -13.3 & 1.53 & 56.27 & 0.1 & -4.915 & 0.006 \\
    3534.7005404 & -9.55  & 1.5  & 6957.6 & 1.47 & -24.7 & 1.47 & 56.59 & 0.1 & -4.894 & 0.006 \rule[-0.8ex]{0pt}{0pt} \\
    \bottomrule
\end{tabular}}}
\end{table*}

\begin{table*}
\centering
\caption[]{TOI-5817 HARPS-N RV data points and activity indices.}\label{tab:5817_RV_HN}
{\tiny\renewcommand{\arraystretch}{.8}
\resizebox{!}{.24\paperheight}{%
\begin{tabular}{llllllllllll}
    \hline\hline \\[-6pt]
    $\mathrm{BJD_{\textsf{TDB}}}$ & RV & $\pm1\upsigma_{\textsf{RV}}$ & FWHM & $\upsigma_{\textsf{FWHM}}$ & BIS & $\upsigma_{\textsf{BIS}}$ & Contrast & $\upsigma_{\textsf{Cont}}$ & $\log R^{\prime}_{\rm HK}$ & $\upsigma_{\log R^{\prime}_{\rm HK}}$ \\ 
    $-2457000\,[d]$ & $\mathrm{~~[m\,s^{-1}]}$ & $\mathrm{~~[m\,s^{-1}]}$ & & & & & & & & & \rule[-0.8ex]{0pt}{0pt} \\ 
    \hline \\
    3076.706816 & -20800.85 & 0.88 & 6.9103 & 0.0018 & -0.0455 & 0.0018 & 53.054 & 0.013 & -4.9467 & 0.0011 \\
    3133.721976 & -20798.09 & 0.80 & 6.9061 & 0.0016 & -0.0427 & 0.0016 & 53.072 & 0.012 & -4.9141 & 0.0009 \\
    3164.626223 & -20799.50 & 0.79 & 6.9108 & 0.0016 & -0.0405 & 0.0016 & 53.068 & 0.012 & -4.9229 & 0.0009 \\
    3190.567237 & -20806.53 & 1.03 & 6.9076 & 0.0021 & -0.0424 & 0.0021 & 53.040 & 0.016 & -4.9269 & 0.0015 \\
    3191.543662 & -20803.66 & 1.24 & 6.9029 & 0.0025 & -0.0436 & 0.0025 & 53.105 & 0.019 & -4.9570 & 0.0022 \\
    3192.540426 & -20805.45 & 1.39 & 6.9160 & 0.0028 & -0.0406 & 0.0028 & 53.096 & 0.021 & -4.9297 & 0.0026 \\
    3193.550663 & -20806.65 & 1.05 & 6.9152 & 0.0021 & -0.0440 & 0.0021 & 53.050 & 0.016 & -4.9262 & 0.0015 \\
    3198.586672 & -20801.94 & 1.02 & 6.9046 & 0.0020 & -0.0435 & 0.0020 & 53.051 & 0.016 & -4.9532 & 0.0015 \\
    3199.592662 & -20798.37 & 0.76 & 6.9085 & 0.0015 & -0.0416 & 0.0015 & 53.027 & 0.012 & -4.9291 & 0.0009 \\
    3203.524407 & -20799.72 & 0.82 & 6.9073 & 0.0016 & -0.0420 & 0.0016 & 53.057 & 0.013 & -4.9374 & 0.0010 \\
    3208.496337 & -20804.29 & 0.73 & 6.9089 & 0.0015 & -0.0432 & 0.0015 & 53.029 & 0.011 & -4.9217 & 0.0008 \\
    3209.550008 & -20803.55 & 0.76 & 6.9110 & 0.0015 & -0.0430 & 0.0015 & 53.010 & 0.012 & -4.9246 & 0.0009 \\
    3213.437564 & -20801.51 & 1.58 & 6.9112 & 0.0032 & -0.0458 & 0.0032 & 53.136 & 0.024 & -4.9128 & 0.0030 \\
    3215.521424 & -20796.12 & 1.19 & 6.9110 & 0.0024 & -0.0419 & 0.0024 & 53.028 & 0.018 & -4.9514 & 0.0020 \\
    3427.731262 & -20805.38 & 0.92 & 6.9072 & 0.0018 & -0.0425 & 0.0018 & 52.994 & 0.014 & -4.9671 & 0.0012 \\
    3430.740254 & -20803.10 & 1.03 & 6.9053 & 0.0021 & -0.0435 & 0.0021 & 52.973 & 0.016 & -4.9524 & 0.0014 \\
    3444.680312 & -20806.38 & 3.01 & 6.9128 & 0.0060 & -0.0519 & 0.0060 & 53.215 & 0.046 & -5.0630 & 0.0120 \\
    3448.717104 & -20801.11 & 0.92 & 6.9085 & 0.0018 & -0.0433 & 0.0018 & 53.031 & 0.014 & -4.9572 & 0.0012 \\
    3451.709225 & -20801.91 & 0.94 & 6.9067 & 0.0019 & -0.0412 & 0.0019 & 53.016 & 0.014 & -4.9534 & 0.0012 \\
    3453.715804 & -20803.28 & 0.93 & 6.9084 & 0.0019 & -0.0453 & 0.0019 & 53.021 & 0.014 & -4.9462 & 0.0012 \\
    3477.720297 & -20800.97 & 1.25 & 6.9096 & 0.0025 & -0.0428 & 0.0025 & 53.063 & 0.019 & -4.9580 & 0.0021 \\
    3481.725265 & -20800.49 & 0.89 & 6.9059 & 0.0018 & -0.0432 & 0.0018 & 53.007 & 0.014 & -4.9534 & 0.0011 \\
    3484.726864 & -20798.72 & 0.87 & 6.9052 & 0.0017 & -0.0452 & 0.0017 & 52.992 & 0.013 & -4.9538 & 0.0011 \\
    3487.67268  & -20803.52 & 0.78 & 6.9014 & 0.0016 & -0.0428 & 0.0016 & 53.018 & 0.012 & -4.9471 & 0.0009 \\
    3489.73175  & -20804.28 & 1.08 & 6.9070 & 0.0022 & -0.0407 & 0.0022 & 53.006 & 0.017 & -4.9702 & 0.0017 \\
    3504.740446 & -20805.77 & 1.48 & 6.9026 & 0.0030 & -0.0452 & 0.0030 & 52.893 & 0.023 & -4.9530 & 0.0027 \\
    3508.71542  & -20803.42 & 0.91 & 6.9096 & 0.0018 & -0.0455 & 0.0018 & 53.030 & 0.014 & -4.9410 & 0.0011 \\
    3514.596052 & -20801.49 & 1.08 & 6.9101 & 0.0022 & -0.0448 & 0.0022 & 53.030 & 0.017 & -4.9411 & 0.0016 \\
    3517.72543  & -20798.57 & 1.37 & 6.9010 & 0.0027 & -0.0429 & 0.0027 & 53.085 & 0.021 & -4.9633 & 0.0025 \\
    3518.688354 & -20804.84 & 1.41 & 6.9119 & 0.0028 & -0.0417 & 0.0028 & 53.095 & 0.022 & -4.9735 & 0.0026 \\
    3519.68936  & -20806.50 & 0.88 & 6.9104 & 0.0018 & -0.0453 & 0.0018 & 53.034 & 0.013 & -4.9468 & 0.0011 \\
    3521.65392  & -20803.41 & 0.86 & 6.9087 & 0.0017 & -0.0445 & 0.0017 & 53.016 & 0.013 & -4.9510 & 0.0010 \\
    3527.583849 & -20801.74 & 1.33 & 6.9096 & 0.0027 & -0.0430 & 0.0027 & 53.077 & 0.020 & -4.9629 & 0.0023 \\
    3528.653431 & -20798.08 & 1.18 & 6.9095 & 0.0024 & -0.0424 & 0.0024 & 53.086 & 0.018 & -4.9555 & 0.0018 \\
    3529.62687  & -20800.45 & 1.52 & 6.9113 & 0.0030 & -0.0389 & 0.0030 & 53.079 & 0.023 & -4.9738 & 0.0031 \\
    3536.657847 & -20805.01 & 1.09 & 6.9056 & 0.0022 & -0.0469 & 0.0022 & 53.064 & 0.017 & -4.9389 & 0.0016 \\
    3537.483723 & -20804.59 & 1.44 & 6.9064 & 0.0029 & -0.0455 & 0.0029 & 53.153 & 0.022 & -4.9622 & 0.0027 \\
    3538.642139 & -20803.62 & 1.84 & 6.9017 & 0.0037 & -0.0479 & 0.0037 & 53.175 & 0.028 & -5.0122 & 0.0048 \\
    3544.632714 & -20799.43 & 0.91 & 6.9077 & 0.0018 & -0.0425 & 0.0018 & 53.011 & 0.014 & -4.9318 & 0.0011 \\
    3545.633157 & -20796.84 & 0.88 & 6.9062 & 0.0018 & -0.0381 & 0.0018 & 53.022 & 0.014 & -4.9473 & 0.0011 \\
    3546.650825 & -20799.11 & 1.08 & 6.9115 & 0.0022 & -0.0409 & 0.0022 & 53.041 & 0.017 & -4.9366 & 0.0016 \\
    3548.62045  & -20801.03 & 1.50 & 6.9159 & 0.0030 & -0.0452 & 0.0030 & 53.089 & 0.023 & -4.9569 & 0.0030 \\
    3550.640185 & -20802.53 & 0.95 & 6.9092 & 0.0019 & -0.0447 & 0.0019 & 53.038 & 0.015 & -4.9339 & 0.0012 \\
    3551.561757 & -20806.36 & 1.20 & 6.9131 & 0.0024 & -0.0444 & 0.0024 & 53.061 & 0.018 & -4.9477 & 0.0020 \\
    3552.552417 & -20804.40 & 0.98 & 6.9094 & 0.0020 & -0.0437 & 0.0020 & 53.043 & 0.015 & -4.9573 & 0.0014 \\
    3564.528943 & -20800.47 & 0.88 & 6.9082 & 0.0018 & -0.0450 & 0.0018 & 53.034 & 0.014 & -4.9500 & 0.0011 \\
    3565.586293 & -20802.35 & 1.49 & 6.9065 & 0.0030 & -0.0450 & 0.0030 & 53.132 & 0.023 & -4.9665 & 0.0030 \\
    3574.61827  & -20798.08 & 1.61 & 6.9102 & 0.0032 & -0.0419 & 0.0032 & 53.091 & 0.025 & -4.9651 & 0.0037 \\
    3575.602886 & -20800.97 & 1.12 & 6.9140 & 0.0022 & -0.0457 & 0.0022 & 53.081 & 0.017 & -4.9457 & 0.0018 \\
    3576.609752 & -20800.39 & 1.88 & 6.9250 & 0.0038 & -0.0392 & 0.0038 & 53.129 & 0.029 & -4.9914 & 0.0051 \\
    3585.519477 & -20807.28 & 2.19 & 6.9102 & 0.0044 & -0.0437 & 0.0044 & 53.188 & 0.034 & -5.0390 & 0.0069 \\
    3592.453168 & -20796.62 & 0.96 & 6.9115 & 0.0019 & -0.0419 & 0.0019 & 53.044 & 0.015 & -4.9440 & 0.0013 \\
    3607.465504 & -20799.58 & 0.97 & 6.9142 & 0.0019 & -0.0449 & 0.0019 & 53.047 & 0.015 & -4.9270 & 0.0013 \\
    3652.299573 & -20800.39 & 0.98 & 6.9060 & 0.0020 & -0.0443 & 0.0020 & 53.062 & 0.015 & -4.9434 & 0.0013 \rule[-0.8ex]{0pt}{0pt} \\
    \bottomrule
\end{tabular}}}
\end{table*}

\begin{table*}[t]
\caption{TOI-5817 SOPHIE and SOPHIE+ RVs, separated by an horizontal line.}\label{tab:5817_RV_SOPHIE}
\begin{center}
\resizebox{!}{.17\paperheight}{%
\begin{tabular}{ccccc}
\hline
BJD$_{\rm UTC}$ -2\,457\,000 & RV (km/s) & $\pm$$1\,\sigma$ (km/s) & exp. (sec) & SNR per pix at 550\,nm \\
\hline
-2439.47494 & -20.8157 & 0.0050 & 718.3 & 45.9    \\   
-2441.49799 & -20.8417 & 0.0049 & 244.3 & 49.7    \\   
-2448.46677 & -20.8341 & 0.0049 & 439.3 & 47.4    \\   
-2467.35902 & -20.8171 & 0.0048 & 335.2 & 51.0    \\   
-2468.32443 & -20.8062 & 0.0049 & 277.8 & 55.2    \\   
-2476.40512 & -20.8461 & 0.0049 & 236.2 & 49.4    \\   
-2477.45085 & -20.8312 & 0.0049 & 523.6 & 50.6    \\   
-2482.43912 & -20.8024 & 0.0105 & 904.6 & 15.1    \\   
-2483.36633 & -20.8339 & 0.0050 & 906.0 & 44.2    \\   
-2483.38011 & -20.8301 & 0.0057 & 900.0 & 32.9    \\   
-2483.38821 & -20.8208 & 0.0049 & 347.7 & 48.0    \\   
\hline
3523.48359 & -20.8312 & 0.0033 & 216.0  & 39.6   \\   
3524.47241 & -20.8271 & 0.0035 & 216.0  & 38.2   \\   
3525.55661 & -20.8117 & 0.0043 & 261.9  & 31.0   \\   
3527.43330 & -20.8195 & 0.0031 & 216.0  & 42.9   \\   
3529.44324 & -20.8169 & 0.0027 & 422.2  & 50.2   \\   
3534.58391 & -20.8177 & 0.0033 & 216.0  & 40.3   \\   
3539.42011 & -20.8183 & 0.0027 & 748.9  & 48.9   \\   
3541.44781 & -20.8154 & 0.0027 & 829.4  & 49.2   \\   
3543.42374 & -20.8120 & 0.0027 & 522.6  & 49.9   \\   
3545.46071 & -20.8239 & 0.0027 & 298.1  & 50.1   \\   
3547.39263 & -20.8235 & 0.0026 & 486.9  & 51.2   \\   
3548.51121 & -20.8187 & 0.0030 & 419.0  & 43.8   \\   
3550.45610 & -20.8212 & 0.0027 & 307.2  & 50.0   \\   
3552.49671 & -20.8246 & 0.0029 & 444.6  & 45.3   \\   
3556.51965 & -20.8270 & 0.0030 & 402.9  & 46.5   \\   
3559.51977 & -20.8028 & 0.0046 & 302.7  & 29.2   \\   
3560.43748 & -20.8227 & 0.0032 & 216.0  & 42.0   \\   
3563.34306 & -20.8115 & 0.0027 & 811.4  & 49.7   \\   
3566.36975 & -20.8188 & 0.0028 & 1863.5 & 48.6   \\   
3575.38454 & -20.7923 & 0.0034 & 1648.2 & 39.6   \\   
3578.34476 & -20.8195 & 0.0027 & 408.4  & 50.1   \\   
3582.42445 & -20.8269 & 0.0027 & 884.1  & 49.7   \\   
3603.31212 & -20.8178 & 0.0027 & 399.4  & 49.4   \\   
3606.39841 & -20.8352 & 0.0033 & 651.1  & 40.7   \\   
3619.36656 & -20.8171 & 0.0026 & 478.5  & 51.3   \\   
3622.23691 & -20.8154 & 0.0026 & 375.6  & 50.9   \\   
3625.29103 & -20.8183 & 0.0030 & 216.0  & 44.8   \\   
3643.22459 & -20.8086 & 0.0039 & 216.0  & 33.8   \\   
3645.25673 & -20.8257 & 0.0027 & 405.8  & 50.4   \\   
\hline
\end{tabular}}
\end{center}
\end{table*}

\begin{table*}
\centering
\caption{Priors utilized for the parameters of the preferred joint model fit of Sect.~\ref{sec:joint_analysis}.}\label{tab:priors}
\renewcommand{\arraystretch}{1.2}\begin{tabular}{lllc}
    \hline\hline
    \multicolumn{2}{l}{Parameter} & Prior distribution & Prior distribution \\
    & & TOI-5800 & TOI-5817 \\
    \hline
    \multicolumn{2}{l}{Keplerian Parameters:} & \rule{0pt}{2.6ex} \\ 
    $\rho_{\star}$ & $\mathrm{[g\,cm^{-3}]}$ & $\mathcal{N}(2.41,\:0.21)$ & $\mathcal{N}(0.47,\:0.05)$ \\
    $T_{0,b}$ & $\mathrm{[BJD_{TDB}-2459000]}$ & $\mathcal{N}(771.71,\:0.01)$ & $\mathcal{N}(799.30,\:0.01)$ \\
    $P_b$ & $\mathrm{[d]}$ & $\mathcal{N}(2.628,\:0.001)$ & $\mathcal{N}(15.61,\:0.01)$ \\
    $e_b^*$ &  & $0$ & $0$ \\
    $\omega_b^*$ &  & $90$ & $90$ \\
    
    \multicolumn{2}{l}{Transit Parameters:} & \rule{0pt}{2.6ex} \\ 
    $R_{\rm p}/R_{\rm \star}$ & & $\mathcal{U}(0.0,\:1.0)$ & $\mathcal{U}(0.0,\:1.0)$ \\
    $D$ & & $1.0$ & $1.0$ \\
    $q_1$ (TESS) & & $\mathcal{U}(0,\:1)$ & $\mathcal{U}(0,\:1)$ \\
    $q_2$ (TESS) & & $\mathcal{U}(0,\:1)$ & $\mathcal{U}(0,\:1)$ \\

    \multicolumn{3}{l}{Light curve GP Hyperparameters:} & \rule{0pt}{2.6ex} \\ 
    $\sigma_{\textsf{TESS}}$ & $\mathrm{[ppt]}$ & $\mathcal{L}(10^{-5},\:1)$ & $\mathcal{L}(10^{-5},\:1)$ \\
    $\rho_{\textsf{TESS}}$ & $\mathrm{[d]}$ & $\mathcal{L}(10^{-1},\:100)$ & $\mathcal{L}(10^{-1},\:100)$ \\
    
    \multicolumn{2}{l}{\footnotesize{RV parameters:}} \rule{0pt}{2.6ex} \\
    $K_b$ & $\mathrm{[m\,s^{-1}]}$   & $\mathcal{U}(0,\:20)$ & $\mathcal{U}(0,\:10)$ \\
    $\overline{\mu}_{\textsf{HARPS}}$ & $\mathrm{[m\,s^{-1}]}$ & $\mathcal{U}(-20,\:20)$ & -- \\
    $\sigma_{\textsf{HARPS}}$ & $\mathrm{[m\,s^{-1}]}$ & $\mathcal{U}(0,\:10)$ & -- \\
    $\overline{\mu}_{\textsf{HARPS-N}}$ & $\mathrm{[m\,s^{-1}]}$ & -- & $\mathcal{U}(-20852,\:-20752)$ \\
    $\sigma_{\textsf{HARPS-N}}$ & $\mathrm{[m\,s^{-1}]}$ & -- & $\mathcal{U}(0,\:10)$ \\
    
    \bottomrule
\end{tabular}
\tablefoot{$\mathcal{U}(a,\:b)$ indicates a uniform distribution between $a$ and $b$, $\mathcal{L}(a,\:b)$ a log-normal distribution, and $\mathcal{N}(a,\:b)$ a normal distribution, where $a$ and $b$ are the mean and standard deviation. $^{(*)}$~In the case of non-null eccentricity, the priors are set as follows: $(\sqrt{e}\,\sin\omega, \sqrt{e}\,\cos\omega)$ in $\mathcal{U}(-1.0,\:1.0)$.} 
\end{table*}

\begin{table*}
\centering
\caption{Prior parameter distribution for the calculation of the internal compositions of TOI-5800\,b and TOI-5817\,b. The parameters $Z$, $L$, and $M_\mathrm{atm}$ are only used in the case of uniformly mixed atmosphere, while the parameter $M_\mathrm{water,\, bulk}$, referring to the bulk water content of a planet, is only used in the pure water case.}\label{tab:interiorprior}
\renewcommand{\arraystretch}{1.2}
\resizebox{0.5\hsize}{!}{%
\begin{tabular}{lcc}
    \toprule
    Parameter & Prior range & Distribution \\
    \midrule
    Fe/Si$_\mathrm{mantle}$\dotfill & $0$--Fe/Si$_\mathrm{star}$ & uniform \\
    Mg/Si$_\mathrm{mantle}$\dotfill & Mg/Si$_\mathrm{star}$ & Gaussian \\
    $M_\mathrm{core}/(M_\mathrm{core}+M_\mathrm{mantle})$\dotfill & $0.1$--$0.9$ & uniform \\[2pt]
    \multicolumn{1}{l}{\large{H$_2$-He-H$_2$O atmosphere}} \\[2pt]
    $Z$\dotfill & $0$--$1$ & uniform \\
    $L$ ($\mathrm{erg}\,\mathrm{s}^{-1}$)\dotfill & $(2.0$--$3.1) \times 10^{21}$ & log-uniform \\
    $M_\mathrm{atm}/M_\mathrm{p}$\dotfill & $0.001$--$0.1$ & log-uniform \\[2pt]
    \multicolumn{1}{l}{\large{Pure water case}}\\[2pt]
    $M_\mathrm{water,\, bulk}/M_\mathrm{p}$\dotfill & $0.03$--$0.5$ & uniform \\
    \bottomrule
\end{tabular}
}
\end{table*}

\end{document}